\documentstyle[12pt]{ihep}
\begin{document}
\makeatletter
\def\@cite#1#2{{#1#2}}
\def\refname{References} 
\def\thebibliography#1{\section*{\refname\@mkboth
 {\uppercase{\refname}}{\uppercase{\refname}}}\list
 {\arabic{enumi}.}{\settowidth\labelwidth{[#1]}\leftmargin\labelwidth
 \advance\leftmargin\labelsep
 \usecounter{enumi}}
 \def\newblock{\hskip .11em plus .33em minus .07em}
 \sloppy\clubpenalty4000\widowpenalty4000
 \sfcode`\.=1000\relax}
\let\endthebibliography=\endlist
\makeatother

\setlength{\unitlength}{0.7mm}\thicklines
\tolerance=2000
\hbadness=2000

\begin{titlepage}
\prepnum{94-81}{}
\author{S.S.Gershtein, V.V.Kiselev, A.K.Likhoded, A.V.Tkabladze}
\title{Physics of $B_c$ mesons}
\end{titlepage}
\begin{abstractpage}[539.1]
\engabs{Gershtein S.S., Kiselev V.V., Likhoded A.K., Tkabladze A.V.}
{Physics of $B_c$ mesons}
In the framework of potential models for heavy quarkonium the mass spectrum
for the system ($\bar b c$) is considered. Spin-dependent splittings,
taking into account a change of a constant for effective coulomb interaction
between the quarks, and widths of radiative transitions between the
($\bar b c$) levels are calculated. In the framework of QCD sum rules,
masses of the lightest vector $B_c^*$ and pseudoscalar $B_c$ states are
estimated, scaling relation for leptonic constants of heavy quarkonia is
derived, and the leptonic constant $f_{B_c}$ is evaluated. The $B_c$ decays
are considered in the framework of both the potential models and the QCD
sum rules, where the significance of Coulomb-like corrections is shown.
The relations, following from the approximate spin symmetry
for the heavy quarks in the heavy quarkonium, are analysed for the form
factors of the semileptonic weak exclusive decays of $B_c$. The $B_c$ lifetime
is evaluated with the account of the corrections to the spectator mechanism of
the decay, because of the quark binding into the meson. The total and
differential cross sections of the $B_c$ production in different interactions
are calculated. The analytic expressions for the fragmentational production
cross sections of $B_c$  are derived. The possibility of the practical $B_c$
search in the current and planning experiments at electron-positron
and hadron colliders is analysed.

\end{abstractpage}

\section{Introduction}

A complete picture for both precise tests of the Standard Model  [\cite{i1}]
and a search of effects from new physics supposes a direct measurement
of the three-boson electroweak vertex, searches of higgs particles
[\cite{i2}], the supermultiplets [\cite{i3}] etc. at colliders of super high
energies (LEP200, LHC) as well as  a study of the $CP$ violation and a
measurement of the fundamental parameters of the electroweak theory
(first of all, in the heavy quark sector).

In the nearest decade, the centre of efforts, directed to the realization
of this program, will certainly be in the field of the heavy quark physics
at both the running colliders (LEP and Fermilab) and the $B$ meson factories,
being planned in SLAC, KEK and at HERA-B. In this case, the extraction of
effects, related with high values of the energy  scale, will be essentially
determined by an accuracy of the theoretical and empirical knowledge on
mechanisms of the quark interactions at non high energy and, first of all,
about effects, caused by the QCD dynamics [\cite{i4}]. Therefore, the
experimental researches of processes with the heavy
$c$-, $ b$-, $t$-quarks take a special importance.

The presence of the small parameter $\Lambda_{QCD}/m_Q$, where $\Lambda_{QCD}$
is the scale of the quark confinement and $m_Q$ is the heavy quark mass, has
allowed one to develop the powerful tools for the study of the QCD
dynamics in the heavy quark interactions, such methods as the
phenomenological potential models
[\cite{2k}--\cite{5k}], the QCD sum rules [\cite{23k}--\cite{i6}]
and Effective Heavy Quark Theory (EHQT) [\cite{13a}],
that is successfully applied for the study of hadrons, containing a
single heavy quark. Thus, the investigation of the processes with the
heavy quarks allows one to  extract and to study nonperturbative QCD effects,
causing the quark hadronization, by means of the use of the heavy quark as
the "marked" atoms. A  successful realization of such program of
studies becomes possible due to the progress in the experimental technique
of the detecting and the identification of particles (mainly,
it is related with the invention and the improvement of the vertex detectors,
allowing one to observe the heavy quark particles due to its running gap
from the primary vertex of the interaction).

Among the heavy quarkonia $(Q\bar Q')$, the $(\bar b c)$  system with the
open charm and beauty takes a particular place.
In contrast to the hidden charm $( c\bar c)$ and beauty $( b\bar b)$ families,
studied in details experimentally [\cite{1k}] and being quite accurately
described theoretically [\cite{i6,i8a,1r}], the heavy quarkonium
$(\bar b c)$, the family of $B_c$ mesons, has some specific production and
decay mechanisms and the spectroscopy, whose study allows one to extend and
to enforce the quantitative understanding of the QCD dynamics as well as
to step forward in the study of the most important parameters of the
electroweak theory.

{}From the spectroscopy viewpoint, the $(\bar b c)$ system
is the heavy quarkonium,
whose spectrum can be quite reliably calculated in the framework of the
QCD-motivated nonrelativistic potential models as well as in the QCD
sum rules. $(\bar b c)$ is the only system, composed of two heavy quarks,
where the description of this system mass spectrum can test the selfconsistency
for the potential models and the QCD sum rules, whose parameters (the quark
masses, for instance) have been fixed from the fitting of the spectroscopic
data on the charmonium and bottomonium. Thus, the study of the
$B_c$ family spectroscopy can serve for the essential improvement of the
quantitative characteristics of the quark models and the QCD sum rules,
which are intensively applied in other  fields of the heavy quark physics
(for example, when one extracts values of elements in the matrix of mixings
of the heavy quark weak charged currents and one estimates contributions,
interfering with the effects of the $CP$ invariance violation, in the heavy
hadron decays [\cite{i9}]).

Moreover, there is a problem of the precise description of the $P$-wave level
splittings in the charmonium and bottomonium, when the experimental measurement
has found an essential deviation from the values, which have been expected
in some well-acknowledged quark models [\cite{i10}]. The study of the
$B_c$ meson family can help in a solution of this problem.

In addition, the $(\bar b c)$  system is interesting due to that it allows one,
in a new way, to use the phenomenological information, obtained from the
detailed experimental study of the charmonium and bottomonium. So, for
example, $(\bar b c)$ takes an intermediate place between the charmonium and
bottomonium in respect to both the system level masses and the values of
average distances between the heavy quarks. As has been clarified,
in the region of the average distances in the $(c\bar c)$ and $(b\bar b)$
systems, the heavy quark potential possesses the simple scaling properties
[\cite{4k,6k,11k}], which state that the kinetic energy of the heavy quarks
is practically a constant value, independent of the quark flavours and the
excitation level in the heavy quarkonium system. Furthermore, this leads to
that the heavy quarkonium level density (the distance between the
$nL$  and $n'L$ levels) does not depend on the flavours of quarks,
composing the heavy quarkonium. This regularity is quite accurately
valid empirically for the
$(c\bar c)$ and $(b\bar b)$ systems and it can be used in the framework of the
QCD sum rules, where a scaling relation, connecting the leptonic constants
of the $S$-wave levels in the different quarkonia
[\cite{7k,i12}], is derived.

Further, having no strong and electromagnetic annihilation channels of
decays, the excited $(\bar b c)$  system levels, being below the
threshold of the decay into the $BD$ meson pair, will decay into the lightest
basic pseudoscalar state $B_c^+(0^-)$ due to the radiative cascade transitions
into the underlying levels. Therefore, the widths of the electromagnetic
($\gamma$) and hadronic ($\pi\pi$, $\eta$, ...) radiative decays of the
given excitation into the other levels will compose its total width. As a
result, the total widths of the excited levels in the $(\bar b c)$  system
turn out two orders of magnitude less than the total widths of the charmonium
and bottomonium excited levels, for which the annihilation channels are
essential. Moreover, maybe, the data on the radiative hadronic decays in the
$(\bar b c)$ family give a possibility for one to solve some problems on the
theory of the hadronic transitions in the heavy quarkonia (for example,
the problem on the anomalous distribution over the
$\pi\pi$ pair invariant mass in the decay of
$\Upsilon'' \to \Upsilon \pi\pi$ [\cite{24k}--\cite{i13}]).

Thus, on the one hand, the methods, applied in the heavy quark physics,
are able quite reliably to point out the spectroscopic characteristics of
the $(\bar b c)$ system for one to make a purposefully-directed experimental
search of the given heavy quarkonium, and, on the other hand, the
measurement of the spectroscopic data in the
$B_c$ family would allow one to improve these methods approaches for the
extraction of the fundamental parameters of the Standard Model from both
the $B_c$ meson physics and the other fields of the heavy quark physics.

Like the other mesons with the open flavour, the basic state of the
$B_c$ meson family, the pseudoscalar meson $B_c^+(0^-)$, is the long-living
particle, decaying due to the weak interaction and having the life time,
comparable with the lifetimes of $B$  and $D$ mesons, so this feature
essentially distinguishes $B_c$ from the heavy quarkonia
$\eta_c$ and $\eta_b$. Therefore, the study of $B_c$ meson decays is the
rich field of the heavy quark physics, where one extracts an important
information about both the QCD dynamics and the weak interactions. The
spectroscopic $B_c$ meson characteristics such as the leptonic constant,
determining the width of the wave package of the $(\bar b c)$ system
in the basic state, essentially determine the description of the $B_c$
decay modes, in which some specific features and effects are observed.

First of all, the presence of the valent heavy quark-spectator leads to a
large probability for the $B_c$ decay modes with the heavy mesons in the
final state, i.e. in the decays
$B_c \to \psi(\eta_c)$ and $B_c \to B_s^{(*)}$
[\cite{i14}--\cite{21k}].
The large $\psi$ particle yield is interesting, in addition, by that the
$\psi$ particle has the perfect experimental signature in the leptonic decay
mode.

Further, in the consideration of the semileptonic
$B_c^+ \to \psi (\eta_c) l^+ \nu$ decays, the nonrelativistic heavy quark
motion inside the quarkonia leads to an essential effect, caused by large
Coulomb-like corrections, which notably change the calculation results
for these decays in the framework of the QCD sum rules [\cite{i14b}].
Only the taking into the account these corrections makes the results of
the QCD sum rules and the potential quark models to be consistent.

Recently, considering the semileptonic transitions of the heavy quarks
$Q \to Q' l \nu$ in the framework of the Effective Heavy Quark Theory
(EHQT) for hadrons with a single heavy quark
($Q\bar q$, $Qqq$), one has stated the universal regularities
[\cite{13a}], which serve, for example, for the model independent extraction
of the Kobayashi--Maskawa matrix element value $|V_{bc}|$. This universality in
the limit of $\Lambda_{QCD}/m_Q \to 0$ is caused by the heavy quark
flavour-independence of the light  quark motion in the gluon field of the
static source (the heavy quark), so that the wave functions of such hadrons
are universal. In the case of the heavy quarkonium with two heavy quarks,
the distances between the quarks depend on the values and ratios of its masses,
i.e. the wave functions of the heavy quarkonia are not universal and
depend on the quark flavours. However, in this case, one can neglect a low
value of the spin-dependent splitting in the heavy quarkonium and suppose
the wave functions of the $nL_J$ quarkonia to be $J$-independent.
This fact finds the expression in an approximate spin-symmetry for the heavy
quarks, so it puts some relations on the form factors of the weak
semileptonic exclusive decays of $B_c$ [\cite{i15}]. Such relations
for the form factors are universal and characteristic for the $B_c$ meson
and reflect the high power of understanding the heavy quark decay dynamics,
needing a direct experimental verification.

Considering the $B_c$ decays with the spectator $b$-quark, one has particularly
to note an essential role of the effects, caused by that the
$c$-quark is not in the free state, but in the bound one. The decrease of the
phase space for the $c$-quark decay within the heavy quarkonium makes the
probability of the decay to be 40 \% less than the probability in the
$D$  and $D_s$ meson decays [\cite{i14e}].
The annihilation channel of the weak $B_c$ meson decay [\cite{7ka}],
allowing one to determine the value of the quark wave function at the origin
$|\Psi(0)|^2$, acquires the important meaning, too.

As in the case of the $(\bar b c)$ system spectroscopy, the heavy quark theory
is able to make the basic predictions on the mechanisms of the $B_c$ meson
decays, whose characteristics measurement would allow one essentially to
develop the methods of its description and also to use these methods
for the precise investigations of the Standard Model as well as possible
deviations from predictions of the latter.

In the case of the $B_c$ meson production, a low value of
the $\Lambda_{QCD}/m_Q$
ratio and, hence, the low value of the quark-gluon coupling
$\alpha_S \sim 1/\ln(m_Q/\Lambda_{QCD}) \ll 1$ allow one to make the
consideration of the pair production of the $b\bar b$ and $c\bar c$ quarks,
from which the $B_c$ meson is formed, in the framework of the perturbative
QCD theory, and also, in a way, to factorize contributions, caused by the
perturbative production of heavy  quarks and forthcoming nonperturbative
binding of the latter into the heavy quarkonium. So, to calculate the
cross sections of the $S$-wave $B_c$ state production in the $Z$ boson peak
is enough to compute the matrix elements for the joint production of the
$b\bar b$ and $c\bar c$ pairs in the colour-singlet state of the $(\bar b c)$
pair with the fixed total spin of quarks ($S=0,1$), when the quarks, being
bound into the meson, move with one and the same velocity, equal to the
meson velocity. After that, one has to multiply these matrix elements
by the nonperturbative factor, whose value is determined by the spectroscopic
characteristics of the bound state (the quark masses and the leptonic
constant, related with the probability of the observation of quarks at the zero
distance between them in the bound state)
[\cite{i17}--\cite{i17f}].
The last notion is caused by that the characteristic virtualities of heavy
quarks inside the heavy quarkonium are much less than its masses, since the
heavy quarks inside the bound states are moving nonrelativistically, otherwise
the quark virtualities in its production are of the order of its masses.
Therefore, considering the $B_c$ production, one can assume, that, inside
the meson, the $\bar b$- and $c$-quarks are close to the mass shell
and practically at rest in respect to each other.
Thus, after the extraction of the nonperturbative factor, the analysis of the
$B_c$ heavy quarkonium production is determined by the consideration of the
matrix elements, calculated in the perturbation theory of QCD.

Note first of all, that the necessity of the two pair production of heavy
quarks in the electromagnetic and strong processes for the $B_c$ yield
leads to that the leading order of the perturbative QCD has an additional
factor of the suppression $\sim\alpha_S^2$ in respect to the leading order
of the perturbation theory for the production of the single flavour
heavy quarks, for example, the $b\bar b$ pair (see Figures
\ref{fp1}, \ref{fp3}),
so $\sigma(B_c)/\sigma(b\bar b)\sim \alpha_S^2|\Psi(0)|^2/m_c^3$.
This causes the low yield of the $B_c$ mesons in respect to the
$B$ meson production.

The analysis of the leading approximation in the perturbative QCD for
the $B_c$ meson production allows one to derive a number of analytical
expressions for the $B_c$ production cross sections
[\cite{i17,i17a}], where one has especially to stress the expressions for the
functions of the heavy  quark fragmentation into the heavy quarkonium in the
scaling limit $M^2/s \to 0$, so these functions are determined by the values of
$\alpha_S$, the quark masses and the leptonic constant of the meson
[\cite{1b}--\cite{3b}]. Thus, the fragmentational $B_c$ production can be
reliably described by the analytical expressions, so this opens new
possibilities in the study of the QCD dynamics, essential in the complete
picture of the heavy quark physics. As one can show, the fragmentational
$B_c$ production certainly dominates in the $Z$ boson decays [\cite{3b}],
so that it can be straightforwardly studied at the LEP facilities.
Moreover, one can analytically study notable spin effects in the
fragmentation into the vector $B_c^*$ meson [\cite{i18}], decaying
electromagnetically $B_c^* \to B_c \gamma$.

In the hadronic $B_c$ production, patron processes at the energies,
comparable with the $B_c$ mass, dominate, so that the processes, having the
character of the fragmentational and also recombinational type
[\cite{i17,i17e}] (see Figure \ref{fp3}), are essential.

Further, the numerical estimates of the $B_c$ meson yield at the colliders
LEP and Tevatron show that the fraction of the $B_c$ mesons in the production
of the beauty hadrons is of the order of $10^{-3}$
[\cite{i17}--\cite{i17f,i19}].
This leads to that at the current experimental facilities, a quite large
number of the $B_c$ mesons are being produced.

Thus, one can point out the expected number of the $B_c$ mesons, being
produced at different colliders, and the differential $B_c$ characteristics,
whose experimental study would significantly clarify the picture of the QCD
interactions of heavy quarks.

A solution of the problem on the experimental discovery and study of the
$B_c$ mesons is determined, first, by the theoretical description of the
characteristics of the $B_c$ meson family (the spectroscopy, the production
and decay mechanisms), so that the present review is devoted to this purpose.
Second, this program is determined by the experimental methodics at the current
detectors, so that the latter would allow one to observe the events
with the $B_c$ production and decays, predicted by the theory.
As for the second part of the problem, at present there is, as mentioned,
the colossal progress, related with the use of the electronic vertex
detectors, possessing the fast operation and allowing one to isolate the
processes with the long living particles ($B$, $B_c$, $D$) from the
production processes (the technique of distinguishing the primary and
secondary vertices), and also accurately to reconstruct the decay
vertices of the particles in space [\cite{i20}].
The presence of distinct signatures in the $B_c$ meson decays and the practical
possibility for the registration of these decay modes have led to the real
chance  of the $B_c$ meson discovery at the LEP and Fermilab detectors
[\cite{16a}]  as well as to the sharp rise of the theoretical interest to the
$(\bar b c)$ system. The latter has reflected in the achievement of a large
number of the essential results in the consideration of the heavy quark
interaction mechanisms at the example of $B_c$ mesons.
So, the present paper is devoted to the review of these results.

\section{Spectroscopy of $B_c$ mesons}

Some preliminary estimates of the bound state masses of the ($\bar b c$)
system have been made in [\cite{2k,5ka}], devoted to
the description of the charmonium and bottomonium properties, as well as in
ref.[\cite{7ka}]. Recently in refs.[\cite{10k}] and [\cite{20k}], the
revised analysis of the $B_c$ spectroscopy has been performed in the
framework of the potential approach and QCD sum rules.

In the present section we consider the ($\bar b c$) spectroscopy with
account of the change of the effective Coulomb interaction constant,
defining spin-dependent splittings of the quarkonium levels. We calculate
the widths of radiative transitions between the levels and analyse the
leptonic constant $f_{B_c}$ in the framework of the QCD sum rules in the
scheme, allowing one to derive scaling relation for the leptonic constants
of the heavy quarkonia.

\subsection{Mass spectrum of $B_c$ mesons}

The $B_c$ meson is the heavy ($\bar b c$) quarkonium with
the open charm and beauty.
It occupies an intermediate place in the mass spectrum of the heavy quarkonia
between the ($\bar c c$) charmonium and the ($\bar b b$) bottomonium.
The approaches, applied to the charmonium and bottomonium study, can be
used to describe the $B_c$ meson properties, and
experimental observation of $B_c$ could serve as a test for these approaches
and it could
be used for the detailed quantitative study of the mechanisms of the heavy
quark production, hadronization and decays.

In the following we obtain the results on the $B_c$ meson spectroscopy.
We will show that below the threshold for the hadronic decay of the
($\bar b c$) system into the $BD$ meson pair, there are 16 narrow bound states,
cascadely decaying into the lightest pseudoscalar $B_c^+(0^-)$  state
with the mass $m(0^-)\approx 6.25$ GeV.

\subsubsection{Potential}

The mass spectra of the charmonium and the bottomonium are experimentally
studied in details [\cite{1k}] and they are properly described
in the framework of phenomenological potential models of
nonrelativistic heavy quarks [\cite{2k}--\cite{4k,5k}]. To describe the mass
spectrum of the ($\bar b c$) system, one would prefer to use the potentials,
whose parameters do not depend on the flavours of the heavy quarks, composing
a heavy quarkonium, i.e. one would use the potentials, which rather accurately
describe the mass spectra of ($\bar c c $) as well as ($\bar b b$), with one
and the same set of potential parameters. The use of such potentials allows
one to avoid an interpolation of the potential parameters from the values,
fixed by the experimental data on the ($\bar c c$) and ($\bar b b$) systems,
to the values in the intermediate region of the ($\bar b c$) system.

As it has been shown in ref.[\cite{6k}],
with an accuracy up to an additive shift,
the potentials, independent of heavy quark flavours
[\cite{2k}--\cite{4k,5k}], coincide with each other in the region of the
average
distances between heavy quarks in the ($\bar c c$) and ($\bar b b$)
systems, so
\begin{equation}
0.1\;{\rm fm} < r < 1\;{\rm fm}\;,
\label{0}
\end{equation}
although those potentials have different asymptotic behaviour in the regions
of very low ($r\to 0$) and very large ($r\to \infty$) distances.

In Cornell model [\cite{2k}] in accordance with asymptotic freedom in QCD,
the potential has  the Coulomb-like behaviour at low distances, and the term,
confining the quarks, rises linearly at large distances
\begin{equation}
V_C(r) = -\frac{4}{3}\;\frac{\alpha_S}{r} + \frac{r}{a^2} + c_0\;,
\end{equation}
so that
\begin{eqnarray}
\alpha_S & = & 0.36\;, \nonumber \\
a & = & 2.34\;\; {\rm GeV}^{-1}\;, \nonumber \\
m_c & = & 1.84\;\; {\rm GeV}\;, \nonumber \\
c_0 & = & -0.25 \;\; {\rm GeV} \;.
\end{eqnarray}

The Richardson potential [\cite{3k}] and its modifications in refs.[\cite{5k}]
and [\cite{6ka}] also correspond to the behaviour,
expected in the framework of QCD, so
\begin{eqnarray}
V_R(r) & = &-\int\frac{{\rm d}^3q}{(2\pi)^3}\;e^{i\bf{r}\bf{q}}\;
\frac{4}{3}\;\frac{48\pi^2}{11 N_c-2n_f}\;
\frac{1}{q^2 \ln(1+q^2/\Lambda^2)} \nonumber\\
& = & -\int\frac{{\rm d}^3q}{(2\pi)^3}\;e^{i\bf{r}\bf{q}}\;
\frac{4}{3}\;\frac{48\pi^2}{27}\;
\biggl(\frac{1}{q^2 \ln(1+q^2/\Lambda^2)}-\frac{\Lambda^2}{q^4}\biggr)
+\frac{8\pi}{27}\;\Lambda^2 r\;,
\end{eqnarray}
with
\begin{equation}
\Lambda  =  0.398\;\; {\rm GeV}\;.
\end{equation}

In the region of the average distances between heavy quarks (\ref{0}),
the QCD-motivated potentials allow the approximations in the forms of
the power (Martin) or logarithmic potentials.

The Martin potential has the form [\cite{4k}]
\begin{equation}
V_M(r) = -c_M + d_M (\Lambda_M r)^k\;,
\label{m}
\end{equation}
so that
\begin{eqnarray}
\Lambda_M & = & 1\;\;{\rm GeV}\;, \nonumber \\
k & = & 0.1\;,
\label{2}\\
m_b & = & 5.174\;\; {\rm GeV}\;, \nonumber \\
m_c & = & 1.8\;\; {\rm GeV}\;, \nonumber \\
c_M & = & 8.064 \;\; {\rm GeV} \;, \nonumber \\
d_M & = & 6.869 \;\; {\rm GeV}\;.\nonumber
\end{eqnarray}

The logarithmic potential is equal to [\cite{11ka}]
\begin{equation}
V_L(r) = c_L + d_L \;\ln (\Lambda_L r)\;,
\label{l}
\end{equation}
so that
\begin{eqnarray}
\Lambda_L & = & 1\;\;{\rm GeV}\;, \nonumber \\
m_b & = & 4.906\;\; {\rm GeV}\;, \nonumber \\
m_c & = & 1.5\;\; {\rm GeV}\;, \\
c_L & = & -0.6635 \;\; {\rm GeV}\;, \nonumber \\
d_L & = & 0.733 \;\; {\rm GeV}\;.\nonumber
\end{eqnarray}

The approximations of the nonrelativistic potential of heavy quarks
in the region of distances (\ref{0}) in the form of the power (\ref{m}) and
logarithmic (\ref{l}) laws, allow one to study its scaling properties.

In accordance with the virial theorem, the average kinetic energy of the quarks
in the bound state is determined by the following expression
\begin{equation}
\langle T\rangle = \frac{1}{2}\; \langle \frac{r {\rm d}V}{{\rm d}r}\rangle\;.
\label{v}
\end{equation}
Then, the logarithmic potential allows one to conclude, that for the
quarkonium states one gets
\begin{equation}
\langle T_L\rangle = {\rm const}
\label{t}
\end{equation}
independently of the flavours of the heavy quarks, composing the heavy
quarkonium,
$$d_L/2={\rm const} \approx 0.367\;\; {\rm GeV}\;.
$$

In the Martin potential, the virial theorem (\ref{v}) allows one to obtain
the expression
\begin{equation}
\langle T_M\rangle = \frac{k}{2+k}\;(c_M+E)\;,
\end{equation}
where $E$ is the binding energy of the quarks in the heavy quarkonium.
Phenomenologically, one has $|E| \ll c_M$ (for example,
$E(1S, c \bar c) \approx -0.5$ GeV),
so that, neglecting the binding energy of the heavy quarks inside the heavy
quarkonium, one can conclude that the average kinetic energy of the heavy
quarks is a constant value, independent of the quark flavours and the number
of the radial or orbital excitation.
The accuracy of such approximation for $\langle T\rangle $ is about 10\%, i.e.
$|\Delta T/T| \sim 30\div 40$ MeV.

{}From the Feynman-Hellmann theorem for the system with the reduced mass $\mu$,
one has
\begin{equation}
\frac{{\rm d}E}{{\rm d}\mu} = -\;\frac{\langle T\rangle}{\mu}\;,
\end{equation}
and, in accordance with condition (\ref{t}), it follows that the difference
of the energies for the radial excitations of the heavy quarkonium levels
does not depend on the reduced mass of the $Q\bar Q'$ system
\begin{equation}
E(\bar n,\mu) - E(n,\mu) = E(\bar n,\mu') - E(n,\mu')\;.
\label{e1}
\end{equation}

Thus, in the approximation of both the low value for the binding energy
of quarks and the zero value for the spin-dependent splittings of the levels,
the heavy quarkonium state density does not depend on the heavy quark flavours
\begin{equation}
\frac{{\rm d}n}{{\rm d}M_n} = {\rm const}\;.
\label{e2}
\end{equation}
The given statement has been also derived in ref.[\cite{7k}] using
the Bohr--Sommerfeld quantization of the S-wave states for the heavy
quarkonium system with Martin potential [\cite{4k}].

Relations (\ref{e1})-(\ref{e2}) are phenomenologically confirmed for the
vector S-levels of the ($b\bar b$), ($c\bar c$), ($s\bar s$) systems
[\cite{1k}]
(see Table \ref{t1}).
\begin{table}[t]
\caption{The mass difference (in MeV) for the two lightest vector states of
different heavy systems,  $\Delta M = M(2S) - M(1S)$}
\label{t1}
\begin{center}
\begin{tabular}{||c|c|c|c|c||}
\hline
System & $\Upsilon$ & $\psi$ & $B_c$ & $\phi$ \\
\hline
$\Delta M$ & 563 & 588 & 585 & 660\\
\hline
\end{tabular}
\end{center}
\end{table}

Thus, the structure of the nonsplitted S-levels of the ($\bar b c$) system
must repeat not only qualitatively, but quantitatively the structure
of the S-levels for the ($\bar b b$) and ($\bar c c$) systems, with an accuracy
up to the overall additive shift of masses.

Moreover, in the framework of the QCD sum rules, the universality of the
heavy quark nonrelativistic potential (the independence on the
flavours and the scaling properties (\ref{t}), (\ref{e1}),
(\ref{e2})) allows one to obtain the scaling relation for the
leptonic constants of the S-wave quarkonia [\cite{7k}]
\begin{equation}
\frac{f^2}{M} = {\rm const}
\label{law1}
\end{equation}
independently of the heavy quark flavours in the regime, when
$$ |m_Q-m_{Q'}|\;\; \mbox{is restricted}\;, \;\;
\frac{\Lambda_{QCD}}{m_{Q,Q'}} \ll 1\;,
$$
i.e., when one can neglect the heavy quark mass difference.
On the other hand, in the regime,
when the mass difference is not low, one has
\begin{equation}
\frac{f^2}{M}\;\biggl(\frac{M}{4\mu}\biggr)^2 = {\rm const}\;,
\label{law}
\end{equation}
where
$$\mu = \frac{m_Q m_{Q'}}{m_Q+m_{Q'}}\;.
$$

Consider the mass spectrum of the ($\bar b c$) system with the Martin
potential [\cite{4k}].

Solving the Schr\" odinger equation with potential (\ref{m})  and
the parameters (\ref{2}), one finds the $B_c$ mass spectrum and the
characteristics of the radial wave functions $R(0)$ and $R'(0)$,
shown in Tables \ref{t2} and \ref{t3}, respectively.
\begin{table}[b]
\caption{The energy levels of the ($\bar b c$)  system, calculated without
taking into account relativistic corrections, in GeV}
\label{t2}
\begin{center}
\begin{tabular}{|c|c|c|c||c|c|c|c||c|c|c|c|}
\hline
$n$ & [\cite{7ka}] & [\cite{8k}] & [\cite{6ka}] &
$n$ & [\cite{7ka}] & [\cite{8k}] & [\cite{6ka}] &
$n$ & [\cite{7ka}] & [\cite{8k}] & [\cite{6ka}] \\
\hline
1S & 6.301 & 6.315 & 6.344 & 2P & 6.728 &6.735&6.763 & 3D & 7.008 &7.145
&7.030  \\
2S & 6.893 & 7.009 & 6.910 & 3P & 7.122 & -- &7.160 & 4D & 7.308 &  --& 7.365
\\
3S & 7.237 & --    & 7.024 & 4P & 7.395 & -- & -- & 5D & 7.532 &  --&  --\\
\hline
\end{tabular}
\end{center}
\end{table}

\begin{table}[t]
\caption{The characteristics of the radial wave functions $R_{nS}(0)$
(in GeV$^{3/2}$) and $R'_{nP}(0)$ (in GeV$^{5/2}$),
obtained from the Schr\" odinger equation}
\label{t3}
\begin{center}
\begin{tabular}{||c|c|c||}
\hline
$n$ & Martin  & ~~~~[\cite{10k}]~~~~\\
\hline
$R_{1S}(0)$  & 1.31 & 1.28 \\
$R_{2S}(0)$  & 0.97 & 0.99 \\
$R'_{2P}(0)$ & 0.55 & 0.45 \\
$R'_{3P}(0)$ & 0.57 & 0.51 \\
\hline
\end{tabular}
\end{center}
\end{table}
The average kinetic energy of the levels, lying below the threshold for
the ($\bar b c$) system decay into the $BD$ pair, is presented in
Table \ref{t4}, wherein one can see that the term, added to the
radial potential due to the orbital rotation,
\begin{equation}
\Delta V_l =\frac{{\bf{L}}^2}{2\mu r^2}
\end{equation}
weakly influences the value of the average kinetic energy, and the binding
energy for the levels with $L\neq 0$ is essentially determined by the orbital
rotation energy, which is approximately independent of the quark flavours
(see Table \ref{t5}), so that the structure of the nonsplitted levels of the
($\bar b c$) system with $L\neq 0$ must quantitatively repeat the structure
of the charmonium and bottomonium levels, too.
\begin{table}[b]
\caption{The average kinetic and orbital energies of the quark motion in the
($\bar b c$) system, in GeV}
\label{t4}
\begin{center}
\begin{tabular}{||c|l|l|l|l|l||}
\hline
$nL          $&$ 1S   $& $2S   $& $2P   $& $3P   $& $3D$\\
\hline
$\langle T\rangle$       & 0.35 & 0.38 & 0.37 & 0.39 & 0.39\\
$\Delta V_l$& 0.00 & 0.00 & 0.22 & 0.14 & 0.29\\
\hline
\end{tabular}
\end{center}
\end{table}

\begin{table}[b]
\caption{The average energy of the orbital motion in the heavy quarkonia,
in the model with the Martin potential, in GeV}
\label{t5}
\begin{center}
\begin{tabular}{||c|l|l|l||}
\hline
System        & $\bar c c$ & $\bar b c$& $\bar b b$\\
\hline
$\Delta V_l(2P)$& 0.23       & 0.22      & 0.21\\
\hline
\end{tabular}
\end{center}
\end{table}
\subsubsection{Spin-dependent splitting of the ($\bar b c$) quarkonium}

In accordance with the results of refs.[\cite{8k,9k}], one introduces
the additional term to the potential
to take into the account the spin-orbital and spin-spin interactions,
causing the splitting of the $nL$ levels ($n$ is the principal quantum number,
$L$ is the orbital momentum), so it has the form
\begin{eqnarray}
   V_{SD}(\bf{r}) & = &\biggl(\frac{{\bf{L}}\cdot{\bf{S}}_c}{2m_c^2} +
\frac{{\bf{L}}\cdot{\bf{S}}_b}{2m_b^2}\biggr)\;
\biggl(-\frac{{\rm d}V(r)}{r{\rm
d}r}+\frac{8}{3}\;\alpha_S\;\frac{1}{r^3}\biggr)
+ \nonumber \\
{}~ & ~ & +\frac{4}{3}\;\alpha_S\;\frac{1}{m_c
m_b}\;\frac{{\bf{L}}\cdot{\bf{S}}}{r^3}
+\frac{4}{3}\;\alpha_S\;\frac{2}{3m_c m_b}\;
{\bf{S}}_c\cdot{\bf{S}}_b\;4\pi\;\delta({\bf{r}}) \label{3} \\
{}~ & ~ & +\frac{4}{3}\;\alpha_S\;\frac{1}{m_c
m_b}\;(3({\bf{S}}_c\cdot{\bf{n}})\;
({\bf{S}}_b\cdot{\bf{n}}) - {\bf{S}}_c\cdot{\bf{S}}_b)\;\frac{1}{r^3}\;,
\;\;{\bf{n}}=\frac{{\bf{r}}}{r}\;.\nonumber
\end{eqnarray}
where $V(r)$ is the phenomenological potential, confining the quarks,
the first term takes into account the relativistic corrections to the
potential $V(r)$; the second, third and fourth terms are the relativistic
corrections, coming from the account of the one gluon exchange between the
$b$ and $c$ quarks; $\alpha_S$ is the effective constant of the quark-gluon
interaction inside the ($\bar b c$) system.

The value of the $\alpha_S$ parameter can be determined in the following way.

The splitting of the S-wave heavy quarkonium ($Q_1\bar Q_2$) is determined
by the expression
\begin{equation}
\Delta M(nS) = \alpha_S\;\frac{8}{9m_1m_2}\;|R_{nS}(0)|^2\;,
\end{equation}
where $R_{nS}(0)$ is the value of the radial wave function of the quarkonium,
at the origin. Using the experimental value of the S-state splitting in the
($c\bar c$) system [\cite{1k}]
\begin{equation}
\Delta M(1S,\;c\bar c) =117\pm2\;\;{\rm MeV}\;,
\end{equation}
and the $R_{1S}(0)$ value, calculated in the potential model for the
($c\bar c$) system, one gets the model-dependent value of the
$\alpha_S(\psi)$ constant for the effective Coulomb interaction of the heavy
quarks (in the Martin potential, one has $\alpha_S(\psi) = 0.44$).

In ref.[\cite{10k}] the effective constant value, fixed in the described way,
has been applied to the description of not only the ($c\bar c$) system, but
also the ($\bar b c)$ and ($\bar b b$) quarkonia.

In the present paper we take into account the variation of the effective
Coulomb interaction constant versus the reduced mass of the system ($\mu$).

In the one-loop approximation at the momentum scale $p^2$, the "running"
coupling constant in QCD is determined by the expression
\begin{equation}
\alpha_S(p^2) = \frac{4\pi}{b\ln(p^2/\Lambda_{QCD}^2)}\;,
\end{equation}
where $b=11-2n_f/3$, and $n_f=3$, when one takes into account the
contribution by the virtual light quarks, $p^2 < m^2_{c,b}$.

In the model with the Martin potential, for the kinetic energy of quarks
($c\bar c$) inside $\psi$, one has
\begin{equation}
\langle T_{1S}(c\bar c)\rangle  \approx 0.357\;\;{\rm GeV}\;,
\end{equation}
so that, using the expression for the kinetic energy,
\begin{equation}
\langle T\rangle  =\frac{\langle p^2\rangle }{2\mu}\;,
\end{equation}
one gets
\begin{equation}
\alpha_S(p^2) = \frac{4\pi}{b\ln(2\langle T\rangle \mu/\Lambda_{QCD}^2)}\;,
\label{al1}
\end{equation}
so that $\alpha_S(\psi) = 0.44$ at
\begin{equation}
\Lambda_{QCD} \approx 164\;\;{\rm MeV} \;.
\label{al2}
\end{equation}

As it has been noted in the previous section, the value of the kinetic energy
of the quark motion weakly depends on the heavy quark flavours, and it,
practically, is constant, and, hence, the change of the effective
$\alpha_S$ coupling is basically determined by the variation of the reduced
mass of the heavy quarkonium. In accordance with eqs.(\ref{al1})-(\ref{al2})
and Table \ref{t4}, for the $(\bar b c)$ system one has
\begin{center}
\begin{tabular}{cccccc}
$nL$       & $1S    $& $2S    $& $2P    $& $3P$    & $3D$\\
$\alpha_S$ & 0.394 & 0.385 & 0.387 & 0.382 & 0.383.
\end{tabular}
\end{center}

Note, the Martin potential leads to the $R_{1S}(0)$ values, which, with the
accuracy up to $15\div20\%$, agrees with  the experimental values of the
leptonic decay constants for the heavy ($c\bar c$) and ($b\bar b$) quarkonia.
The leptonic constants are determined by the expression
\begin{equation}
\Gamma(Q\bar Q \to l^+l^-) = \frac{4\pi}{3}\;e_Q^2\;\alpha^2_{em}\;
\frac{f_{Q\bar Q}^2}{M_{Q\bar Q}}\;,
\end{equation}
where $e_Q$ is the heavy quark charge.

In the nonrelativistic model one has
\begin{equation}
f_{Q\bar Q} = \sqrt{\frac{3}{\pi M_{Q\bar Q}}}\;R_{1S}(0)\;.
\label{27s}
\end{equation}
For the effective Coulomb interaction of the heavy quarks in the basic
1S-state one has
\begin{equation}
R_{1S}^{C}(0) = 2 \;\biggl(\frac{4}{3}\;\mu\;\alpha_S\biggr)^{3/2}\;.
\end{equation}
One can see from Table \ref{t7}, that, taking into account the variation
of the effective $\alpha_S$ constant versus the reduced mass of the heavy
quarkonium (see eq.(\ref{al1})), the Coulomb wave functions give the values
of the leptonic constants for the heavy 1S-quarkonia, so that in the framework
of the accuracy of the potential models, those values agree with the
experimental values and the values, obtained by the solution of the
Schr\" odinger equation with the given potential.
\begin{table}[t]
\caption{The leptonic decay constants of the heavy quarkonia, the values,
measured experimentally and obtained in the model with the Martin potential,
in the model with the effective Coulomb interaction and from the scaling
relation (SR), in MeV}
\label{t7}
\begin{center}
\begin{tabular}{||c|c|c|c|c||}
\hline
Model    & Exp.[\cite{1k}] & Martin     & Coulomb      & SR\\
\hline
$f_\psi$  & $410\pm15$     & $547\pm80$ & $426\pm60$ & $410\pm40$\\
$f_{B_c}$ & --             & $510\pm80$ & $456\pm70$ & $460\pm60$\\
$f_\Upsilon$ & $715\pm15$  & $660\pm90$ & $772\pm120$& $715\pm70$\\
\hline
\end{tabular}
\end{center}
\end{table}

The consideration of the variation of the effective Coulomb interaction
constant becomes especially essential for the $\Upsilon$ particles, for which
$\alpha_S(\Upsilon)\approx 0.33$ instead of the fixed value $\alpha_S = 0.44$.

Thus, calculating the splitting of the ($\bar b c$) levels, we take into
account the $\alpha_S$ dependence on the reduced mass of the heavy
quarkonium.

As one can see from eq.(\ref{3}), in contrast to the $LS$-coupling
in the $(\bar c c)$ and $(\bar b b)$ systems, there is the $jj$-coupling
in the heavy quarkonium, where the heavy quarks have different masses
(here, $ {\bf{L}}{\bf{S}}_c$ is diagonalized at the given ${\bf{J}}_c$
momentum, (${\bf{J}}_c ={\bf{L}}+{\bf{S}}_c$,
${\bf{J}}={\bf{J}}_c +{\bf{S}}_b$), $\bf{J}$ is the total spin of the system).
We use the following spectroscopic notations for the splitted levels of the
($\bar b c$) system, -- $n^{2j_c}L_J$.

One can easily show, that independently of the total spin $J$ projection
one has
\begin{eqnarray}
|^{2L+1}L_{L+1}\rangle & = & |J=L+1,\;S=1\rangle\;, \nonumber \\
|^{2L-1}L_{L-1}\rangle & = & |J=L-1,\;S=1\rangle\;,\label{3.c}  \\
|^{2L+1}L_{L}\rangle & = & \sqrt{\frac{L}{2L+1}}|J=L,\;S=1\rangle +
                    \sqrt{\frac{L+1}{2L+1}}|J=L,\;S=0\rangle\;,\nonumber \\
|^{2L-1}L_{L}\rangle & = & \sqrt{\frac{L+1}{2L+1}}|J=L,\;S=1\rangle -
                    \sqrt{\frac{L}{2L+1}}|J=L,\;S=0\rangle\;,\nonumber
\end{eqnarray}
where $|J,\;S\rangle $ are the state vectors
with the given values of the total
quark spin  ${\bf{S}} = {\bf{S}}_c+{\bf{S}}_b$, so that the potential terms of
the order of $1/m_c m_b$, $1/m_b^2$ lead, generally speaking, to the mixing
of the levels with the different $J_c$ values at the given $J$ values.
The tensor forces (the last term in eq.(\ref{3})) are equal to zero at
$L=0$  or  $S=0$.

To calculate values of the level shifts, appearing due to the spin-spin and
spin-orbital interactions, one has to take the averaged expression
(\ref{3}) over the wave functions of the corresponding states.

The averaging over the angle variables can be performed in the following
standard way. Let us represent the matrix element of the unit vector
${\bf{n}}={\bf{r}}/r$ pair in the form
\begin{equation}
\langle L,\;m| n^p n^q |L,\;m'\rangle = a(L^p L^q + L^q L^p)_{mm'}+
b \delta^{pq}\delta_{mm'}\;,
\label{3l}
\end{equation}
where $\bf{L}$ are the orbital momentum matrices in the corresponding
irreducible representation.

{}From the conditions of
the normalization of the unit vector,
$
\langle n^p n^q\rangle  \delta^{pq} = 1
$,
the orthogonality of the radius-vector to the orbital momentum,
$
n^p L^p = 0
$,
the commutation relations for the angle momentum,
$
[L^p; L^q] = i\epsilon^{pql} L_l
$,
one finds the values of constants $a$ and $b$ in eq.(\ref{3l})
\begin{eqnarray}
a & = & - \frac{1}{4{\bf{L}}^2-3}\;,\\
b & = & \frac{2{\bf{L}}^2-1}{4{\bf{L}}^2-3}\;.
\end{eqnarray}
Note further, that from the condition for the quark spins
$S_Q^p S_Q^q + S_Q^q S_Q^p = \frac{1}{2} \delta^{pq}$ it follows, that
\begin{equation}
3 (n^p n^q -\frac{1}{3}\delta^{pq}) S_c^p S_b^q =
\frac{3}{2} (n^p n^q -\frac{1}{3}\delta^{pq}) S^p S^q\;.
\end{equation}

Thus, (see also ref.[\cite{11k}])
\begin{equation}
\langle 6 (n^p n^q -\frac{1}{3}\delta^{pq}) S_c^p S_b^q\rangle =
- \frac{1}{4{\bf{L}}^2-3} (6({\bf{L}} {\bf{S}})^2 + 3 ({\bf{L}} {\bf{S}}) -2
{\bf{L}}^2 {\bf{S}}^2)\;.
\label{3.t}
\end{equation}

\begin{figure}[t]
\begin{center}
\begin{picture}(175,150)
\put(15,30){\line(1,0){20}}
\put(15,33){$1S$}
\put(36,25){\line(1,0){20}}
\put(36,32){\line(1,0){20}}

\put(57,21){$^{0^-}$}
\put(57,29){$^{1^-}$}

\put(15,89){\line(1,0){20}}
\put(15,92){$2S$}
\put(36,87){\line(1,0){20}}
\put(36,90){\line(1,0){20}}

\put(57,83){$^{0^-}$}
\put(57,87){$^{1^-}$}

\put(15,124){\line(1,0){20}}
\put(15,127){$3S$}

\put(70,73){\line(1,0){20}}
\put(70,76){$2P$}
\put(91,68){\line(1,0){20}}
\put(91,72){\line(1,0){20}}
\put(91,74){\line(1,0){20}}
\put(91,76){\line(1,0){20}}

\put(112,63){$^{0^+}$}
\put(112,75){$^{2^+}$}
\put(112,67){$^{1^+}$}
\put(112,71){$^{1'^+}$}

\put(70,112){\line(1,0){20}}
\put(70,115){$3P$}
\put(91,109){\line(1,0){20}}
\put(91,111){\line(1,0){20}}
\put(91,112){\line(1,0){20}}
\put(91,114){\line(1,0){20}}

\put(112,102){$^{0^+}$}
\put(112,114){$^{2^+}$}
\put(112,106){$^{1^+}$}
\put(112,110){$^{1'^+}$}

\put(70,140){\line(1,0){20}}
\put(70,143){$4P$}

\put(125,101){\line(1,0){20}}
\put(125,104){$3D$}
\put(146,099){\line(1,0){20}}
\put(146,100){\line(1,0){20}}
\put(146,101){\line(1,0){20}}
\put(146,102){\line(1,0){20}}

\put(167,099){$^{1^-}$}
\put(167,095){$^{3^-}$}
\put(167,091){$^{2^-}$}
\put(167,103){$^{2'^-}$}

\put(125,131){\line(1,0){20}}
\put(125,134){$4D$}

\put(10,0){\framebox(165,150)}
\put(0,0){$6.0$}
\put(10,50){\line(1,0){3}}
\put(0,50){$6.5$}
\put(10,100){\line(1,0){3}}
\put(0,100){$7.0$}
\put(0,150){$7.5$}
\put(10,115){\line(1,0){55}}
\put(10,115.3){\line(1,0){55}}
\put(120,115){\line(1,0){55}}
\put(120,115.3){\line(1,0){55}}
\put(130,117){$BD$ $threshold$}
\end{picture}
\end{center}
\caption{The mass spectrum of the $(\bar b c)$ system with account
of splittings.}
\label{f1k}
\end{figure}
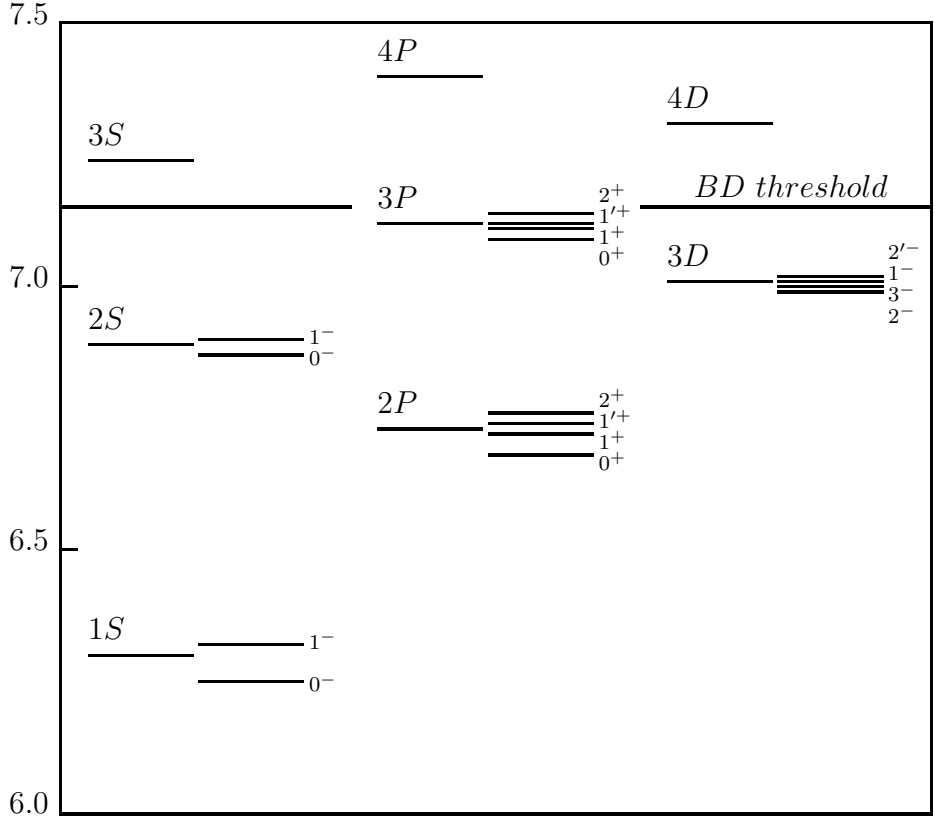

Using eqs.(\ref{3.c}), (\ref{3.t}), for the level shifts, calculated in the
perturbation theory at $S=1$, one gets the following formulae
\begin{eqnarray}
   \Delta E_{n^1S_0} & = &- \alpha_S\;\frac{2}{3m_cm_b}\;|R_{nS}(0)|^2\;,\\
   \Delta E_{n^1S_1} & = & \alpha_S\;\frac{2}{9m_cm_b}\;|R_{nS}(0)|^2\;,
            \label{5}\\
   \Delta E_{n^3P_2} & = & \alpha_S\;\frac{6}{5m_cm_b}\;\langle
\frac{1}{r^3}\rangle +
\frac{1}{4}\;\biggl(\frac{1}{m_c^2} + \frac{1}{m_b^2}\biggr)\;
\langle -\frac{{\rm d}V(r)}{r{\rm
d}r}+\frac{8}{3}\;\alpha_S\frac{1}{r^3}\rangle\;, \\
   \Delta E_{n^1P_0} & = & - \alpha_S\;\frac{4}{m_cm_b}\;\langle
\frac{1}{r^3}\rangle -
\frac{1}{2}\;\biggl(\frac{1}{m_c^2} + \frac{1}{m_b^2}\biggr)\;
\langle -\frac{{\rm d}V(r)}{r{\rm
d}r}+\frac{8}{3}\;\alpha_S\frac{1}{r^3}\rangle\;,\\
   \Delta E_{n^5D_3} & = &  \alpha_S\;\frac{52}{21m_cm_b}\;\langle
\frac{1}{r^3}\rangle +
\frac{1}{2}\;\biggl(\frac{1}{m_c^2} + \frac{1}{m_b^2}\biggr)\;
\langle -\frac{{\rm d}V(r)}{r{\rm
d}r}+\frac{8}{3}\;\alpha_S\frac{1}{r^3}\rangle\;,\\
   \Delta E_{n^3D_1} & = & - \alpha_S\;\frac{92}{21m_cm_b}\;\langle
\frac{1}{r^3}\rangle -
\frac{3}{4}\;\biggl(\frac{1}{m_c^2} + \frac{1}{m_b^2}\biggr)\;
\langle -\frac{{\rm d}V(r)}{r{\rm
d}r}+\frac{8}{3}\;\alpha_S\frac{1}{r^3}\rangle\;,
\end{eqnarray}
where $R_{nS}(0)$ are the radial wave functions at $L=0$,  $\langle ...\rangle
$  denote
the average values, calculated under the wave functions $R_{nL}({r})$.
The mixing matrix elements have the forms
\begin{eqnarray}
\langle ~^3P_1|\Delta E|~^3P_1\rangle & = &
-\alpha_S\;\frac{2}{9m_cm_b}\;\langle \frac{1}{r^3}\rangle +
\nonumber \\
 & &\biggl(\frac{1}{4m_c^2} - \frac{5}{12m_b^2}\biggr)\;
\langle -\frac{{\rm d}V(r)}{r{\rm
d}r}+\frac{8}{3}\;\alpha_S\frac{1}{r^3}\rangle\;,\\
\langle ~^1P_1|\Delta E|~^1P_1\rangle & = &
-\alpha_S\;\frac{4}{9m_cm_b}\;\langle \frac{1}{r^3}\rangle +
\nonumber \\
&&\biggl(-\frac{1}{2m_c^2} + \frac{1}{6m_b^2}\biggr)\;
\langle -\frac{{\rm d}V(r)}{r{\rm
d}r}+\frac{8}{3}\;\alpha_S\frac{1}{r^3}\rangle\;, \\
\langle ~^3P_1|\Delta E|~^1P_1\rangle & = &
-\alpha_S\;\frac{2\sqrt{2}}{9m_cm_b}\;\langle \frac{1}{r^3}\rangle
\nonumber \\
&&-\frac{\sqrt{2}}{6m_b^2}\;
\langle -\frac{{\rm d}V(r)}{r{\rm
d}r}+\frac{8}{3}\;\alpha_S\frac{1}{r^3}\rangle\;,\\
\langle ~^5D_2|\Delta E|~^5D_2\rangle & = &
-\alpha_S\;\frac{4}{15m_cm_b}\;\langle \frac{1}{r^3}\rangle +
\nonumber \\
&&\biggl(\frac{1}{2m_c^2} - \frac{1}{5m_b^2}\biggr)\;
\langle -\frac{{\rm d}V(r)}{r{\rm
d}r}+\frac{8}{3}\;\alpha_S\frac{1}{r^3}\rangle\;,\label{6}\\
\langle ~^3D_2|\Delta E|~^3D_2\rangle & = &
-\alpha_S\;\frac{8}{15m_cm_b}\;\langle \frac{1}{r^3}\rangle +
\nonumber \\
&&\biggl(-\frac{3}{4m_c^2} + \frac{9}{20m_b^2}\biggr)\;
\langle -\frac{{\rm d}V(r)}{r{\rm
d}r}+\frac{8}{3}\;\alpha_S\frac{1}{r^3}\rangle\;, \\
\langle ~^5D_2|\Delta E|~^3D_2\rangle & = &
-\alpha_S\;\frac{2\sqrt{6}}{15m_cm_b}\;\langle \frac{1}{r^3}\rangle
\nonumber \\
&&-\frac{\sqrt{6}}{10m_b^2}\;
\langle -\frac{{\rm d}V(r)}{r{\rm
d}r}+\frac{8}{3}\;\alpha_S\frac{1}{r^3}\rangle\;,
\end{eqnarray}

As one can see from eq.(\ref{5}), the S-level splitting is essentially
determined by the $|R_{nS}(0)|$ value, which can be related to the leptonic
decay constants of the S-states ($0^-$, $1^-$). Section 2.3 is devoted to the
calculation of these constants in different ways.
We only note here, that with enough accuracy, the predictions of different
potential models on the $|R_{1S}(0)|$ value are in agreement with each other
as well as with predictions in other approaches.
\begin{table}[t]
\caption{The masses (in GeV) of the lightest pseudoscalar $B_c$ and vector
$B_c^*$ states in different models ($\ast$ is the present paper)}
\label{t6}
\begin{center}
\begin{tabular}{|c|c|c|c|c|c|c|c|}
\hline
State & $\ast$ & [\cite{8k}] & [\cite{6ka}]& [\cite{12k}] & [\cite{5ka}]
& [\cite{13k}]& [\cite{19k,7k}] \\
\hline
$0^-$ &6.253&6.249&6.314&6.293&6.270&6.243&6.246\\
$1^-$ &6.317&6.339&6.354&6.346&6.340&6.320&6.319\\
\hline
\hline
State & [\cite{10k}] & [\cite{14k}] & [\cite{15k}]& [\cite{16k}] & [\cite{17k}]
& [\cite{18k}]& [\cite{20k}] \\
\hline
$0^-$ &6.264&6.320&6.256&6.276&6.286&--&6.255\\
$1^-$ &6.337&6.370&6.329&6.365&6.328&6.320&6.330\\
\hline
\end{tabular}
\end{center}
\end{table}

For the $2P$, $3P$ and $3D$ levels, the mixing matrices of the states with the
total
quark spin $S=1$ and $S=0$ have the forms
\begin{eqnarray}
|2P,\;1'^+\rangle & = & 0.294 |S=1\rangle + 0.956 |S=0\rangle\;,\\
|2P,\;1^{+}\rangle & = & 0.956 |S=1\rangle - 0.294 |S=0\rangle\;,
\end{eqnarray}
so that in the $1^+$ state the probability of the total quark spin value
$S=1$ is equal to
\begin{equation}\label{w.2p}
w_1(2P) =0.913\;.
\end{equation}
For the $3P$ level one has
\begin{eqnarray}
|3P,\;1'^+\rangle & = & 0.371 |S=1\rangle + 0.929 |S=0\rangle\;,\\
|3P,\;1^{+}\rangle & = & 0.929 |S=1\rangle - 0.371 |S=0\rangle\;,
\end{eqnarray}
so that
\begin{equation}\label{w.3p}
w_1(3P) =0.863\;,
\end{equation}
For the $3D$ level one gets
\begin{eqnarray}
|3D,\;2'^-\rangle & = & -0.566 |S=1\rangle + 0.825 |S=0\rangle\;,\\
|3D,\;2^{-}\rangle & = & 0.825 |S=1\rangle + 0.566 |S=0\rangle\;,
\end{eqnarray}
so that
\begin{equation}\label{w.3d}
w_2(3D) =0.680\;.
\end{equation}

With account of the calculated splittings,
the $B_c$ mass spectrum is shown in Figure \ref{f1k} and Table \ref{t8}.
\begin{table}[t]
\caption{The masses (in GeV) of the bound $(\bar b c$) states below
the threshold of the decay into the $BD$ meson pair ($\ast$ is
the present paper)}
\label{t8}
\begin{center}
\begin{tabular}{||l|c|c|c||}
\hline
State  & $\ast$  & [\cite{10k}] & [\cite{6ka}]\\
\hline
$1^1S_0$    & 6.253       & 6.264      & 6.314     \\
$1^1S_1$    & 6.317       & 6.337      & 6.355     \\
$2^1S_0$    & 6.867       & 6.856      & 6.889     \\
$2^1S_1$    & 6.902       & 6.899      & 6.917     \\
$2^1P_0$    & 6.683       & 6.700      & 6.728     \\
$2P\; 1^+$  & 6.717       & 6.730      & 6.760     \\
$2P\; 1'^+$ & 6.729       & 6.736      & --        \\
$2^3P_2$    & 6.743       & 6.747      & 6.773     \\
$3^1P_0$    & 7.088       & 7.108      & 7.134     \\
$3P\; 1^+$  & 7.113       & 7.135      & 7.159     \\
$3P\; 1'^+$ & 7.124       & 7.142      & --        \\
$3^3P_2$    & 7.134       & 7.153      & 7.166     \\
$3D\; 2^-$  & 7.001       & 7.009      & --        \\
$3^5D_3$    & 7.007       & 7.005      & --        \\
$3^3D_1$    & 7.008       & 7.012      & --        \\
$3D\; 2'^-$ & 7.016       & 7.012      & --        \\
\hline
\end{tabular}
\end{center}
\end{table}

The masses of the $B_c$ mesons have been also calculated in papers of
ref.[\cite{k20k}].

As one can see from Tables \ref{t2} and \ref{t6}, the place of the 1S-level in
the ($\bar b c$) system ($m(1S) \approx 6.3$ GeV) is predicted by the potential
models with the rather high accuracy $\Delta m(1S) \approx 30$ MeV, and
the 1S-level splitting into the vector and pseudoscalar states is about
$m(1^-) - m(0^-) \approx 70$ MeV.

\subsubsection{$B_c$ meson masses from QCD sum rules}

Potential model estimates for the masses of the lightest ($\bar b c$)
states are in agreement with the results of the calculations
for the vector and pseudoscalar ($\bar b c$) states in the framework of the
QCD sum rules [\cite{20k,21k,22k}], where the calculation accuracy is lower,
than the accuracy of the potential models, because the results essentially
depend on both the modelling of the nonresonant hadronic part of the current
correlator (the continuum threshold) and the parameter of the sum rule
scheme (the moment number for the spectral density of the current
correlator or the Borel transformation parameter),
\begin{equation}
m^{SR}(0^-)\approx m^{SR}(1^-) \approx 6.3\div 6.5\;{\rm GeV}\;.
\end{equation}

As it has been shown in [\cite{23k}], for the lightest vector
quarkonium, the following QCD sum rules take place
\begin{equation}
\frac{f_V^2 M_V^2}{M_V^2- q^2} = \frac{1}{\pi}\;
\int_{s_i}^{s_{{\rm th}}} \frac{\Im m \Pi_V^{\rm QCD(pert)}(s)}{s-q^2}\;
{\rm d}s + \Pi_V^{\rm QCD(nonpert)}(q^2)\;.
\label{4.1}
\end{equation}
where $f_V$ is the leptonic constant of the vector ($\bar b c$) state with
the mass $M_V$,
\begin{eqnarray}
i f_V M_V \epsilon^{\lambda}_\mu \exp({ipx}) & = &
\langle 0|J_\mu (x)|V(p,\;\lambda)\rangle\;,\label{59s}\\
J_\mu (x) & = & \bar c(x) \gamma_\mu b(x)\;,\label{60s}
\end{eqnarray}
where $\lambda$, $p$ are the $B^*_c$ polarization and momentum, respectively,
and
\begin{eqnarray}
\int {\rm d}^4x\;\exp{(iqx)} \langle 0|TJ_\mu(x) J_\nu(0)|0\rangle & = &
\biggl(-g_{\mu\nu}+
\frac{q_\mu q_\nu}{q^2}\biggr)\;\Pi_V^{\rm QCD} + q_\mu q_\nu\;\Pi_S^{\rm
QCD}\;,\\
\Pi_V^{\rm QCD}(q^2) & = & \Pi_V^{\rm QCD(pert)} + \Pi_V^{\rm
QCD(nonpert)}(q^2)\;,\\
\Pi_V^{\rm QCD(nonpert)}(q^2) & = & \sum C_i(q^2) O^i\;,
\end{eqnarray}
where $O^i$ are the vacuum expectation values of the composite operators such
as $\langle m\bar \psi \psi\rangle $, $\langle \alpha_S\;G_{\mu\nu}^2\rangle $,
etc. The Wilson
coefficients are calculable in the perturbation theory of QCD.
$s_i=(m_c+m_b)^2$ is the kinematical threshold of the perturbative
contribution,  $M_V^2 > s_i$, $s_{th}$ is the threshold of the nonresonant
hadronic contribution, which is considered to be equal to the perturbative
contribution at $s > s_{th}$.

Considering the respective correlators, one can write down the sum rules,
analogous to eq.(\ref{4.1}), for the scalar and pseudoscalar states.

One believes that the sum rule (\ref{4.1}) must rather accurately be valid at
$q^2 <0$.

For the $n$-th derivative of eq.(\ref{4.1}) at $q^2=0$ one gets
\begin{equation}
f_V^2 (M_V^2)^{-n} = \frac{1}{\pi}\;
\int_{s_i}^{s_{{\rm th}}} \frac{\Im m \Pi_V^{\rm QCD(pert)}(s)}{s^{n+1}}\; {\rm
d}s +
\frac{(-1)^n}{n!}\;\frac{{\rm d}^n}{{\rm d}(q^2)^n} \Pi_V^{\rm
QCD(nonpert)}(q^2)\;,
\label{4.4}
\end{equation}
so, considering the ratio of the $n$-th derivative to the $n+1$-th
one, one can obtain the value of the vector $B_c^*$ meson mass. The
calculation result depends on the $n$ number in the sum rules (\ref{4.4}),
because of taking into the account both the finite number of terms
in the perturbation theory expansion and the restricted set of composite
operators.

The analogous procedure can be performed in the sum rule scheme with the
Borel transform, leading to the dependence of the results on the transformation
parameter.

As one can see from eq.(\ref{4.4}), the result, obtained in the framework
of the QCD sum rules, depends on the choice of the values for the hadronic
continuum threshold energy and the current masses of quarks. Then, this
dependence causes large errors in the estimates of the masses for the lightest
pseudoscalar, vector and scalar  $(\bar b c)$ states.

Thus, the QCD sum rules give the estimates of the quark binding energy in the
quarkonium, and the estimates are in agreement with the results of the
potential models, but sum rules involve a considerable parametric uncertainty.

\subsection{Radiative transitions in the $B_c$ family}

The $B_c$ mesons have no annihilation channels for the decays due to QCD and
electromagnetic interactions. Therefore, the mesons, lying below the threshold
for the $B$ and $D$ mesons production, will, in a cascade way, decay into
the $0^-(1S)$ state by emission of $\gamma$ quanta and $\pi$ mesons.
Theoretical estimates of the transitions between the levels with the emission
of the $\pi$ mesons have uncertainties, and the electromagnetic transitions
are quite accurately calculable.

\subsubsection{Electromagnetic transitions}

The formulae for the radiative E1-transitions have form [\cite{1r,2r}]
\begin{eqnarray}
\Gamma(\bar nP_J\to n^1S_1 +\gamma) & = & \frac{4}{9}\;\alpha_{{\rm
em}}\;Q^2_{{\rm eff}}\;
       \omega^3\;I^2(\bar nP;nS)\;w_J(\bar nP) \;,\nonumber\\
\Gamma(\bar nP_J\to n^1S_0 +\gamma) & = & \frac{4}{9}\;\alpha_{{\rm
em}}\;Q^2_{{\rm eff}}\;
       \omega^3\;I^2(\bar nP;nS)\;(1-w_J(\bar n P)) \;,\nonumber\\
\Gamma(n^1S_1\to \bar n P_J +\gamma) & = & \frac{4}{27}\;\alpha_{{\rm
em}}\;Q^2_{{\rm eff}}\;
       \omega^3\;I^2(nS;\bar n P)\;(2J+1)\;w_J(\bar n P)\;, \label{7} \\
\Gamma(n^1S_0\to \bar n P_J +\gamma) & = & \frac{4}{9}\;\alpha_{{\rm
em}}\;Q^2_{{\rm eff}}\;
       \omega^3\;I^2(nS;\bar n P)\;(2J+1)\;(1-w_J(\bar n P))\;,\nonumber\\
\Gamma(\bar n P_J\to n D_{J'} +\gamma) & = & \frac{4}{27}\;\alpha_{{\rm
em}}\;Q^2_{{\rm eff}}\;
       \omega^3\;I^2(nD;\bar n P)\;(2J'+1)\;w_J(\bar n P)) w_{J'}(nD)
        S_{JJ'}\;,\nonumber\\
\Gamma(n D_J\to \bar n P_{J'} +\gamma) & = & \frac{4}{27}\;\alpha_{{\rm
em}}\;Q^2_{{\rm eff}}\;
       \omega^3\;I^2(nD;\bar n P)\;(2J'+1)\;w_{J'}(\bar n P)) w_{J}(nD)
        S_{J'J}\;,\nonumber
\end{eqnarray}
where $\omega$ is the photon energy, $\alpha_{\rm em}$ is the electromagnetic
fine structure constant.

In eq.(\ref{7}) one uses
\begin{equation}
   Q_{{\rm eff}}=\frac{m_c Q_{\bar b} - m_b Q_c }{m_c +m_b}\;,
\end{equation}
where $Q_{c,b}$ are the electric charges of the quarks. For the
$B_c$ meson with the parameters from the Martin potential, one gets
$Q_{eff}=0.41$.

$w_J(nL)$ is the probability that the spin $S=1$
in the $nL$ state, so that
$w_0(nP)=w_2(nP)=1$, $w_1(nD)=w_3(nD)=1$, and the $w_1(nP)$, $w_2(nD)$
values have been presented in the previous section (see
eqs.(\ref{w.2p}), (\ref{w.3p}), (\ref{w.3d}) ).

The statistical factor $S_{JJ'}$ takes values [\cite{2r}]

\begin{center}
\begin{tabular}{ccl}
$J$  & $J'$  & $S_{JJ'}$\\
0 & 1 & 2\\
1 & 1 & 1/2\\
1 & 2 & 9/10\\
2 & 1 & 1/50\\
2 & 2 & 9/50\\
2 & 3 & 18/25.
\end{tabular}
\end{center}

The $I(\bar nL;nL')$ value is expressed through the radial wave functions,
\begin{equation}
   I(\bar n L;nL') = |\int R_{\bar n L}(r) R_{nL'}(r) r^3 {\rm d}r|\;.
\label{7aa}
\end{equation}
For the set of the transitions one obtains
\begin{eqnarray}
 I(1S,2P) & = & 1.568\; {\rm GeV}^{-1}\;,\;\;\;
 I(1S,3P)  =    0.255\; {\rm GeV}^{-1}\;,\nonumber\\
 I(2S,2P) & = & 2.019\; {\rm GeV}^{-1}\;,\;\;\;
 I(2S,3P)  =    2.704\; {\rm GeV}^{-1}\;, \\
 I(3D,2P) & = & 2.536\; {\rm GeV}^{-1}\;,\;\;\;
 I(3D,3P)  =    2.416\; {\rm GeV}^{-1}\;.\nonumber
\end{eqnarray}

For the dipole magnetic transitions one has [\cite{2k,1r,2r}]
\begin{equation}
\Gamma(\bar n^1S_i\to n^1S_f +\gamma) =  \frac{16}{3}\;\mu^2_{{\rm
eff}}\;\omega^3\;
(2f+1)\;A_{if}^2\;,
\label{10}
\end{equation}
where
$$ A_{if} = \int R_{\bar n S}(r) R_{nS}(r) j_0(\omega r/2) r^2\; {\rm d}r\;,
$$
and
\begin{equation}
\mu_{{\rm eff}}=\frac{1}{2}\;\frac{\sqrt{\alpha_{{\rm em}}}}{2m_c m_b}\;
(Q_c m_b - Q_{\bar b} m_c)\;.
\label{11}
\end{equation}
Note, in contrast to the $\psi$  and $\Upsilon$ particles, the total width of
the $B_c^*$ meson is equal to the width of its radiative decay into
the $B_c(0^-)$ state.

\begin{table}[t]
\caption{The energies (in MeV) and widths (in keV) of the electromagnetic
E1-transitions in the $(\bar b c$) family ($\ast$ is the present paper)}
\label{tr1}
\begin{center}
\begin{tabular}{||l|c|r|r||}
\hline
Transition & $\omega$ & $\Gamma$[$\ast$] & $\Gamma$[\cite{10k}]\\
\hline
$2P_2\to 1S_1 +\gamma$ & 426  & 102.9  & 112.6 \\
$2P_0\to 1S_1 +\gamma $& 366  & 65.3  & 79.2 \\
$2P\;1'^+\to 1S_1 +\gamma$ & 412  & 8.1  & 0.1\\
$2P\;1^+\to 1S_1 +\gamma$ & 400  & 77.8  & 99.5\\
$2P\;1'^+\to 1S_0 +\gamma$ & 476  & 131.1  & 56.4 \\
$2P\;1^+\to 1S_0 +\gamma$ & 464  & 11.6 & 0.0 \\
$3P_2\to 1S_1 +\gamma$ & 817  & 19.2  & 25.8 \\
$3P_0\to 1S_1 +\gamma $& 771  &  16.1 & 21.9 \\
$3P\;1'^+\to 1S_1 +\gamma$ & 807  & 2.5 & 2.1 \\
$3P\;1^+\to 1S_1 +\gamma$ & 796  & 15.3  & 22.1\\
$3P\;1'^+\to 1S_0 +\gamma$ & 871  & 20.1  & --~~ \\
$3P\;1^+\to 1S_0 +\gamma$ & 860  & 3.1  & --~~ \\
$3P_2\to 2S_1 +\gamma$ & 232  & 49.4  & 73.8 \\
$3P_0\to 2S_1 +\gamma $& 186  & 25.5  & 41.2 \\
$3P\;1'^+\to 2S_1 +\gamma$ & 222  & 5.9  & 5.4 \\
$3P\;1^+\to 2S_1 +\gamma$ & 211  & 32.1  & 54.3 \\
$3P\;1'^+\to 2S_0 +\gamma$ & 257  & 58.0 & --~~ \\
$3P\;1^+\to 2S_0 +\gamma$ & 246  & 8.1  & --~~ \\
$2S_1\to 2P_2 +\gamma$ & 159  & 14.8  & 17.7   \\
$2S_1\to 2P_0 +\gamma$ & 219  & 7.7  & 7.8   \\
$2S_1\to 2P\;1'^+ +\gamma$ & 173  & 1.0  & 0.0   \\
$2S_1\to 2P\;1^+ +\gamma$ & 185  & 12.8  & 14.5   \\
$2S_0\to 2P\;1'^+ +\gamma$ & 138  & 15.9  & 5.2   \\
$2S_0\to 2P\;1^+ +\gamma$ & 150  & 1.9  & 0.0   \\
\hline
\end{tabular}
\end{center}
\end{table}

The electromagnetic widths, calculated with accordance of
eqs.(\ref{7}),(\ref{10}),
and the frequencies of the emitted photons are presented in Tables \ref{tr1},
\ref{tr2}, \ref{tr3}.

Note, E0-transitions with the conversion of virtual $\gamma$-quantum
into the lepton pair can take place. Moreover, due to the tensor forces,
the states with $J>0$ and $S=1$ can, in addition to the $L$-wave, have the
admixture of $|L\pm 2|$-waves, giving the quadrupole moment to the
corresponding states and causing the E2-transitions. However, the
mentioned transitions are suppressed by the additional factor $\alpha_{\rm em}$
in the first case, and by the small value of amplitude, determining, say,
the probability of the admixture appearance of the $D$-wave in the
$1^-(nS)$ state.
\begin{table}[t]
\caption{The energies (in MeV) and widths (in keV) of the electromagnetic
E1-transitions in the $(\bar b c$) family ($\ast$ is the present paper)}
\label{tr2}
\begin{center}
\begin{tabular}{||l|c|r|r||}
\hline
Transition & $\omega$ & $\Gamma$[$\ast$] & $\Gamma$[\cite{10k}]\\
\hline
$3P_2\to 3D_1 +\gamma$ & 126  & 0.1 & 0.2 \\
$3P_2\to 3D\;2'^- +\gamma$ & 118  & 0.5  & --~~\\
$3P_2\to 3D\;2^- +\gamma$ & 133  & 1.5 & 3.2\\
$3P_2\to 3D_3 +\gamma$ & 127 & 10.9  & 17.8 \\
$3P_0\to 3D_1 +\gamma $& 80  & 3.2 & 6.9 \\
$3P\;1'^+\to 3D_1 +\gamma$ & 116  & 0.3 & 0.4 \\
$3P\;1^+\to 3D_1 +\gamma$ & 105  & 1.6  & 0.3 \\
$3P\;1'^+\to 3D\;2'^- +\gamma$ & 108  & 3.5  & --~~ \\
$3P\;1^+\to 3D\;2^- +\gamma$ & 112  & 3.9 & 9.8\\
$3P\;1'^+\to 3D\;2^- +\gamma$ & 123  & 2.5  & 11.5 \\
$3P\;1^+\to 3D\;2'^- +\gamma$ & 97  & 1.2 & --~~\\
$3D_3 \to 2P_2 + \gamma$ & 264  & 76.9  & 98.7  \\
$3D_1 \to 2P_0 + \gamma$ & 325  & 79.7  & 88.6  \\
$3D_1 \to 2P\;1'^+ + \gamma$   & 279  & 3.3  & 0.0  \\
$3D_1 \to 2P\;1^+ + \gamma $  & 291  & 39.2  & 49.3  \\
$3D_1 \to 2P_2 + \gamma $  & 265  & 2.2  & 2.7  \\
$3D\;2'^- \to 2P_2 + \gamma$ & 273  & 6.8  & --~~   \\
$3D\;2'^- \to 2P_2 + \gamma$ & 258  & 12.2  & 24.7  \\
$3D\;2'^- \to 2P\;1'^+ + \gamma$ & 287  & 46.0  & 92.5  \\
$3D\;2'^- \to 2P\;1^+ + \gamma$ & 301  & 25.0  & --~~  \\
$3D\;2^- \to 2P\;1'^+ + \gamma$ & 272  & 18.4  & 0.1  \\
$3D\;2^- \to 2P\;1^+ + \gamma$ & 284  & 44.6  & 88.8  \\
\hline
\end{tabular}
\end{center}
\end{table}

Thus, the registration of the cascade electromagnetic transitions in the
$(\bar b c)$ family can be used for the observation of the higher
$(\bar b c)$ excitations, having no annihilation channels of the decays.
\begin{table}[t]
\caption{The energies (in MeV) and widths (in keV) of the electromagnetic
M1-transitions in the $(\bar b c$) family ($\ast$ is the present paper)}
\label{tr3}
\begin{center}
\begin{tabular}{||l|r|r|r||}
\hline
Transition & $\omega$ & $\Gamma$[$\ast$] & $\Gamma$[\cite{10k}]\\
\hline
$2S_1\to 1S_0 +\gamma$ & 649  & 0.098  & 0.123   \\
$2S_0\to 1S_1 +\gamma$ & 550  & 0.096  & 0.093   \\
$1S_1\to 1S_0 +\gamma$ & 64  & 0.060  & 0.135   \\
$2S_1\to 2S_0 +\gamma$ & 35  & 0.010  & 0.029   \\
\hline
\end{tabular}
\end{center}
\end{table}

\subsubsection{Hadronic transitions}

In the framework of QCD the consideration of the hadronic
transitions between the states of the
heavy quarkonium family is built on the basis of the multipole expansion
for the gluon emission by the heavy nonrelativistic quarks [\cite{24k}], with
forthcoming hadronization of gluons, independently of the heavy
quark motion.

In the leading approximation over the velocity of the heavy quark motion,
the action, corresponding to the heavy quark coupling to the external
gluon field,
\begin{equation}
S_{\rm int} = -g\;\int {\rm d}^4x\;A_\mu^a(x)\cdot j^\mu_a(x)\;,
\end{equation}
can be expressed in the form
\begin{equation}
S_{\rm int}  =  g\;\int {\rm d}t \;
r^k E_k^a(t,{\bf{x}})\;\frac{\lambda_a^{ij}}{2} \Psi_n({\bf{r}})
\Psi_f^{ji}({\bf{r}}) \;K(s_n,f) \;{\rm d}^3{\bf{r}}\;,
\end{equation}
where $\Psi_n(\bf{r})$ is the wave function of the quarkonium, emitting
gluon, $\Psi_f^{ij}(\bf{r})$ is the wave function of the colour-octet
state of the quarkonium, $K(s_n,f)$  corresponds to the spin factor (in the
leading approximation, the heavy quark spin is decoupled from the interaction
with the gluons).

Then the matrix element for the E1-E1 transition of the quarkonium
$nL_J \to n'L'_{J'} + gg$ can be written in the form
\begin{eqnarray}
M(nL_J \to n'L'_{J'}+gg) & = & 4\pi \alpha_S\;E_k^a E_m^b\;
\cdot \nonumber \\
{}~ & ~ & \int {\rm d}^3r {\rm d}^3r'\;r_k r'_m\;G^{ab}_{s_{n'},s_n}(r,r')\;
\Psi_{nL_J}(r) \Psi_{n'L'_{J'}}(r'),
\label{h1}
\end{eqnarray}
where $G^{ab}_{s_{\bar n},s_n}(r,r')$ corresponds to the propagator of the
colour-octet state of the heavy quarkonium
\begin{equation}
G = \frac{1}{\epsilon - H_{Q\bar Q}^c}\;,
\end{equation}
where $H_{Q\bar Q}^c$ is the hamiltonian of the coloured state.

One can see from eq.(\ref{h1}), that the determination of the transition
matrix element depends on both the wave function of the quarkonium and the
hamiltonian $H_{Q\bar Q}^c$.
Thus, the theoretical consideration of the hadronic transitions in the
quarkonium family is model dependent.

In a number of papers of ref.[\cite{25k}], for the calculation of the values
such as (\ref{h1}), the potential approach has been developed.

In papers of ref.[\cite{26k}] it is shown that nonperturbative conversion
of the gluons into the $\pi$ meson pair allows one to give a consideration
in the framework of the low-energy theorems in QCD, so that this consideration
agrees with the papers, performed in the framework of PCAC and soft pion
technique [\cite{27k}].

However, as it follows from eq.(\ref{h1}) and the Wigner-Eckart theorem,
the differential width for the E1-E1 transition allows the representation
in the form [\cite{25k}]
\begin{equation}
\frac{{\rm d}\Gamma}{{\rm d}m^2}(nL_J \to n'L'_{J'} + h) = (2J'+1)\;
\sum_{k=0}^{2} \left\{\begin{array}{ccc}
k& L& L'\\ s & J'& J \end{array}\right\}^2 A_k(L,L')\;,
\label{h2}
\end{equation}
where $m^2$ is the invariant mass of the light hadron system $h$, $\{~\}$ are
6j-symbols, $A_k(L,L')$ is the contribution by the irreducible tensor of the
rang, equal to $k < 3$,
$s$ is the total quark spin inside the quarkonium.

In the limit of soft pions, one has $A_1(L,L')=0$.

{}From eqs.(\ref{h1}), (\ref{h2}) it follows, that, with the accuracy up to the
difference in the phase spaces, the widths of the hadronic transitions in the
$(Q\bar Q)$ and $(Q\bar Q')$ quarkonia are related to the following expression
[\cite{24k,25k}]
\begin{equation}
\frac{\Gamma  (Q \bar Q')}{\Gamma (Q \bar Q)} = \frac{\langle r^2(Q\bar
Q')\rangle^2}
{\langle r^2(Q\bar Q)\rangle^2}\;.
\label{h3}
\end{equation}
Then the experimental data on the transitions of $\psi ' \to J/\psi + \pi\pi$,
$\Upsilon ' \to \Upsilon + \pi\pi$, $\psi(3770) \to J/\psi + \pi\pi$
[\cite{28k}] allow one to extract the values of $A_k(L,L')$ for the transitions
$2S \to 1S + \pi\pi$ and $3D \to 1S + \pi\pi$ [\cite{10k}].

The invariant mass spectrum of the $\pi$ meson pair has the universal form
[\cite{26k,27k}]
\begin{equation}
\frac{1}{\Gamma}\;\frac{{\rm d}\Gamma}{{\rm d}m} = B\;\frac{|{\bf k}_{\pi\pi}|}
{M^2}\;(2x^2 -1 )\;\sqrt{x^2-1}\;,
\label{h4}
\end{equation}
where $x=m/2m_\pi$, $|{\bf k}_{\pi\pi}|$ is the $\pi\pi$ pair momentum.

The estimates for the widths of the hadronic transitions in the $(\bar b c)$
family have been made in ref.[\cite{10k}]. The hadronic transition widths,
having the values comparable with the electromagnetic transition width values,
are presented in Table \ref{th1}.
\begin{table}[b]
\caption{The widths (in keV) of the radiative hadronic transitions in the
$(\bar b c)$ family}
\label{th1}
\begin{center}
\begin{tabular}{||c|c||}
\hline
Transition & $\Gamma$ [\cite{10k}] \\
\hline
$2S_0 \to 1S_0 + \pi\pi$ & 50 \\
$2S_1 \to 1S_1 + \pi\pi$ & 50 \\
$3D_1 \to 1S_1 + \pi\pi$ & 31 \\
$3D_2 \to 1S_1 + \pi\pi$ & 32 \\
$3D_3 \to 1S_1 + \pi\pi$ & 31 \\
$3D_2 \to 1S_0 + \pi\pi$ & 32 \\
\hline
\end{tabular}
\end{center}
\end{table}

The transitions in the $(\bar b c)$ family with the emission of $\eta$ mesons
are suppressed by the low value of the phase space.

Thus, the registration of the hadronic transitions in the $(\bar b c)$ family
with the emission of the $\pi$ meson pairs can be used to observe the higher
2S- and 3D-excitations of the basic state.

\subsection{Leptonic constant of $B_c$ meson}

As we have seen in Section 2.1, the value of the leptonic constant of the
$B_c$ meson determines the splitting of the basic 1S-state of the
($\bar b c$) system. Moreover, the higher excitations in the ($\bar b c$)
system transform, in a cascade way, into the lightest $0^-$ state of $B_c$,
whose widths of the decays are essentially determined by the value of
$f_{B_c}$, too. In the quark models
[\cite{1a}--\cite{3a}], used to calculate the
weak decay widths of mesons, the leptonic constant, as the parameter,
determines the quark wave package inside the meson (generally,
the wave function is chosen in the oscillator form), therefore, the practical
problem for the extraction of the value for the weak charged current mixing
matrix element $|V_{bc}|$ from the data on the weak $B_c$ decays can
be only solved at the known value of $f_{B_c}$.

Thus, the leptonic constant $f_{B_c}$ is the most important quantity,
characterizing the bound state of the ($\bar b c$) system.

In the present Section we calculate the value of $f_{B_c}$ in different ways.

To describe the bound states of the quarks, the use of the
nonperturbative approaches is required.
The bound states of the heavy quarks allow one to
consider simplifications, connected to both large values of the quark masses
$\Lambda_{QCD}/m_Q \ll 1$ and the nonrelativistic quark motion $v \to 0$.
Therefore the value of $f_{B_c}$ can be quite reliably determined in the
framework of the potential models and the QCD sum rules [\cite{23k}].

\subsubsection{$f_{B_c}$ from potential models}

In the framework of the nonrelativistic potential models, the leptonic
constants of the pseudoscalar and vector mesons (see eqs.(\ref{59s}),
(\ref{60s}))
\begin{eqnarray}
\langle 0|\bar c(x) \gamma_\mu b(x)|B_c^*(p,\epsilon)\rangle & = & i f_V\; M_V
\;
\epsilon_\mu\; \exp({ipx})\;, \\
\langle 0|\bar c(x) \gamma_5 \gamma_\mu b(x)|B_c(p)\rangle & = & i f_P\; p_\mu
\;
 \exp({ipx})\;,
\end{eqnarray}
are determined by expression (\ref{27s})
\begin{equation}
f_V = f_P = \sqrt{\frac{3}{\pi M_{B_c(1S)}}} R_{1S}(0)\;,
\label{3a}
\end{equation}
where $R_{1S}(0)$ is the radial wave function of the $1S$ state of the
($\bar b c$) system, at the origin. The wave function is calculated by
solving the Schr\" odinger equation with  different potentials
[\cite{2k}--\cite{4k,5k,6ka}], in the quasipotential approach [\cite{5a}] or
by solving the Bethe-Salpeter equation with instant potential and in the
expansion up to the second order over the quark motion
velocity $v/c$ [\cite{6a,7a}].

The values of the leptonic $B_c$ meson constant, calculated in different
potential models and effective Coulomb potential with the "running"
$\alpha_S$ constant, determined in Section 2.1, are presented in
Table \ref{t1a}.
\begin{table}[t]
\caption{The leptonic $B_c$ meson constant (in MeV), calculated in the
different
potential models (the accuracy $\sim 15 \%$)}
\label{t1a}
\begin{center}
\begin{tabular}{||c|c|c|c|c|c|c|c||}
\hline
Model & Martin & Coulomb & [\cite{5ka}] & [\cite{10k}]& [\cite{5a}] &
[\cite{6a,7a}]&[\cite{8a}] \\
\hline
$f_{B_c}$ & 510 & 460 & 570 & 495 & 410 & 600 & 500 \\
\hline
\end{tabular}
\end{center}
\end{table}

Thus, in the approach accuracy, the potential quark models give the $f_{B_c}$
values, which are in a good agreement with each other, so that
\begin{equation}
f^{\rm pot}_{B_c} = 500 \pm 80\;\;{\rm MeV}\;.
\label{4a}
\end{equation}

\subsubsection{$f_{B_c}$ from QCD sum rules}

In the framework of the QCD sum rules [\cite{23k}], expressions
(\ref{4.1})-(\ref{4.4}) have been derived for the vector states. The
expressions are considered at $q^2 <0$ in the schemes of the spectral density
moments (\ref{4.4}) or with the application of the Borel transform
[\cite{23k}].
As one can see from eqs.(\ref{4.1}) - (\ref{4.4}), the result of the QCD sum
rule calculations is determined not only by physical parameters such as
the quark and meson masses, but also by the unphysical parameters of the
sum rule scheme such as the number of the spectral density moment or the Borel
transformation parameter. In the QCD sum rules, this unphysical dependence
of the $f_{B_c}$ value is due to the consideration being performed
with the finite number of terms in the expansion of the QCD perturbation
theory for the Wilson coefficients of the unit and composite operators.
In the calculations, the set of the composite operators is also restricted.

Thus, the ambiguity in the choice of the hadronic continuum threshold and the
parameter of the sum rule scheme essentially reduces the reliability of the
QCD sum rule predictions for the leptonic constants of the vector and
pseudoscalar $B_c$ states.

Moreover, the nonrelativistic quark motion inside the heavy quarkonium
$v \to 0$ leads to the $\alpha_S/v$-corrections, becoming the most important,
to the perturbative part of the quark current correlators, where
$\alpha_S$ is the effective Coulomb coupling constant in the heavy quarkonium.
As it is noted in refs.[\cite{23k,7k,9a}], the Coulomb
$\alpha_S/v$-corrections can be summed up and represented in the form of the
factor, corresponding to the Coulomb wave function of the heavy quarks, so
that
\begin{equation}
F(v) = \frac{4\pi \alpha_S}{3v}\; [1-\exp(-4\pi \alpha_S/3v)]^{-1}\;,
\label{5a}
\end{equation}
where $2v$ is the relative velocity of the heavy quarks inside the quarkonium.
The expansion of the factor (\ref{5a}) in the first order over $\alpha_S/v$
\begin{equation}
F(v) \approx 1+\frac{2\pi \alpha_S}{3v}\;,
\label{6a}
\end{equation}
gives the expression, obtained in the first order of the QCD perturbation
theory [\cite{23k}].

Note, the $\alpha_S$ parameter in eq.(\ref{5a}) should be at the scale of the
characteristic quark virtualities in the quarkonium (see Section 2.1),
but at the scale of the quark or quarkonium masses, as sometimes
one makes it thereby decreasing the value of factor (\ref{5a}).

The choice of the $\alpha_S$ parameter essentially determines the spread of
the sum rule predictions for the $f_{B_c}$ value (see Table \ref{t2a})
\begin{equation}
f^{SR}_{B_c} = 160 \div 570\;\;{\rm MeV}\;.
\label{7a}
\end{equation}

As one can see from eq.(\ref{7a}), the ambiguity in the choice of the QCD
sum rule parameters leads to the essential deviations in the results from
the $f_{B_c}$ estimates (\ref{4a}) in the potential models.
\begin{table}[t]
\caption{The leptonic $B_c$ constant (in MeV), calculated in the QCD sum rules
(SR -- the scaling relation)}
\label{t2a}
\begin{center}
\begin{tabular}{||c|c|c|c|c|c|c|c|c||}
\hline
Model & [\cite{9a}]& [\cite{20k}] & [\cite{21k}] & [\cite{22k}]& [\cite{10a}] &
[\cite{11a}]&[\cite{12a}]& SR[\cite{7k}] \\
\hline
$f_{B_c}$ & 375 & 400 & 360 & 300 & 160 & 300 & 450& 460\\
\hline
\end{tabular}
\end{center}
\end{table}

However, as it has been noted in Section 2.1,

1) the large value of the heavy quark masses $\Lambda_{QCD}/m_Q \ll 1$,

2) the nonrelativistic heavy quark motion inside the heavy quarkonium
$v \to 0$, and

3) the universal scaling properties of the potential in the
heavy quarkonium, when
the kinetic energy of the quarks and the quarkonium state density do not depend
on the heavy quark flavours (see eqs.(\ref{v}) - (\ref{e2})),

allow one to state the scaling relation (\ref{law}) for the leptonic constants
of the S-wave quarkonia
$$ \frac{f^2}{M}\;\biggl(\frac{M}{4\mu}\biggr)^2 = {\rm const}\;.
$$

Indeed,
at $\Lambda_{QCD}/m_Q \ll 1$ one can neglect the quark-gluon condensate
contribution, having the order of magnitude $O(1/m_b m_c)$ (the contribution
into the $\psi$ and $\Upsilon$ leptonic constants is less than 15\%).
At $v\to 0$ one has to take into account the Coulomb-like
$\alpha_S/v$-corrections in the form of factor (\ref{5a}), so that the
imaginary part of the correlators for the vector and axial quark currents
has the form
\begin{equation}
\Im m \Pi_V(q^2) \approx \Im m \Pi_P(q^2) = \frac{\alpha_S}{2}\;q^2\;
\biggl(\frac{M}{4\mu}\biggr)^2\;,
\label{8a}
\end{equation}
where
$$ v^2 = 1- \frac{4m_b m_c}{q^2-(m_b-m_c)^2}\;,\;\; v\to 0\;.
$$
Moreover, condition (\ref{e2}) can be used in the specific QCD sum rule scheme,
so that this scheme excludes the dependence of the results on the parameters
such as the number of the spectral density moment or the Borel parameter.

Indeed, for example, the resonance contribution into the hadronic part
of the vector current correlator, having the form
\begin{equation}
\Pi_V^{({\rm res})}(q^2)  =  \int \frac{{\rm d}s}{s-q^2}\;\sum_n f^2_{Vn}
M^2_{Vn}
\delta(s-M_{Vn}^2)\;,
\end{equation}
can be rewritten as
\begin{equation}
\Pi_V^{({\rm res})}(q^2) = \int \frac{{\rm d}s}{s-q^2}\; s
f^2_{Vn(s)}\;\frac{{\rm d}n(s)}{{\rm d}s}\;
\frac{{\rm d}}{{\rm d}n} \sum_k \theta(n-k)\;.
\end{equation}
where $n(s)$ is the number of the vector S-state versus the mass, so that
\begin{equation}
n(m_k^2) = k\;.
\end{equation}
Taking the average value for the derivative of the step-like function, one gets
\begin{equation}
\Pi_V^{({\rm res})}(q^2) = \langle \frac{{\rm d}}{{\rm d}n} \sum_k
\theta(n-k)\rangle\; \int \frac{{\rm d}s}{s-q^2}
s f^2_{Vn(s)} \frac{{\rm d}n(s)}{{\rm d}s}\;,
\end{equation}
and, supposing
\begin{equation}
\langle \frac{{\rm d}}{{\rm d}n} \sum_k \theta(n-k)\rangle \approx 1\;,
\end{equation}
one can, in average, write down
\begin{equation}
\Im m \langle \Pi^{\rm (hadr)}(q^2)\rangle = \Im m \Pi^{\rm (QCD)}(q^2)\;,
\end{equation}
so, taking into the account the Coulomb factor and neglecting power
corrections over $1/m_Q$, at the physical points $s_n =M_n^2$ one obtains
\begin{equation}
\frac{f_n^2}{M_n}\; \biggl(\frac{M}{4\mu}\biggr)^2 =
\frac{\alpha_S}{\pi} \; \frac{{\rm d}M_n}{{\rm d}n}\;,
\label{14a}
\end{equation}
where one has supposed that
\begin{eqnarray}
m_b +m_c & \approx & M_{B_c}\;,\label{15a}\\
f_{Vn} \approx f_{Pn} & = & f_n\;.
\end{eqnarray}
Further, as it has been shown in Section 2.1, in the heavy quarkonium the value
of ${\rm d}n/{\rm d}M_n$ does not depend on the
quark masses (see eq.(\ref{e2})),
and, with the accuracy up to the logarithmic corrections, $\alpha_S$ is the
constant value (the last fact is also manifested in the flavour independence
of the Coulomb part of the potential in the Cornell model). Therefore, one can
draw the conclusion that, in the leading approximation, the right hand side
of eq.(\ref{14a}) is the constant value, and there is the scaling relation
(\ref{law}) [\cite{7k}]. This relation is valid in the resonant region,
where one can neglect the contribution by the hadronic continuum.

Note, scaling relation (\ref{law}) is in a good agreement with  the
experimental data on the leptonic decay constants of the $\psi$  and
$\Upsilon$ particles (see Table \ref{t7}), for which one has $4\mu/M = 1$
[\cite{7k}].

The value of the constant in the right hand side of eq.(\ref{law}) is in
agreement with the estimate, when we suppose
\begin{equation}
\langle \frac{{\rm d}M_{\Upsilon}}{{\rm d}n}\rangle \approx
\frac{1}{2}[(M_{\Upsilon'}-M_{\Upsilon}) + (M_{\Upsilon''}- M_{\Upsilon'})]\;,
\end{equation}
and $\alpha_S= 0.36$, as it is in the Cornell model.

Further, in the limit case of $B$  and $D$ mesons,
when the heavy quark mass is much greater than the light quark mass
$m_Q \gg m_q$, one has
$$ \mu \approx m_q
$$
and
\begin{equation}
f^2\cdot M = \frac{16 \alpha_S}{\pi}\;\frac{{\rm d}M}{{\rm d}n}\;\mu^2\;.
\label{18a}
\end{equation}
Then it is evident that at one and the same $\mu$ one gets
\begin{equation}
f^2\;M = {\rm const}\;.
\label{19a}
\end{equation}
Scaling law (\ref{19a}) is very well known in EHQT
[\cite{13a}] for mesons with a single heavy quark ($Q\bar q$),
and it follows, for example, from the identity of the $B$  and $D$ meson
wave functions within the limit, when an infinitely heavy quark can
be considered as a static source of gluon field (then eq.(\ref{19a})
follows from eq.(\ref{3a})).

In our derivation of eqs.(\ref{18a}) and (\ref{19a}) we have neglected
power corrections over the inverse heavy quark mass.
Moreover, we have used the notion about the light constituent quark
with the mass equal to
\begin{equation}
m_q \approx 330\;\;{\rm MeV}\;,
\label{20a}
\end{equation}
so that this quark has to be considered as nonrelativistic one $v \to 0$,
and the following conditions take  place
\begin{eqnarray}
m_Q +m_q & \approx & M^{(*)}_{(Q\bar q)}\;,\;\;m_q \ll m_Q\;,
\label{21a} \\
f_{V} & \approx & f_{P} = f\;.
\end{eqnarray}

In agreement with eqs.(\ref{18a}) and (\ref{20a}), one finds the
estimates\footnote
{In ref.[\cite{7k}] the dependence of the S-wave state density
${\rm d}n/{\rm d}M_n$ on the reduced mass of the system with the Martin
potential
has been found by the Bohr-Sommerfeld quantization, so that at the step from
($\bar b b$) to ($\bar b q$), the density changes less than about 15\%.}
\begin{eqnarray}
f_{B^{(*)}} = 120 \pm 20\;\;{\rm MeV}\;, \\
f_{D^{(*)}} = 220 \pm 30\;\;{\rm MeV}\;,
\end{eqnarray}
that is in an agreement with the estimates in the other schemes of the QCD sum
rules [\cite{23k,14a}].

Thus, in the limits of $4\mu/M=1$ and $\mu/M \ll 1$, scaling relation
(\ref{law}) is consistent.

The $f_{B_c}$ estimate from eq.(\ref{law}) contains the uncertainty,
connected to the choice of the ratio for the $b$- and $c$-quark masses,
so that (see Table \ref{t2a})
\begin{equation}
f_{B_c} = 460 \pm 60\;\;{\rm MeV}\;.
\label{24a}
\end{equation}
In ref.[\cite{9a}] the sum rule scheme with the double Borel transform was
used. So, it allows one to study effects, related to the power corrections from
the gluon condensate, corrections due to nonzero quark velocity and nonzero
binding energy of the quarks in the quarkonium.

Indeed, for the set of narrow pseudoscalar states, one has the sum rules
\begin{equation}
\sum_{k=1}^\infty \frac{M_k^4 f_{Pk}^2}{(m_b+m_c)^2 (M_k^2-q^2)} =
\frac{1}{\pi} \int \frac{{\rm d}s}{s-q^2}\;\Im m \Pi_P(s) + C_G(q^2)\;
\langle \frac{\alpha_S}{\pi} G^2\rangle\;,
\label{a1}
\end{equation}
where
\begin{equation}
C_G(q^2) = \frac{1}{192 m_b m_c}\;\frac{q^2}{\bar q^2}\;
\biggl(\frac{3(3v^2+1)(1-v^2)^2}
{2v^5} \ln \frac{v+1}{v-1} - \frac{9v^4+4v^2+3}{v^4}\biggr)\;,
\end{equation}
and
\begin{equation}
\bar q^2 = q^2-(m_b-m_c)^2\;,\;\;\;v^2=1-\frac{4m_bm_c}{\bar q^2}\;.
\end{equation}
Acting by the Borel operator $L_\tau(-q^2)$ on eq.(\ref{a1}), one gets
\begin{equation}
\sum_{k=1}^\infty \frac{M_k^4 f_{Pk}^2}{(m_b+m_c)^2}\;\exp({-M_k^2\tau}) =
\frac{1}{\pi} \int {{\rm d}s}\;\Im m \Pi_P(s) \;\exp({-s\tau})+ C'_G(\tau)\;
\langle \frac{\alpha_S}{\pi} G^2\rangle\;,
\label{a2}
\end{equation}
where
\begin{eqnarray}
L_\tau (x) & = & \lim_{n,x\to \infty} \frac{x^{n+1}}{n!}\;
\biggl(-\frac{{\rm d}}{{\rm d}x}\biggr)^n\;,\;\;\;\frac{n}{x}=\tau\;,\\
C'_G(\tau) & = & L_\tau(-q^2)\;C_G(q^2)\;.
\end{eqnarray}
For the exponential set in the left hand side of eq.(\ref{a2}),
one uses the Euler-MacLaurin formula
\begin{eqnarray}
\sum_{k=1}^\infty \frac{M_k^4 f_{Pk}^2}{(m_b+m_c)^2}\;\exp({-M_k^2\tau}) &=&
\int_{m_n}^{\infty} {{\rm d}M_k}\;
\frac{{\rm d}k}{{\rm d}M_k}\;M_k^4 f_{Pk}^2 \;
\exp({-M_k^2\tau})+ \nonumber\\
&&\sum_{k=0}^{n-1} {M_k^4 f_{Pk}^2}\;\exp({-M_k^2\tau})+\cdots\;.
\label{a3}
\end{eqnarray}
Making the second Borel transform $L_{M_k^2}(\tau)$ on eq. (\ref{a2})
with account of eq.(\ref{a3}), one finds the expression for the
leptonic constants of the pseudoscalar ($\bar b c$) states, so that
\begin{equation}
f_{Pk}^2 = \frac{2(m_b+m_c)^2}{M_k^3}\;\frac{{\rm d}M_k}{{\rm d}k} \;
\bigg\{\frac{1}{\pi} \Im m \Pi_P(M_k^2) + C''_G(M_k^2)\;
\langle \frac{\alpha_S}{\pi} G^2\rangle\bigg\}\;,
\label{a4}
\end{equation}
where we have used the following property of the Borel operator
\begin{equation}
L_\tau (x)\;x^n \exp({-bx}) \to \delta_+^{(n)}(\tau-b)\;.
\end{equation}
Explicit form for the spectral density and Wilson coefficients can be found in
ref.[\cite{9a}].

Expression (\ref{a4}) is in the agreement with the above performed derivation
of scaling relation (\ref{law}).

The numerical effect from the mentioned corrections is considered to be
not large (the power corrections are of the order of 10\%), and the
uncertainty, connected to the choice of the quark masses, dominates in
the error of the $f_{B_c}$ value determination (see eq.(\ref{24a})).

Thus, we have shown that, in the framework of the QCD sum rules, the most
reliable estimate of the $f_{B_c}$ value (\ref{24a}) is coming from
the use of the scaling relation (\ref{law}) for the leptonic decay constants of
the quarkonia, and this relation agrees very well with the results of the
potential models.


\section{Decays of $B_c$ mesons}

\subsection{Life time of $B_c$ mesons}

The processes of the $B_c$ meson decay can be subdivided into three classes
a) the $\bar b$-quark decay with the spectator $c$-quark, b) the $c$-quark
decay with the spectator $\bar b$-quark, c) the annihilation channel
$B_c^+\rightarrow l^+\nu_l (c\bar s, u\bar s)$, $l=e,\; \mu,\; \tau$.

The total width is summed from three partial widths
\begin{equation}
\Gamma (B_c\rightarrow X)=\Gamma (b\rightarrow X)
+\Gamma (c\rightarrow X)+\Gamma \mbox{(ann)}\;.
\end{equation}
The simplest estimates with no account for the quark binding inside the
$B_c$ meson and in the framework of the spectator mechanism of the decay for
the first and second cases, lead to the expressions
\begin{eqnarray}
\Gamma (b\rightarrow X) & = & \frac{G^2_F|V_{bc}|^2m^5_b}{192\pi^3}\cdot 9 \;,
 \nonumber \\
\Gamma (c\rightarrow X) & = & \frac{G^2_F|V_{cs}|^2m^5_c}{192\pi^3}\cdot 5  \;,
\label{d2}
\end{eqnarray}
So, that $m_b$ and $m_c$ are chosen for to represent correctly the spectator
parts of the total widths for the $B$  and $D$ mesons.

The width of the annihilation channel equals
\begin{equation}
\Gamma \mbox{(ann)} =\sum_i\frac{G^2_F}{8\pi}
|V_{bc}|^2f^2_{B_c}M_{bc} m^2_i \biggl(1-\frac{m^2_i}{m^2_{Bc}}\biggr)^2\cdot
C_i\;,
\label{d3}
\end{equation}
where $C_i = 1$ for the $\tau\nu_\tau$ channel and $C_i =3|V_{cs}|^2 $ for
the $\bar c s$ channel, and $m_i$ is the mass of the most heavy fermion
($\tau$ or $c$).
\begin{figure}[t]
\vspace*{7cm}
\caption{The diagrams of the $B_c$ meson decays: (a) the
$c$-spectator decay; (b) the $b$-spectator decay; (c) the annihilation}
\label{fd11}
\end{figure}

Note that in the case of the nonleptonic decays, an account of the strong
interaction results in the multiplicative factor of enhancement to formulae
(\ref{d2})-(\ref{d3}) (see section 3.2).

The mentioned widths, calculated with the use of the know values of parameters
$m_q,\;\; |V_{bc}|=0.046,\;\; |V_{cs}|=0.96$ etc., are presented in
Table \ref{td2.1}.
\begin{table}[b]
\caption{ The widths ($10^{-6}$ {\rm eV}) of the inclusive decays of $b$-
and $c$-quarks in free and bound states (in the $B_c$ meson) and the
branching ratios ($BR$ in $\%$) of inclusive $B_c$ decays}
\label{td2.1}
\begin{center}
\begin{tabular}{|l|r|r|r||l|r|r|r|}\hline
Decay& free  & $B^+_c$ & $BR$ &Decay & free & $B^+_c$ & $BR$\\
mode & quarks &       &       & mode & quarks & &  \\ \hline
$\bar b\rightarrow \bar c +e^+\nu_e$ & 62 & 62 & 4.7  &
$c\rightarrow s+e^++\nu_e$ & 124 & 74 & 5.6  \\
$\bar b\rightarrow \bar c +\mu^+\nu_\mu$ & 62 & 62 & 4.7  &
$c\rightarrow s+\mu^++\nu_\mu$ & 124 & 74 & 5.6  \\
$\bar b\rightarrow \bar c +\tau^+\nu_\tau$ & 14 & 14 & 1.0  &
$c\rightarrow s+u +\bar d$ & 675 & 405 & 30.5  \\
$\bar b\rightarrow \bar c +\bar d+u$ & 248 & 248 & 18.7  &
$c\rightarrow s+u +\bar s$ & 33 & 20 & 1.5  \\
$\bar b\rightarrow \bar c +\bar s+u$ & 13 & 13 & 1.0  &
$c\rightarrow d+e^+\nu$ & 7 & 4 & 0.3  \\
$\bar b\rightarrow \bar c +\bar s+c$ & 87 & 87 & 6.5  &
$c\rightarrow d+\mu^++\nu_\mu$ & 7 & 4 & 0.3  \\
$\bar b\rightarrow \bar c +\bar d+c$ & 5 & 5 & 0.4  &
$c\rightarrow d+u+\bar d$ & 39 & 23 & 1.7  \\
$B^+_c\to \tau^++\nu_\tau$ & -- & 63 & 4.7  &
$B^+_c\to c +\bar s$ & -- & 162 & 12.2  \\
$B^+_c\to c +\bar d$ & -- & 8 & 0.6  &  $B^+_c \to \mbox{all}$ &-- & 1328 &100
  \\ \hline
\end{tabular}
\end{center}
\end{table}

Thus, the rough estimate of the life time leads to
$\tau_{B_c} \approx (2 - 5)\cdot 10^{-13}$ s. So, the fraction of the
$c$-quark decay is approximately 50 \%, the $b$-quark one is 45 \%,
and the annihilation channel is 5 \%.
However, these estimates do not take into account a quite strong binding
of the quarks inside the $B_c$ meson: corresponding corrections to the
estimates can reach about 40 \%.

Let us consider this effect in the semileptonic modes of decays with the
spectator $\bar b$-quark. The final state of such decays generally contains
the $B_s^{(*)}$ mesons with the more small phase space of the lepton pair.

The effect of the phase space decrease is shown on Figure \ref{fd1},
where the kinematical borders of Dalitz plot for the
$B^+_c\rightarrow B_s e^+\nu$ decay are compared with the borders for the
$c$-quark and calculated at different values of the $c$-quark mass.
As one can see from Figure \ref{fd1}, the end-point of the leptonic spectrum
is approximately one and the same in the different decays
\begin{equation}
E^{\rm max} =\frac{M^2_{B_c}-M^2_{B_S}}{2M_{B_c}}\;.
\end{equation}
However, the maximum values of the leptonic pair masses $q^2_{max}$ are
different.

One can easily find that the spectator model better describes the
semileptonic decay $D\to K$.
In the case of the $B_c$ meson decay, the admissible kinematical region
is strongly reduced. With the account for the phase space reduction in
the spectator model, one can get [\cite{i14e}]:
\begin{equation}
\Gamma (B_c^+ \rightarrow X_be^+\nu)\approx 0.71\; \Gamma (D^+\rightarrow
X_s e^+\nu)\;.
\end{equation}

The effect of the phase space reduction does not sizably appear in the case of
decays with the spectator $c$-quark. For such decays, as one can see from
Figure \ref{fd2}, the spectator model well describes the $B$ meson
decays as well as the $B_c$ meson decays, and one can believe that
\begin{equation}
\Gamma (B_c^+ \rightarrow X_ce^+\nu)\approx \Gamma (B^+\rightarrow
X_c e^+\nu)\;.
\end{equation}
\begin{figure}[t]
\vspace*{8cm}
\caption{The Dalitz diagrams for the semileptonic decays:
($1$) $B_c \to B_s^*l\nu$, ($2$)  $B_c\to B_sl\nu$, ($3$) $D\to K^*l\nu$,
($4$) $D\to Kl\nu$, ($5$) $c\to sl\nu$  ($m_c=1.7$ {\rm GeV}, $m_s=0.55$
{\rm GeV}),
($6$) $c\to sl\nu$ ($m_c=1.5$ {\rm GeV}, $m_s = 0.15$ {\rm GeV}); $E$ is the
lepton energy,
$q^2$ is the square of the lepton pair mass}
\label{fd1}
\end{figure}

Another possible way of the estimate is related with the summation of the
exclusive decays into the channels $B_s e^+\nu$ and $B_s^\ast e^+\nu$.
In agreement with the same kinematical arguments, their sum is the main
fraction
of the semileptonic decays [\cite{Suzuki}]. If one neglects the decaying
quark momentum inside the $B_c$ meson, the admissible region of masses in the
inclusive semileptonic decay $Q\rightarrow Q'e\nu$ is varied within the limits
\begin{equation}
(m_{q'} +m_{{\rm sp}})^2 < (M^2_x) <
m^2_{q'} +m^2_{{\rm sp}} + m_{{\rm sp}}\frac{m^2_{q'}}{m_q}\;.
\label{appr}
\end{equation}
{}From approximate formula (\ref{appr}) with the use of the constituent quark
masses, one can see, that the admissible $M_x$ region
in the decay $B_c\rightarrow X_c$ is varied in the limit of 200 MeV and, hence,
the final state is saturated by the lowest states. For the considered case
($m_q=m_c=1.7$ {\rm GeV}, $m_{q'}=m_s= 0.55$ {\rm GeV}
and $m_{{\rm sp}}=m_b=5.1$ {\rm GeV}), this region has the widths, equal to
340 {\rm MeV}, that is less than the expected difference of masses
between the basic state and the first orbital excitation of the
($\bar bs$) system.
\begin{figure}[b]
\vspace*{8cm}
\caption{The Dalitz diagrams for the semileptonic decays:
($1$) $B_c \to \psi l\nu$, ($2$)  $B_c\to \eta_c l\nu$, ($3$) $B\to Dl\nu$,
($4$) $B\to D^*l\nu$, 5) $b\to cl\nu$, $E$ is the lepton energy,
$q^2$ is the square of the lepton pair mass}
\label{fd2}
\end{figure}

Thus, one can consider that
\begin{equation}
\Gamma (B_c^+ \rightarrow X_be^+\nu)\approx
\Gamma (B_c\rightarrow B_s+e\nu)+\Gamma(B_c\rightarrow B_s^{*}+e\nu)\;.
\end{equation}

The results of different quark models for the semileptonic $B_c$ decays
(see section 3.2) lead to the following sum of the widths of decays into
$B_s$ and $B_s^*$
\begin{equation}
\Gamma (B_c\rightarrow B_s+e\nu)+\Gamma(B_c\rightarrow B_s^{*}+e\nu)\approx
(60\pm 7) 10^{-15}\;{\rm GeV}\approx 0.5\; \Gamma(D^+\rightarrow X_se^+\nu)\;.
\label{d9}
\end{equation}

Accounting for the current theoretical uncertainties, one can calculate
\begin{equation}
\Gamma (B_c\rightarrow X_{\bar b} + e^+\nu) =
(0.6 \pm 0.2)\; \Gamma (D^+ \rightarrow X_se^+\nu)\;.
\end{equation}

For the $c$-spectator decays, the calculations in quark models and QCD
sum rules show, that the semileptonic decays are saturated by the transitions
into the lowest $\eta_c$  and $J/\psi$ states, i.e.
\begin{equation}
\Gamma (B_c^+ \rightarrow X_ce^+\nu)\approx
\Gamma(B_c^+\rightarrow(\eta_c+J/\psi)
e^+\nu)\approx \Gamma (B^+\rightarrow X_c e^+\nu)\;.
\end{equation}

With the account for these factors, the probabilities of inclusive decays
are presented in Table \ref{td2.1}.  The widths of the hadronic inclusive
decays,
which are in details discussed in Section 3.3, are also shown.

The compact sizes of $B_c$ meson lead to the large value of the weak
decay constant ($f_{B_c}\approx $ 500 {\rm MeV}),  that enforces the role of
the annihilation channel into the massive fermions $c, \tau$.
The decays of $B_c$ meson into the light fermions are suppressed due to
the forbidding over the spirality. Although the use of the effective masses
for the $u$- and $d$-quarks instead of the current masses can increase the
width of the annihilation channel into $u\bar d$, the latter will yet be
much less than the width into the heavy massive fermions.
In agreement with eq. (\ref{d3}), the conservative estimates of the
annihilation decay probabilities are presented in Table \ref{td2.1}.

Thus, the consideration of three types of processes for the $B_c$ meson decay
leads to the life time estimate
$$ \tau_{B_c} \sim 5\cdot 10^{-13}\;\;{\rm s}
$$
with the following approximate sharing of branching fractions:
37 \%, 45 \% and 18 \% correspond to the $c$-spectator mechanism,
$\bar b$-spectator one and the annihilation, respectively.

The uncertainty in the estimation of the $B_c$ meson life time is
generally related to the choice of quark masses. The mass of $b$-quark
$m_b=4.9$ {\rm GeV} is chosen so that one can describe the
$B$ meson life time in the framework of the spectator mechanism. Note,
that the differences of the life times for the charged and neutral
$B$ mesons are unessential and, hence, the given choice of the mass
is quite certain. For the $D$ mesons, this is not the case, since the
life times of $D^+$  and $D^0$ mesons differ twice.

Nevertheless, there is another way, that is more reliable for the
extraction of the $c$-quark mass, this is the consideration of the semileptonic
decays of $D$ mesons. Indeed, the value $m_c=1.5$ {\rm GeV}
in the spectator mechanism well describes the decays
$D^+\rightarrow \bar K^0 e^+ \nu$ and $D^0\rightarrow \bar K^- e^+ \nu$,
whose widths approximately are equal to each other.
Note, at any other reasonable choice of $m_c$ (from the total widths, say),
the error in the $B_c$ meson life time will be not large, since
the summed branching ratio of the $B_c$ meson decays due to the
$c$-quark decays is about 40 \%.

\subsection{Semileptonic decays of $B_c$ mesons}
\subsubsection{Quark models}

In the framework of quark models, the semileptonic decays of $B_c$ mesons are
considered in refs.[\cite{i14a,i14c,i14e}].
The detailed study of the $B_c$ meson decays in the quark model of the
relativistic oscillator WSB [\cite{3a}] has been first made in
ref.[\cite{i14c}] and further in ref.[\cite{i14e}], where the quark model
ISGW [\cite{2a}] has also been used. The covariant description approach,
early offered for the composed quarkonium model, is developed in
ref.[\cite{i14a}].
\begin{figure}[t]
\begin{center}
\begin{picture}(90,60)
\put(25,20){\circle*{3}}
\put(40,5){\circle*{3}}
\put(40,35){\circle*{3}}
\put(25,20){\line(1,1){15}}
\put(40,35){\line(0,-1){30}}
\put(40,5){\line(-1,1){15}}
\put(10,21){\line(1,0){15}}
\put(10,19){\line(1,0){15}}
\put(40,6){\line(1,0){15}}
\put(40,4){\line(1,0){15}}
\put(40,35){\line(1,0){5}}
\put(48,35){\line(1,0){5}}
\put(56,35){\line(1,0){5}}
\put(61,35){\line(1,1){15}}
\put(61,35){\line(1,-1){15}}
\put(61,35){\circle*{3}}

\put(10,25){$B_c$}
\put(78,48){$l$}
\put(78,18){$\nu$}
\put(50,42){$W$}
\put(30,30){$1$}
\put(42,18){$2$}
\put(57,5){$\psi, \;\eta_c, \; B_s^{(*)}$}
\end{picture}
\end{center}
\caption{The diagram of the semileptonic decay of $B_c$ meson}
\label{fd3}
\end{figure}
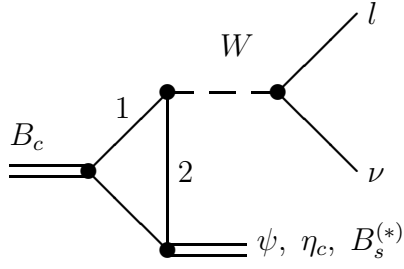

Consider the amplitude of the $B_c^+ \to M_X e^+ {\nu}_e$ transition with the
weak decay of the quark $1$ into the quark $2$ (Figure \ref{fd3})
\begin{equation}
A = \frac{G_{\rm F}}{\sqrt{2}}\;V_{12}\;l_{\mu}\;H^{\mu}\;,
\label{d12}
\end{equation}
where $G_{\rm F}$ is the Fermi constant, $V_{12}$ is the element of the
Kobayashi--Maskawa matrix. The lepton current $l_{\mu}$ is determined
by the expression
\begin{equation}
l_{\mu} = \bar e(q_1)\gamma_{\mu} (1 - \gamma_5) \nu(q_2)\;,
\end{equation}
where $q_{1}$ and $q_2$ are the lepton and neutrino momenta, respectively,
$(q_1+q_2)^2 = t$.

The $H_\mu$ quantity in eq.(\ref{d12}) is the matrix element of the
hadronic current $J_\mu$
\begin{equation}
J_{\mu} = V_\mu - A_\mu = \bar Q_1 \gamma_{\mu} (1 - \gamma_5) Q_2\;.
\end{equation}

The matrix element for the $B_c$ meson decay into the pseudoscalar state $P$
can be written down in the form
\begin{equation}
\langle B_c(p)| A_{\mu} |P(k)\rangle = F_+(t) (p+k)_{\mu} + F_-(t)
(p-k)_{\mu}\;,
\label{d16}
\end{equation}
and for the transition into the vector meson $V$ with the mass $M_V$
and the polarization $\lambda$,  one has
\begin{eqnarray}
\langle B_c(p)| J_{\mu} |V(k,\lambda)\rangle & = &
-(M+M_V) A_1(t)\;\epsilon_{\mu}^{(\lambda)} \nonumber \\
& ~ & +  \frac{A_2(t)}{M+M_V}\;(\epsilon^{(\lambda)} p)\;(p+k)_{\mu} \nonumber
\\
& ~ &  + \frac{A_3(t)}{M+M_V}\;(\epsilon^{(\lambda)} p)\;(p-k)_{\mu}  \nonumber
\\
& ~ &  + i \frac{2 V(t)}{M+M_V}\;\epsilon_{\mu \nu \alpha \beta}\;
     \epsilon_{(\lambda)}^{\nu} p^\alpha k^\beta.
\label{d17}
\end{eqnarray}

Relations (\ref{d16}), (\ref{d17}) define the form factors of the
$B_c^+ \to M_X e^+ {\nu}_e$ transitions, so, for the massless leptons,
$F_-$ and $A_3$ do not give contributions into matrix element (\ref{d12}).

In the Covariant model of the quarkonium (see Appendix I), one can easily find
\begin{eqnarray}
F_+(t) & = & \frac{1}{2}\; (m_1+m_2) \sqrt{\frac{M_P}{M}} \frac{1}{m_2}
      \xi _P (t)\;,\\
\label{d18}
F_-(t) & = & -\frac{1}{2}\; (m_1-m_2+2 m_{{\rm sp}}) \sqrt{\frac{M_P}{M}}
       \frac{1}{m_2} \xi _P (t)\;,
\end{eqnarray}
Here $m_{{\rm sp}}$ is the mass of the spectator quark (see Figure \ref{fd3}),
and the function $\xi (t)$ has the form
\begin{eqnarray}
\xi _X(t) & = &
\biggl(\frac{2 \omega \omega_X}{\omega^2+\omega_X^2}\biggr)^{3/2}\;
\exp\bigg\{-\frac{m_{{\rm sp}}^2}{\omega^2+\omega_X^2}\;\frac{t_{\rm max}-t}{M
M_X} \nonumber \\
& ~ & \biggl(1 + \frac{\omega^2}{\omega_X^2}
\biggl(1-\frac{t_{\rm max}-t}{4 M M_X}\biggr)\biggr)\bigg\}\;,
\label{d19}
\label{d20}
\end{eqnarray}
where $M_X$ is the recoil meson mass,  $\omega_X$ is the wave function
parameter (I.6)-(I.8) for the recoil meson.
\begin{equation}
t_{\rm max} = (M-M_X)^2\;,
\end{equation}
$t_{\rm max}$ is the maximal square of the lepton pair mass.

For the vector state one has $M_V=M_X$, and we obtain
\begin{eqnarray}
V(t) & = & \frac{1}{2}\; (M+M_V) \sqrt{\frac{M_V}{M}} \frac{1}{m_2} \xi _V
(t)\;,
\label{d22} \\
A_1(t) & = & \frac{1}{2}\; \frac{M^2+M_V^2-t+2 M (m_2-m_{{\rm sp}})}{M+M_V}
\sqrt{\frac{M_V}{M}} \frac{1}{m_2} \xi _V (t)\;,
\label{d23}\\
A_2(t) & = & \frac{1}{2}\; (M+M_V) \biggl(1-\frac{2 m_{{\rm sp}}}{M}\biggr)
            \sqrt{\frac{M_V}{M}} \frac{1}{m_2} \xi _V (t)\;,
\label{d24}\\
A_3(t) & = & -\frac{1}{2}\; (M+M_V) \biggl(1+\frac{2 m_{{\rm sp}}}{M}\biggr)
            \sqrt{\frac{M_V}{M}} \frac{1}{m_2} \xi _V (t)\;.
\label{d25}
\end{eqnarray}
It is interesting to note, that the exponential form of the form factor
dependence on $t$ (\ref{d20})  can be quite accurately represented, in the
admissible kinematical region, by the form, corresponding to the model
of the meson dominance
\begin{equation}
\xi _k(t) = \xi _k(0)\;\frac{1}{1-t/m_k^2}\;,
\label{d26}
\end{equation}
where $m_k$ are presented in Table \ref{td2.2}. One can see from
eqs.(\ref{d18})--(\ref{d25}), that the form factors (excepting $A_1(t)$) are
also representable in form (\ref{d26}), and the degeneration takes place
\begin{equation}
m_V = m_{A_2} = m_{A_3} \approx m_+\;,
\end{equation}
if $\omega_P \approx \omega_V$, $M_P \approx M_V$.
\begin{table}[t]
\caption{The $m_k$ parameters (in GeV) for the $\xi (t)$ representation
in eq.(137)}
\label{td2.2}
\begin{center}
\begin{tabular}{||l|c|c|c|c||}
\hline
{}~ & $B_c^+ \to \psi e^+ \nu_e$ & $B_c^+ \to \eta_c e^+ \nu_e$ & $B_c^+ \to
B_s e^+ \nu_e$ & $B_c^+ \to B_s^* e^+ \nu_e$ \\
\hline
$m_k\;,\;{\rm GeV}$ & $6.3$ & $6.45$ & $1.9$ & $1.95$\\
\hline
\end{tabular}
\end{center}
\end{table}

As for the $A_1(t)$ form factor, it can be represented in the form
\begin{equation}
A_1(t) = \varphi (t)\;\frac{1}{1-t/m_{A_1}^2} = a_1+\frac{
A'_1(0)}{1-t/m_{A_1}^2}\;,
\end{equation}
where
\begin{eqnarray}
m_{A_1} & = & m_V\;,\\
 A'_1(0) & = & \frac{1}{2}\; \frac{M^2+M_V^2-m_{A_1}^2+2 M
       (m_2-m_{{\rm sp}})}{M+M_V} \sqrt{\frac{M_V}{M}}
       \frac{1}{m_2} \xi _V (0)\;,\\
a_1 & = & A_1(0) -  A'_1(0)\;.
\end{eqnarray}

The values of the transition form factors at zero mass of the lepton pair are
shown in Table \ref{td2.3}. The numerical calculations in ref.[\cite{i14a}]
have been performed for the mass values
\begin{equation}
m_b = 4.9\;{\rm GeV}\;,\;\;m_c =1.6\;{\rm GeV}\;\;\;m_s = 0.5-0.55\;{\rm
GeV}\;.
\end{equation}
The element of the Kobayashi--Maskawa matrix has been taken equal to
$V_{bc}= 0.046$.
\begin{table}[b]
\caption{The form factors of the semileptonic $B_c$ decays}
\label{td2.3}
\begin{center}
\begin{tabular}{||l|c|c|c|c|c||}
\hline
mode & $F_+(0)$ & $A_1(0)$ & $ A'_1(0)$ & $A_2(0)$ & $V(0)$ \\
\hline
$B_c^+ \to \psi e^+ \nu_e$ & -- & $0.73$ & $0.14$ & $0.67$ & $1.31 $\\
$B_c^+ \to \eta_c e^+ \nu_e$ & $0.89$ & -- & -- & -- & -- \\
$B_c^+ \to B_s e^+ \nu_e$ & $0.61$ & -- & -- & --& -- \\
$B_c^+ \to B_s^* e^+ \nu_e$ & -- & $0.52$ & --  & $-2.79$ & $5.03$\\
\hline
\end{tabular}
\end{center}
\end{table}

The $f_{B_c}$ constant in ref.[\cite{i14a}] has been varied in the limits
\begin{equation}
f_{B_c} = 360-570\;{\rm MeV}\;,
\end{equation}
where the upper limit corresponds to the values, obtained in the
nonrelativistic potential model [\cite{i14e,7ka}], in the parton model
[\cite{8a}], and in the QCD sum rules [\cite{21k,9a}].  The lower limit
corresponds to the value, obtained in the Borel sum rules of QCD
[\cite{21k,9a}].  Note, that for the $B_c^+ \to \psi e^+ {\nu}_e$ decay,
the result weakly depends on the $f_{B_c}$ choice (3 \%).

One has also supposed
\begin{equation}
f_{\eta_c} = f_{\psi}\;,
\end{equation}
and one has varied the values
\begin{eqnarray}
f_{B_S} & = & 100-110\;{\rm MeV}\;, \\
f_{B_S^*} & = & 160-180\;{\rm MeV}\;,
\end{eqnarray}
that does not contradict the estimates, made in the QCD sum rules [\cite{14a}].

Note, that for the semileptonic $B_c$ meson decays
$B_c^+ \to M_X e^+ {\nu}_e$, where $M_X$ is the recoil meson, the explicit
covariance of the model allows one to take into account corrections on the
velocity of the $M_X$ meson. As for  corrections due to the quark motion
inside the meson, they are taken into account due to the difference
between the constituent and current masses of the quark.

In the ISGW model for the meson state vector, one uses the following
nonrelativistic expression
\begin{eqnarray}
|X({\bf p}_X;s_x)\rangle & = & \sqrt{2 m_x}\int {\rm d}^3p\sum C_{m_L m_S}^{s_x
LS}
\phi_X({\bf p})_{L m_L} \chi_{s\bar s}^{S m_S} \nonumber \\
&&|q\biggl(\frac{m_q}{m_x}{\bf p}_X+{\bf p},s\biggr)\;
\bar q\biggl(\frac{m_{\bar q}}{m_x}{\bf p}_X-{\bf p},\bar s\biggr)\rangle.
\end{eqnarray}
where $\chi _{s\bar s}^{S m_S}$ is the spin wave function of the
quark-antiquark pair in the state with the total spin $S$ and the spin
projection $m_S$,  $C_{m_L m_S}^{s_x LS}$ is the coupling between the orbital
momentum $L$ and the total spin $S$ of the system with the total momentum
$s_x$; $\phi_X({\bf p})_{L m_L}$ is the corresponding nonrelativistic
wave function, ${\bf p}_X$ is the meson momentum, $\bf p$
is the relative momentum of quarks. In the considered model, the meson mass is
equal to the sum of quark masses in the approximation of infinitely narrow wave
package, only.

As the probe functions, the nonrelativistic oscillator wave functions have
been chosen
\begin{eqnarray}
\Psi^{1 S} &=& \frac{\beta_s^{3/2}}{\pi^{3/4}} \exp\biggl(-\frac{\beta_s^2
r^2}{2}\biggr),\nonumber\\
\Psi^{1P}_{11} &=& -\frac{\beta_P^{5/2}}{\pi^{3/4}}r
\exp\biggl(-\frac{\beta_P^2 r^2}{2}\biggr),\nonumber \\
\Psi^{2S} &=& \bigl(\frac{2}{3}\bigr)^{1/2}\frac{\beta_S^{7/2}}{\pi^{3/4}}
(r^2-\frac{3}{2}\beta_S^{-2}) \exp\biggl(-\frac{\beta_S^2
r^2}{2}\biggr).\nonumber
\end{eqnarray}
The $\beta$ parameters have been determined by the variational principle
and the Cornell potential [\cite{2k}].

In the WSB model, the mesons are considered as a relativistic bound state of
a quark $q_1$ and an antiquark $\bar q_2$ in the system of infinitely
large momentum [\cite{3a}]:
\begin{eqnarray}
|P,m,j,j_z\rangle &=& \sqrt{2} (2\pi)^{3/2}\sum_{s_1,s_2}\int {\rm d}^3p_1\;
{\rm d}^3p_2\;
\delta^3(\bf p-\bf p_1 -\bf p_2)\nonumber \\
&~& L_m^{j,j_z}({\bf p}_{1t},x,s_1,s_2) a_1^{{s_1}^+}({\bf p}_1)
b_2^{{s_2}^+}({\bf p}_2)|0\rangle,\nonumber
\end{eqnarray}
where $P_{\mu}=(P_0,0,0,P)$, and at $P\rightarrow\infty$, $x=p_{1z}/p$
corresponds to the momentum fraction, carried out by the nonspectator
quark,  $p_{1t}$ is the transverse momentum.

For the orbital part of wave function, the solution of the relativistic
oscillator is used
\begin{eqnarray}
L_m({\bf p}_{\rm t},x) & = & N_m\sqrt{x(1-x)}\;
\exp\biggl(-\frac{{\bf p}_{t}^2}{2 \omega^2}\biggr) \nonumber\\
&& \exp\biggl[-\frac{m^2}{2\omega^2}\biggl(x-\frac{1}{2}
-\frac{m_{q_1}^2-m_{q_2}^2}{2 m^2}\biggr)^2\biggr].
\label{d.38}
\end{eqnarray}

In the both models, the calculation of hadronic matrix elements
$\langle B_c(p)|J_\mu|X(k)\rangle$ comes to the calculation of the
matrix elements of the quark currents between the quark states
and the overlapping the corresponding wave functions.

In the potential models, the bound state of two particles is described
by the wave function with the argument, being the relative momentum
of the particle motion in the meson system of the mass centre. However,
in the case of the decays with large recoil momenta, one can not chose the
system, where the both mesons (the initial one and the decay product)
would be at rest, so that one has a kinematical uncertainty in the form
factor values.

For instance, in the ISGW model the form factor dependence on the invariant
mass of lepton pair $t$ is determined by the function $\xi(t)$:
\begin{equation}
\xi_{\rm IGSW}(t) = \biggl(\frac{2\beta\beta_1}{\beta^2+\beta_1^2}\biggr)^{3/2}
\exp\biggl(-\frac{m_{{\rm sp}}^2}{2 \tilde M{\tilde M}_1}\frac{t_{\rm
max}-t}{k^2(\beta^2+\beta_1^2)}
\biggr);
\label{dop1}
\end{equation}
where $\beta$ and $\beta_1$ are the parameters of wave functions for the
initial and final mesons,
$m_{{\rm sp}}$ is the spectator quark mass, $\tilde M$ and ${\tilde M}_1$
are the model parameters (the masses of the initial and final "mock"-mesons)
[\cite{2a}]. The $k$ parameter in eq.(\ref{dop1}) is introduced synthetically
for the correct description of the electromagnetic form factor of
$\pi$ meson ($k=0.7$). So, the authors of ref.[\cite{2a}]
related this factor with possible relativistic corrections at
large recoil momenta.

Recently in ref.[\cite{kitay}], the model for the description of
the heavy quarkonium decays has been offered. In this model, the required
behaviour of form factors (at $k=0.7$) is automatical with no introduction
of additional parameters. In contrast to the above mentioned approaches
(the Covariant quark model and ISGW model), the nonrelativistic
approximation is performed for the hadronic matrix element as a whole,
but it is not performed separately for the wave functions of the initial
and final states, only. At small recoil momenta, this formalism practically
repeats the ISGW model, but at large momenta, there are some differences in
the structure of the spin part of the wave function and the argument of the
wave function of the final meson. So, the latter change is the most important
and it leads to the difference in the form factor dependence on
$t$ [\cite{close}].

The transition form factors in the ISGW model depend on $\beta_{B_c}$ and
$\beta_{B_S}$. For its values,
$\beta_{B_c}=0.82$ and $\beta_{B_S}=0.51$ are obtained from the variational
principle. Since the considered model is the nonrelativistic approximation,
the form factors are the most accurately predicted at
$q^2 = q^2_{\rm max}=(M_{B_c}-M_X)^2$ (at the maximal value of the lepton pair
invariant mass).

One can calculate the form factors in the region of low $q^2$ values
in two different ways: by the use of the exponential dependence on
$q^2$ as in ISGW or in the pole model of meson dominance. The results for the
decay widths, calculated in these ways, are presented in Table \ref{td2.4}.
The additional parameter in the ISGW model is $k=1$ (\ref{dop1}).

The results, obtained in ref.[\cite{kitay}], is also presented in the
same Table. In the constituent quark model, the exponential dependence
of the form factors can be represented in the pole form. As one can see
from Table \ref{td2.4}, in the ISGW model for the decays, where the
$c$-quark is the spectator, the exponential dependence and the pole
model give the different results.

In the WSB model, the form factor values at $q^2=0$  are predicted
dependently of the $\omega$ parameter (see eq.(\ref{d.38})), that
corresponds to the average transverse momentum of quarks inside the meson.
In ref.[\cite{i14e}] the $\omega$ values were equal to the average
$p_{\rm t}^2$  values, estimated in the ISGW model ($\omega_X\approx \beta_X$).
Note, that the $\omega$ parameter is external for the WSB model.

The results of the mentioned approaches are presented in Table \ref{td2.4}.

Note, that the relative yield of the pseudoscalar states in respect to
the vector states is much greater in ref. [\cite{kitay}], where
$\Gamma^*/\Gamma \approx 2$ in comparison with
$\Gamma^*/\Gamma \approx$ 3--4 in the ISGW model.
This leads to that, for example, the exclusive decay modes
$B_c^+ \to \psi(\eta_c) e^+ {\nu}_e$ practically saturate the
$b \to c e \nu$ transition. This feature is analogous to the consideration of
the $B \to D^{(*)} e {\nu}$ decay, that also
saturates free $b$-quark decay. The decays into the excited states and
manyparticle modes are suppressed.

As one can see, these three models for the decays with the spectator $b$-quark,
give the close values
$$
\Gamma(B_c\rightarrow B_s+e+\nu)+\Gamma(B_c\rightarrow B_s^*+e+\nu)=
(60\pm7)\cdot 10^{-6}\; {\rm eV}.
$$

Note also, that in the case of the heavy quarkonium $B_c$,
the application of the nonrelativistic wave function instead of the wave
function of the relativistic oscillator in the meson of the WSB model,
seems to be more acceptable. This circumstance and uncertainty in
$\omega$, maybe, explain, why the WSB model gives the underestimated value
for the width of the $B_c^+\rightarrow J/\psi+e+\nu$ decay.

\subsubsection{$B_c^+\rightarrow J/\Psi(\eta_c)e^+\nu$ decay in
QCD sum rules}

The most suitable for the registration modes of the $B_c$ decays are
the semileptonic or hadronic transitions with the $J/\psi$ particle
in the final state. But in the QCD sum rules (SR) [\cite{i14a,21k,20k}]
and in the quark models, one found different results for both the widths of the
corresponding decays and the form factors of the transitions, although
in the framework of the separate approach, the calculations, performed in
different ways, coincided with each other. Recently in ref. [\cite{i14b}],
one has shown that the existing discrepancy can be cancelled by means of
account for the higher QCD corrections in SR.

The widths of the semileptonic $B_c$ decays are defined, in general, by
the form factors $F_+$, $V$, $A_1$ and $A_2$
(see eqs.(\ref{d16}), (\ref{d17})). Following the notation of
ref.[\cite{i14b}], redefine the form factors (\ref{d16}), (\ref{d17})
in the way
$$~~~~~~~~~~~~f_+ = F_+,~~~~~~~~~~~~~~~F_0^A = (M_{B_c}+M_V) A_1,
$$
$$~~~~~~~~~~~~F_+^A =
-\frac{A_2}{M_{B_c}+M_V},~~~~~~~~~F_V=\frac{V}{M_{B_c}+M_V}.
$$
For the calculation of these form factors in the QCD SR, let us consider
the three-point functions
\begin{eqnarray}
\Pi_\mu(p_1,p_2,q^2)& =&   i^2\int {\rm d}x\; {\rm d}y\; \exp\{i(p_2x-p_1y)\}
\nonumber \\ &&  \langle 0|T\{\bar c(x)\gamma_5 c(x),V_\mu(0),
\bar b(y)\gamma_5 c(y)\}|0\rangle\;,\\
\Pi_{\mu\nu}^{V,A}(p_1,p_2,q^2) &=& i^2\int {\rm d}x\; {\rm d}y\;
            \exp\{i(p_2x-p_1y)\} \nonumber \\
&&\langle 0|T\{\bar c(x)\gamma_\nu c(x),J_\mu^{V,A}(0),
\bar b(y) \gamma_5 c(y)\}|0\rangle\;,
\end{eqnarray}

Introduce the Lorentz structures in the correlators
\begin{eqnarray}
\Pi_\mu & = & \Pi_+(p_1+p_2)_\mu+\Pi_-q_\mu, \\
\Pi_{\mu\nu}^V & = & i \Pi_V \epsilon_{\mu\nu\alpha\beta} p_2^\alpha
p_1^\beta,\\
\Pi_{\mu\nu}^A & =& i\Pi^A_0 g_{\mu\nu}+\Pi_1^A p_2^\mu p_1^\nu+\Pi_2^A p_1^\mu
p_1^\nu+
\Pi_3^A p_2^\mu p_2^\nu + \Pi_4^A p_1^\mu p_2^\nu.
\end{eqnarray}

The form factors $f_+$, $F_V$, $F_0^A$ and $F_+^A$
are determined by the amplitudes $\Pi_+$, $\Pi_V$, $\Pi_0^A$ and
$\Pi_+^A=\frac{1}{2} (\Pi_1+\Pi_2)$, respectively. For the amplitudes,
one can write down the double dispersion relation
\begin{eqnarray}
\Pi_i(p_1^2,p_2^2,q^2) & = & -\frac{1}{(2\pi)^2}\int\frac{\rho_i(s_1,s_2,Q^2)}
{(s_1-p_1^2)(s_2-p_2^2)}{\rm d}s_1 {\rm d}s_2,
\label{Q6}
\end{eqnarray}
where $Q^2 = -q^2 > 0$.

The integration region in eq.(\ref{Q6}) is determined by the condition
\begin{eqnarray}
  -1<\frac{2 s_1s_2+(s_1+s_2-q^2)(m_b^2-m_c^2-s_1)}{\lambda^{1/2}(s_1,s_2,q^2)
\lambda^{1/2}(m_c^2,s_1,m_b^2)}<1,
\label{Q7}
\end{eqnarray}
where $\lambda(x_1,x_2,x_3)=(x_1+x_2-x_3)^2-4x_1x_2$.

In accordance with the general ideology of the QCD sum rules
[\cite{23k}], the right hand, theoretical, side of eq.(\ref{Q6}) can be
calculated at large euclidian $p_1^2$ and $p_2^2$ values
by the use of the operator product expansion (OPE). The perturbative parts
of the corresponding spectral densities (the unit operator in OPE) to the
one loop approximation are presented in Appendix II.
Since we consider the systems, composed of the heavy quarks, one can neglect
the
power corrections [\cite{21k}].

Consider the physical part of SR. As has been already mentioned in the
consideration of the axial constant  of $B_c$, there are two approaches.
In the first one, one assumes that the physical part includes the contribution
of the lowest mesons and the continuum, that is approximated by the
perturbative part of the spectral function from some threshold values
$s_0^1$ and $s_0^2$ [\cite{20k,21k}]. The contribution of the higher
excitations and the continuum is suppressed due to the Borel transformations
over two variables $(-p_1^2)$ and $(-p_2^2)$. The numerical results, obtained
in such way of ref.[\cite{20k,21k}], are presented below.

In the second way, one saturates the spectral density by infinite number
of narrow resonances [\cite{i14a}], so that

\begin{eqnarray}
\rho_+(s_1,s_2,Q^2) &=&
(2\pi)^2\sum_{i,j=1}^{\infty}f_{B_c}^i\frac{{M_{B_c}^i}^2}{m_b+m_c}
f_{\eta_c}^j\frac{{M_{\eta_c}^j}^2}{2m_c}f_+^{ij}(Q^2) \nonumber \\
& ~ & \delta(s_1-{M_{B_c}^i}^2)\delta(s_2-{M_{\eta_c}^j}^2)
\label{new1}
{}~ \nonumber \\
\rho_V(s_1,s_2,Q^2) &=&
2(2\pi)^2\sum_{i,j=1}^{\infty}f_{B_c}^i\frac{{M_{B_c}^i}^2}{m_b+m_c}
\frac{{M_\psi^j}^2}{g_\psi}F_V^{ij}(Q^2) \nonumber \\
& ~ & \delta(s_1-{M_{B_c}^i}^2)\delta(s_2-{M_{\psi}^j}^2)
\label{new2}
{}~ \nonumber \\
\rho_{0,+}^A(s_1,s_2,Q^2) &=&
(2\pi)^2\sum_{i,j=1}^{\infty}f_{B_c}^i\frac{{M_{B_c}^i}^2}{m_b+m_c}
\frac{{M_{\psi}^j}^2}{g_\psi}F_{0,+}^{ij}(Q^2) \nonumber \\
& ~ & \delta(s_1-{M_{B_c}^i}^2)\delta(s_2-{M_{\psi}^j}^2)
\label{new3}
\end{eqnarray}

Substituting the expressions for the
spectral densities (\ref{new1})--(\ref{new3})
in the dispersion relations for correlators (\ref{Q6})
to the one hand, and their perturbative values to the other hand,
one gets the corresponding sum rules.

Applying the procedure, described in Appendix III, for the both sums over the
resonances, one obtains for the form factors under consideration
\begin{eqnarray}
f_+^{kl}(Q^2) = \frac{8m_c(m_b+m_c)}{M_{B_c}^k M_{\eta_c}^lf_{B_c}^k
f_{\eta_c}^l}\frac{{\rm d}M_{B_c}^k}{{\rm d}k}\frac{{\rm d}M_{\eta_c}^l}{{\rm
d}l}\frac{1}{(2\pi)^2}
\rho_+({M_{B_c}^k}^2,{M_{\eta_c}^l}^2,Q^2)\;,
\label{Q11}
\end{eqnarray}

\begin{eqnarray}
F_V^{kl}(Q^2) & = & \frac{2(m_b+m_c)g_\psi^l}{M_{B_c}^k M_\psi^lf_{B_c}^k}
\frac{{\rm d}M_{B_c}^k}{{\rm d}k}\frac{{\rm d}M_\psi^l}{{\rm
d}l}\frac{1}{(2\pi)^2}
\rho_V({M_{B_c}^k}^2,{M_\psi^l}^2,Q^2)
\label{Q12}\;,
    ~~ \nonumber \\
F_{0,+}^{kl}(Q^2) & = & \frac{4(m_b+m_c)g_\psi^l}{M_{B_c}^k M_\psi^lf_{B_c}^k}
\frac{{\rm d}M_{B_c}^k}{{\rm d}k}\frac{{\rm d}M_\psi^l}{{\rm
d}l}\frac{1}{(2\pi)^2}
\rho_{0,+}({M_{B_c}^k}^2,{M_\psi^l}^2,Q^2)\;.
\label{Q13}
\end{eqnarray}

Choosing the $k$ and $l$ values, one can extract the transitions between the
given resonances. At $k=l=1$, one gets the required
form factors for the  $B_c^+\rightarrow J/\psi(\eta_c)e^+\nu$ decays.

Thus, we use the phenomenological parameters ${\rm d}M_k/{\rm d}k$ instead of
the additional parameters such as the continuum thresholds. As has been
mentioned, the former is, in a sense, the density of the quarkonium states
with the given quantum numbers. One can quite accurately
calculate these factors. The masses of the radial excitations of
$\psi$ are known experimentally [\cite{1k}], and for the $B_c$ and $\eta_c$
systems, composed of the heavy quarks, one can use the predictions of the
potential models [\cite{2k}--\cite{5k,7ka,11k}--\cite{k20k}].

The ${\rm d}M_k/{\rm d}k$ values at $k=1$ for the systems under the
consideration, are presented in Table \ref{td2.5}.
\begin{table}[t]
\caption{The derivatives ${\rm d}M_k/{\rm d}k$ (in {\rm GeV}) for the lowest
states}
\label{td2.5}
\begin{center}
\begin{tabular}{||c|c|c|c||}
\hline
Quarkonium & $B_c$ & $J/\psi$ & $\eta_c$ \\
\hline
${\rm d}M_k/{\rm d}k\mid_{k=1}$ & 0.75 & 0.75 & 0.76 \\
\hline
\end{tabular}
\end{center}
\end{table}

Let us chose the following values of parameters
$f_{B_c}=360$ {\rm MeV}, $f_{\eta_c}=330$ {\rm MeV},
[\cite{21k,9a}],  $m_b=4.6\pm 0.1$ {\rm GeV},
$m_c=1.4\pm 0.05$ {\rm GeV}, $g_{J/\psi}=8.1$
(from the data on $\Gamma(J/\psi\rightarrow e^+e^-)$).
For the axial constant, we chose $360\; {\rm MeV}$ [\cite{i14a}] instead of
$460$, to compare the form factor values with the results of ref.[\cite{21k}].
The $B_c$ meson mass will be varied from 6.245 to 6.284 {\rm GeV}
(the data of the different potential models). Note, that at such choice of
the parameters, we do not go out from the integration region
(\ref{Q7}). In ref.[\cite{21k}] one used $M_{B_c} = 6.35$ {\rm GeV}.
The form factors values, obtained in ref.[\cite{i14a,21k,20k}]
at $Q^2=0$, are show in Table \ref{td2.6}.
The deviation from the central values in Table \ref{td2.6} corresponds to
the variation of the quark and $B_c$ meson masses within
the limits, mentioned above (for [\cite{i14a}]).
As in the case of the potential models, the SR predictions agree with
each other.
\begin{figure}[b]
\setlength{\unitlength}{1mm}\thicklines
\begin{center}
\begin{picture}(120,60)
\put(10,15){\circle*{3}}
\put(110,15){\circle*{3}}
\put(60,40){\circle*{3}}
\put(10,15){\line(1,0){100}}
\put(10,15){\line(2,1){50}}
\put(60,40){\line(2,-1){50}}
\put(30,15){\line(0,1){2}}
\put(30,19){\line(0,1){2}}
\put(30,23){\line(0,1){2}}
\put(30,19){\vector(0,1){2}}
\put(32,19){$k_1$}
\put(39,19){$\cdots$}
\put(50,15){\line(0,1){2}}
\put(50,19){\line(0,1){2}}
\put(50,23){\line(0,1){2}}
\put(50,19){\vector(0,1){2}}
\put(50,27){\line(0,1){2}}
\put(50,31){\line(0,1){2}}

\put(52,19){$k_n$}

\put(90,15){\line(0,1){2}}
\put(90,19){\line(0,1){2}}
\put(90,23){\line(0,1){2}}
\put(90,19){\vector(0,1){2}}
\put(92,19){$q_m$}
\put(70,15){\line(0,1){2}}
\put(70,19){\line(0,1){2}}
\put(70,23){\line(0,1){2}}
\put(70,19){\vector(0,1){2}}
\put(70,27){\line(0,1){2}}
\put(70,31){\line(0,1){2}}
\put(72,19){$q_1$}
\put(79,19){$\cdots$}
\put(21,15){\vector(-1,0){2}}
\put(41,15){\vector(-1,0){2}}
\put(61,15){\vector(-1,0){2}}
\put(81,15){\vector(-1,0){2}}
\put(101,15){\vector(-1,0){2}}
\put(20,12){$p_1$}
\put(40,12){$p_n$}
\put(60,12){$p$}
\put(80,12){$q_1$}
\put(100,12){$q_m$}
\put(20,20){\vector(2,1){2}}
\put(40,30){\vector(2,1){2}}
\put(54,37){\vector(2,1){2}}
\put(64,38){\vector(2,-1){2}}
\put(80,30){\vector(2,-1){2}}
\put(98,21){\vector(2,-1){2}}
\put(13,24){$p_1+P_1$}
\put(33,34){$p_n+P_1$}
\put(46,41){$p+P_1$}
\put(62,41){$p+P_2$}
\put(75,34){$q_1+P_2$}
\put(95,24){$q_m+P_2$}
\put(0,14){\line(1,0){10}}
\put(0,16){\line(1,0){10}}
\put(3,17){\vector(1,0){6}}
\put(5,19){$P_1$}
\put(110,14){\line(1,0){10}}
\put(110,16){\line(1,0){10}}
\put(113,17){\vector(1,0){6}}
\put(115,19){$P_2$}
\put(5,10){$B_c$}
\put(112,10){$\eta_c$,$J/\psi$}
\end{picture}
\end{center}
\setlength{\unitlength}{0.5mm}\thicklines
\caption{The Coulomb corrections in the semileptonic
$B_c$ meson decay}
\label{fd4}
\end{figure}
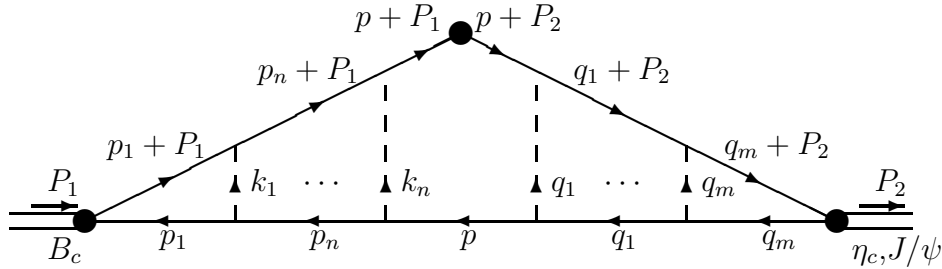

In ref.[\cite{i14a}], the form factors have the following pole behaviour
\begin{equation}
   F_i(Q^2) = \frac{F_i(0)}{1+Q^2/m^2_{\rm pole}}\phi_i(Q^2)\;,
\end{equation}
where $m_{\rm pole}=$ 6.3--6.4 {\rm GeV}, and $\phi_i(Q^2)=1+a_i Q^2$.
The representation of $f_+$, $F_V$, $F^A_0$ and $F^A_+$ by the form
(\ref{Q11})--(\ref{Q13}) gives the following $a_i$ values, which are
quite low and equal to $ -0.025$, $-0.007$, $-0.012$, $-0.02$
respectively. The behaviour, considered above, does not practically differ
from the ordinary pole behaviour [\cite{i14a}], where $a_i=0$.
The results for the transition widths are presented in Table \ref{td2.7}.

As one can see from Tables \ref{td2.6} and \ref{td2.7}, the results of the
Borel
SR are, in general, in a good agreement with the results of the considered
approach, within the model errors. The widths, obtained in
ref.[\cite{20k}], are greater, than in ref.[\cite{i14a,21k}], since in
ref.[\cite{20k}], the $q^2$-dependence of the transition form factors
strongly differs from the behaviour, expected in the meson dominance model.
\begin{table}[t]
\caption{The form factors of the $B_c\rightarrow J/\psi(\eta_c)e\nu$
transitions
at $Q^2=0$}
\label{td2.6}
\begin{center}
\begin{tabular}{||c|c|c|c|c||}
\hline
 $f_+(0)$ & $F_V(0)$ & $F_+^A(0)$ & $F_0^A(0)$ & Reference \\
 & ${\rm GeV}^{-1}$ & ${\rm GeV}^{-1}$   & ${\rm GeV}$ & \\
\hline
 $0.23\pm0.01$ & $0.035\pm0.03$ & $-0.024\pm0.002$ & $2\pm0.2$ &
[\cite{i14a}]\\
 $0.2\pm0.02$ & $0.04\pm0.01$ & $-0.03\pm0.01$ & $2.5\pm0.3$ & [\cite{21k}] \\
 $0.55\pm0.1$ & $0.048\pm0.007$ & $-0.030\pm0.003$ & $3.0\pm0.5$ & [\cite{20k}]
\\\hline
\end{tabular}
\end{center}
\end{table}

The deviation from the quark models is related, from our point of view,
with that in the calculations of the transition form factors in the
QCD sum rules, one has to account for $\alpha_S/v$-corrections, where
$v$ is the relative velocity of the quarks inside the meson. For the heavy
quarkonia, where the velocity of the quark motion is small, such corrections,
corresponding to the Coulomb-like interactions (Figure  \ref{fd4}), can play
an essential role [\cite{i14b}].
\begin{table}[b]
\caption{The widths (in units $10^{-6}$ eV) of the semileptonic $B_c$
decays in the QCD sum rules with no account of $\alpha_S/v$-corrections}
\label{td2.7}
\begin{center}
\begin{tabular}{||c|c|c|c||}
\hline
mode & [\cite{i14a}] & [\cite{21k}]& [\cite{20k}] \\
\hline
 $B_c^+\rightarrow J/\psi e^+\nu$ & 4.6 & 7 & 10.5\\
 $B_c^+\rightarrow \eta_c e^+\nu$ & 1.4 & 1 & 9. \\
\hline
\end{tabular}
\end{center}
\end{table}

Indeed, the spectral densities $\rho_i(s_1, s_2, Q^2)$, determining the
$B_c$ decay form factors, are calculated near the threshold
$s_1 = M_{B_c}^2,\;\;s_2 = M_{\eta_c,\;\psi}^2$. When the recoil meson momentum
is small, the calculation of the ladder diagrams in the
formalism of the nonrelativistic quantum mechanics (see [\cite{1r}],
Figure  \ref{fd4}), leads to the finite renormalization of $\rho$, so that
\begin{equation}
\bar \rho_i(s_1, s_2, Q_{\rm max}^2) = C\;\rho_i(s_1, s_2, Q_{\rm max}^2)\;,
\end{equation}
where the factor $C$ has the form
\begin{equation}
C = \bigg\vert\frac{\Psi^C_{B_c}(0) \Psi^C_{\eta_c,\;\psi}(0)}
{\Psi^{free}_{B_c}(0) \Psi^{free}_{\eta_c,\;\psi}(0)}\bigg\vert\;,
\label{Q16}
\end{equation}
and $\Psi^{C,\;free}(0)$ are the Coulomb and free wave functions of quarks,
so that
\begin{equation}
\bigg\vert\frac{\Psi^C(0)}{\Psi^{free}(0)}\bigg\vert^2 = \frac{4 \pi
\alpha_S}{3 v}\;
[1 - \exp(-\frac{4 \pi \alpha_S}{3 v})]^{-1}\;.
\label{Q17}
\end{equation}
For the two-point quark correlators, determining the decay constants $f$ of
the heavy quarkonia $\psi,\;\Upsilon,\;B_c$, the account for factor (\ref{Q16})
led to the essential enhancement of $f$, so that one observed the agreement
with the experimental data on $f_\psi$ and $f_\Upsilon$. Note, that
the expansion in (\ref{Q17}) over $\alpha_S/v \to 0$ exactly leads to the
dominant term, appearing in account for the one-loop $\alpha_S$-corrections
to the two-point correlator of currents. Moreover, these corrections
have been taken into account in the evaluation of $f$ for the three-point
correlators, but one did not take into account the loop corrections in the
determination of the three-point spectral densities.

One would, in a logics, not to account for the $\alpha_S$-corrections
in the $f$ evaluation as well as in the $\rho$ determination, or
one would take into account these corrections in both cases.
As one can see, for example, from (\ref{Q11}), one can write down
\begin{eqnarray}
f_+^{kl}(Q^2) = \frac{8m_c(m_b+m_c) C}{M_{B_c} M_{\eta_c} f_{B_c}^{(0)}
f_{\eta_c}^{(0)} C^{{1}/{2}}_{B_c} C^{{1}/{2}}_{\eta_c} }
\frac{{\rm d}M_{B_c}^k}{{\rm d}k}\frac{{\rm d}M_{\eta_c}^l}{{\rm
d}l}\frac{1}{(2\pi)^2}
\rho_+^{(0)}({M_{B_c}^k}^2,{M_{\eta_c}^l}^2,Q^2))\;,
\label{4.4s}
\end{eqnarray}
where the $f^{(0)}$ and $\rho^{(0)}$ values are calculated with no
account for the $\alpha_S$-corrections, and the factors $C$ appear
due to the Coulomb-like corrections and defined in eqs.(\ref{Q16}),
(\ref{Q17}). It is evident, that
\begin{equation}
\frac{C}{C^{{1}/{2}}_{B_c} C^{{1}/{2}}_{\eta_c}} = 1\;.
\label{4.5}
\end{equation}
Thus, in the determination of the transition form factors, we can use the
"bare" $f$ and $\rho$ quantities, calculated in zero approximation over
$\alpha_S$ [\cite{KLT24}], instead of that, say, was done in ref. [\cite{21k}],
where $\rho^{(0)}$ was used without the $C$ factor and the $f$ constants with
the account for the $\alpha_S$-corrections, i.e. with the factors $C_{B_c}$
and $C_{\eta_c,\;\psi}$, were instantaneously used.

As the result, one gets the following values for the form factors
$f_+$ and $F_0^A$  [\cite{i14b}]
$$ f_+(0)=0.85\pm 0.15\;, ~~~~~~~~~~~~~~~~~~~~F^A_0=6.5\pm1\; {\rm GeV}.
$$
For the corresponding widths, one has found  [\cite{i14b}]
$$ \Gamma(B_c^+ \to \psi e^+ \nu)\approx44\cdot10^{-6}\; {\rm eV}\;,~~~~~~~~
\Gamma(B_c^+ \to \eta_c e^+\nu)\approx15\cdot10^{-6}\; {\rm eV}.
$$
Note, that we have neglected the contributions of the form factors
$F_V$ and $F_+^A$ in the decay $B_c\rightarrow J/\psi e\nu$. This
can result in the overestimation of the widths values up to
10--20 \%. One can make the agreement between the obtained values of the widths
and the results of the quark models (see Table  \ref{td2.4}) within the
limits of the theoretical uncertainties of the used methods.
\begin{table}[t]
\caption{The partial widths (in $10^{-6}$ {\rm eV}) of the
semileptonic $B_c$ decays (ISGW1 and ISGW2 are the results
of the ISGW model with the exponential dependence of the form factors and
the pole model, respectively)}
\label{td2.4}
\begin{center}
\begin{tabular}{||l|c|c|c|c|c||}
\hline
mode & ISGW1 [\cite{i14e}] & ISGW2 [\cite{i14e}]& WSB [\cite{i14e}] &
[\cite{i14a}]& [\cite{kitay}]\\
\hline
$B_c^+ \to \psi e^+ \nu_e$ & $38.5$  & $53.1$  & $21.8$ & $37.3$ &  $34.4$ \\
$B_c^+ \to \eta_c e^+ \nu_e$ & $10.6$& $16.1$  & $16.5$ & $20.4$ &  $14.2$  \\
$B_c^+ \to D^0 e^+ \nu_e$ & $0.033$   & $0.12$  & $0.002$ & -- & $0.094$ \\
$B_c^+ \to D^{0*} e^+ \nu_e$ & $0.13$ & $0.32$  & $0.011$ & -- & $0.268$\\
$B_c^+ \to \psi(2S) e^+ \nu_e$ & --  & --  & -- & -- &  $1.45$ \\
$B_c^+ \to \eta_c' e^+ \nu_e$ & -- & --  & -- & -- &  $0.727$  \\
$B_c^+ \to B_s e^+ \nu_e$ & $16.4$   & $17.9$  & $11.1$ & $16\pm 4$ & $26.6$\\
$B_c^+ \to B_s^* e^+ \nu_e$ & $40.9$ & $46.3$  & $43.7$ & $41\pm 6$ & $44.0$\\
$B_c^+ \to B_d e^+ \nu_e$ & $1.0$   & $1.1$  & $0.5$ & -- &  $2.30$ \\
$B_c^+ \to B_d^* e^+ \nu_e$ & $2.5$ & $3.0$  & $2.9$ & -- &  $3.32$ \\
\hline
\end{tabular}
\end{center}
\end{table}

Comparing the results of the QCD SR and the quark models, one can accept
as a central value of the $B_c \to J/\psi e \nu$ decay width (with the accuracy
about 40 \%)
$$ \Gamma(B_c \to J/\psi e \nu)\approx 40\cdot10^{-6}\; {\rm eV},
$$
that corresponds to the branching fraction, equal to 3 \%. Then the
relative probability of the three lepton yield in the
$B_c$ decays, when two of them reconstruct $J/\psi$, equals
$$ Br(B_c^+\rightarrow {(l^+l^-)}_{J/\psi}l^{\prime+}\nu)\approx 8\cdot10^{-3},
$$
where $l,l^{\prime}$ denotes $e$ or $\mu$.

\subsubsection{Approximate spin symmetry}

In the bound state, the heavy quark virtualities are much less than
their masses, i.e. the following kinematical expansion for the quark
momentum $p_Q$ is accessible
\begin{equation}
p_Q^\mu = m_Q\cdot v^\mu+k^\mu\;,
\label{k.1}
\end{equation}
so that
\begin{equation}
v\cdot k \approx 0\;,\;\;|k^2|\ll m_Q^2\;.
\label{k.2}
\end{equation}
Then in the system, where $v=(1,\bf{0})$, the heavy quark hamiltonian
in a gluon field of an external source has the form
\begin{equation}
H = m_Q + V({\bf{r}})+\frac{{\bf{k}}^2}{2m_Q}+g\frac{{\bf{\sigma}}\cdot
{\bf{B}}}
{2m_Q}+O\biggl(\frac{1}{m_Q^2}\biggr)\;,
\label{k.3}
\end{equation}
so that in the limit $\Lambda_{QCD}\ll m_Q$, the spin-flavour
symmetry EHQT [\cite{13a}] takes place for hadrons with a single heavy quark.

For the heavy quarkonium, one has purely phenomenologically, that
the kinetic energy does practically not depend on their flavours,
however, the value of the potential energy term $V(\bf{r})$ is determined
by the average distance between the heavy quarks. This distance depends on the
quark masses, i.e. the flavours. Therefore, there is no flavour symmetry
of the wave functions in the heavy quarkonium. However, the magnetic field
of the heavy quark is determined by its motion velocity
(as well as magnetic moment). The quark motion is nonrelativistic in the
heavy quarkonium, so that
\begin{equation}
{\bf{B}}\sim O({\bf{v}})\sim O\biggl(\frac{1}{m_Q}\biggr)\;.
\label{k.4}
\end{equation}
{}From eqs.(\ref{k.3}), (\ref{k.4}) it follows, that the spin-dependent
potential
in the heavy quarkonium appears in the second order over the inverse
heavy quark masses (see Section 2)
\begin{equation}
V_{SD}\sim O\biggl(\frac{1}{m_Q^2}\biggr)\;.
\label{k.5}
\end{equation}
Thus, in the leading approximation for the heavy quarkonium, one can neglect
the spin-dependent forces in comparison with the kinetic energy and the
nonrelativistic potential. This means, that in this approximation, the
quark spin is decoupled from the interaction with the gluons of low
virtualities, therefore the masses of the $nL_J$ quarkonium
states are degenerated over $J$, and these states have the identical
wave functions.

Thus, there is the approximate spin symmetry for the heavy quarks in the
heavy quarkonium.

Further, let us consider the matrix element
\begin{equation}
M=\langle n^SL_J(Q\bar Q')|\Gamma|h\rangle\;,
\label{k.6}
\end{equation}
where $\Gamma$ is the operator of quark currents, $h$ is a state. Then
the spin symmetry means that the action of spin operators of the heavy quark
is factorized, and the matrix element $\bar M$, obtained by the action
of the quark  $Q$ spin (or by the antiquark $\bar Q'$ spin)
\begin{equation}
S_\mu^Q = \frac{1}{4}\;\epsilon_{\mu\nu\alpha\beta}\;
v_Q^\nu \sigma^{\alpha\beta}\;,\;\;\;
\sigma^{\alpha\beta} = \frac{i}{2}[\gamma^\alpha; \gamma^\beta]\;,
\label{k.7}
\end{equation}
is related with the matrix element $M$ by the equation
\begin{equation}
\bar M=\langle n^SL_J(Q\bar Q')|S_\mu^Q \Gamma|h\rangle =
\sum C_{SS'}^{JJ'}\;\langle n^{S'}L_{J'}(Q\bar Q')|\Gamma|h\rangle\;,
\label{k.8}
\end{equation}
where $\bar M$ is the sum of the matrix elements with $J'$, and
$C_{SS'}^{JJ'}$ are defined by the rules for the spin operator action.

For the semileptonic $B_c^+ \to \eta_c(\psi) l^+\nu$ decays, the spin symmetry
is valid in the point of zero recoil of $\eta_c(\psi)$. Indeed, in this case
the spectator $c$-quark and the $\bar c$-quark, produced in the weak decay of
the $\bar b$-quark, are practically at rest in respect to each other, so
binding into the state, they interact with low virtualities, characteristic for
the heavy quarkonium. At a nonzero velocity of the $\bar c$-quark, it must
exchange with the $c$-quark by a momentum, comparable with its mass, for
to make the bound state, where their velocities are close. Thus, at the nonzero
meson recoil, the gluons with high virtualities can shift the heavy quark spin,
and the spin symmetry does already not take place.

At the zero recoil of the charmonium $v_{B_c}=v_{\eta_c(\psi)}$, in the
covariant amplitude of the weak current, the nonzero contributions are given by
$A_1(t)$, $F_\pm(t)$ at $t=t_{max}$, and the heavy quark spin symmetry means,
that
\begin{equation}
(M_{B_c}+M_{\eta_c})\;F_+ + (M_{B_c}-M_{\eta_c})\;F_- =
(M_{B_c}+M_\psi)\;A_1\;,\;\;t=t_{\rm max}\;,\;\;M_{\eta_c} = M_\psi\;.
\label{k.9}
\end{equation}

Thus, in the approximation of zero spin-dependent splitting of the
heavy quarkonium, one derives the specific relation for the form factors
of the semileptonic exclusive $B_c$ decays into the charmonium.

Note now, that the covariant model, considered above, gives the
semileptonic form factor values for the $B_c$ decay into the charmonium, so
that these quantities satisfy the symmetry relation (\ref{k.9}). In contrast
to the decays of the heavy hadrons with a single heavy quark, where
the form factor normalization at zero recoil is fixed due to the
flavour symmetry, the normalization of form factors for the weak semileptonic
transitions between the heavy quarkonia is determined by the overlapping
of their wave functions, which depend on the quarkonium model.
For the oscillator wave functions in the considered potential model, we get
\begin{equation}
(M_{B_c}+M_\psi)\;A_1(t_{\rm max})=\sqrt{2M_{B_c}2M_\psi}\;\xi(t_{\rm max})\;,
\label{k.10}
\end{equation}
where
\begin{equation}
\xi(t_{\rm max}) = \biggl(\frac{2\omega_{B_c}\omega_\psi}
{\omega_{B_c}^2+\omega_\psi^2}\biggr)^{3/2}\;.
\label{k.11}
\end{equation}
In ref.[\cite{i15}] the factor $\xi(t_{\rm max})$ was determined in the
quarkonium model with the Coulomb potential, that is quite rough
approximation.

Note further, that in the semileptonic $B_c$ decay,
the lepton pair kinematically has, in average, large invariant masses
$m(l^+\nu)\approx 1.9$ {\rm GeV}, so that the $A_1$ form factor contribution
dominates, and in accordance with the meson dominance of the
$t$-dependence of the form factors, relation (\ref{k.9}), giving
$A_1(t_{\rm max })$, determines, in a sense, the matrix element of the
semileptonic $B_c^+ \to \psi(\eta_c) l^+\nu$ decay. This feature
can be used for the determination of the $B_c$ meson mass from the
$\psi l^+$ mass spectrum as well as the element $|V_{bc}|$ of the
Kobayashi--Maskawa matrix.


\subsection{Hadronic decays of $B_c$ mesons}

Although the semileptonic $B_c^+\rightarrow J/\psi\mu^+(e^+)\nu_{\mu}(\nu_e)$
decays can serve as a good trigger for the $B_c$ registration, the complete
$B_c$  reconstruction needs a large statistics because of the
neutrino presence in the decay products. The direct measurement
of the $B_c$ meson mass is possible only in the hadronic exclusive decays.
The preliminary estimates of some nonleptonic decay widths with the
$J/\psi$ particle in the final state were made in refs.
[\cite{i14,i14d,17a}] in the framework of the potential models.
The hadronic decays were in details considered in
refs.[\cite{i14c,i14e,kitay}]. In ref.[\cite{i14e}]  for the calculations
of the transition form factors, one used the WSB and ISGW models,
mentioned above. In the calculations of the decay widths, the phase space
reduction was accounted for the $c$-spectator decays (see Section 3.1), in
contrast to some other calculations [\cite{i14d,17a}].
In the forthcoming consideration of the hadronic decays of the
$B_c$ meson, we will follow the results of the latter paper.

The effective four-fermion hamiltonian for the nonleptonic decays of
the $c$- and $b$-quarks has the form [\cite{okun}]
\begin{eqnarray}
{\cal H}_{{\rm eff}}^c & = & \frac{G}{2\sqrt{2}}V_{uq_1}V_{cq_1}^*
[C_+^c(\mu)O_+^c+C_-^c(\mu)O_-^c]+h.c.\;,
\label{ad 1} \\
{\cal H}_{{\rm eff}}^b & = & \frac{G}{2\sqrt{2}}V_{q_1b}V_{q_2q_3}^*
[C_+^b(\mu)O_+^b+C_-^b(\mu)O_-^b]+h.c. \;,
\label{ad 2}
\end{eqnarray}
where
$$
O^c_{\pm} = (\bar q_{1\alpha}\gamma_{\nu}(1-\gamma_5)c_\beta)
(\bar u_\gamma \gamma^\nu(1-\gamma_5)q_{2\delta})
(\delta_{\alpha\beta}\delta_{\gamma\delta}\pm\delta_{\alpha\delta}
\delta_{\gamma\beta}),
$$
$$
O^b_{\pm} = (\bar q_{1\alpha}\gamma_\nu(1-\gamma_5)b_\beta)
(\bar q_{3\gamma}\gamma^{\nu}(1-\gamma_5)q_{2\delta})
(\delta_{\alpha\beta}\delta_{\gamma\delta}\pm\delta_{\alpha\delta}
\delta_{\gamma\beta}).
$$

The factors $C_{\pm}^{c,b}(\mu)$ account for the strong corrections to the
corresponding four-fermion operators because of hard gluons [\cite{i14e,okun}].

The transition amplitudes must not depend on the subtraction point
$\mu$, if one consistently calculates them in the perturbation theory, i.e.
one constructs the corresponding functions of the initial and final hadronic
states in the perturbation theory, in accordance to the operators.

The problem is complicated, when one deals with the factorization
approximation,
used for the calculation of the matrix elements. In this approximation,
one assumes that the current is proportional to a single stable or
quasistable hadronic field, and one calculates its matrix element between
the vacuum and the corresponding asymptotic hadronic state, so this procedure
gives  a value, proportional to the decay constant of hadron. After that,
the amplitude of the weak decay is factorized and it is completely
determined by the hadronic matrix element of an other current, that can be
calculated by use of a model, as in the case of the semileptonic decays.
In this approximation, the interaction in the final state is neglected.

Note, that the exact factorization takes place in the leading order of
the $1/N_c$ expansion [\cite{Ruckl}]. In this approximation, one has to
be careful in the choice of the subtraction point, since the matrix
elements depend on $\mu$. (The dependence of coefficients for the four-fermion
operators of effective hamiltonian on the subtraction point does not
compensated by the functions of the initial and final states.) The
most suitable choice is $\mu\approx m_c$, since the radius of the
$B_c$ meson is determined by the $c$-quark mass, and the transferred momenta
in the decays are about $m_c$ [\cite{i14e}].

The anomalous dimensions of the $O_+^c$ and $O_-^c$ operators at $\mu=m_c$
have the form
\begin{eqnarray}
\gamma_{\pm} = -\frac{\alpha_S}{2\pi}\frac{3}{N_c}(1\mp N_c)
\label{ad 3}
\end{eqnarray}
In the leading logarithm approximation at $\mu > m_c$, one has [\cite{Bcd19}]
\begin{eqnarray}
C_+^c(\mu) & = &{\biggl(\frac{\alpha_S(M_W^2)}{\alpha_S(m_b^2)}\biggr)}^{6/23}
 {\biggl(\frac{\alpha_S(m_b^2)}{\alpha_S(\mu^2)}\biggr)}^{6/25}, \nonumber \\
C_-^c(\mu) & = & [С_+^c(\mu)]^{-2}.
\label{ad 4}
\end{eqnarray}
at $\alpha_S(m_c^2)=0.27$, $\alpha_S(m_b^2)=0.19$, $\alpha_S(M_W^2)=0.11$,
one has the values $C_+^c(m_c)=0.80$ and $C_-^c(m_c)=1.57$.

At $\mu > m_b$, the anomalous dimensions of the $C_{\pm}^b$ operators are
determined by eq.(\ref{ad 3}),  but at $m_c < \mu < m_b$ one finds
\begin{eqnarray}
\gamma_{\pm} = - \frac{\alpha_S}{2\pi}\bigl[3\frac{N_c^2-1}{4N_c}+
\frac{3}{2N_c}(1\mp N_c)\bigr],
\end{eqnarray}
\begin{eqnarray}
C_+^b(\mu) & = &{\biggl(\frac{\alpha_S(M_W^2)}{\alpha_S(m_b^2)}\biggr)}^{6/23}
 {\biggl(\frac{\alpha_S(m_b^2)}{\alpha_S(\mu^2)}\biggr)}^{-3/25},  \\
C_-^b(\mu) & = &
{\biggl(\frac{\alpha_S(M_W^2)}{\alpha_S(m_b^2)}\biggr)}^{-12/23}
 {\biggl(\frac{\alpha_S(m_b^2)}{\alpha_S(\mu^2)}\biggr)}^{-12/25}.
\end{eqnarray}
The numerical values are $C_+^b(m_c)=0.90$ and $C_-^b(m_c)=1.57$.

For the nonleptonic inclusive spectator decays of $B_c$ meson, the
enhancement factor due to the "dressing" of the four-fermion operators by
hard gluons, is equal to
\begin{equation}
3 \bigl[C_+^2\frac{N_c+1}{2N_c} + C_-^2\frac{N_c-1}{2N_c}\bigr],
\label{ad 5}
\end{equation}
where $3$ is the colour factor. For the annihilation decays, the
corresponding factor equals
\begin{equation}
3 \bigl[C_+\frac{N_c+1}{2N_c} + C_-\frac{N_c-1}{2N_c}\bigr]^2.
\label{ad 6}
\end{equation}

The widths of the annihilation and inclusive spectator decays are presented in
Table \ref{td2.1}. As has been mentioned, the quark masses have the following
values $m_c=1.5$ {\rm GeV}, $m_b=4.9$ {\rm GeV} and $m_s=0.15$ {\rm GeV},
i.e. one takes the choice for the good description of both the semileptonic
decays of $B$  and $D$ mesons and the total $B$ meson width.
The enhancement factors (\ref{ad 5}), (\ref{ad 6}) are calculated in the
large $N_c$ limit (this approximation gives a good description of $B$  and
$D$ meson decays [\cite{3a,Bcd14}]). For the $b$-spectator decays, one
accounts for the phase space reduction, in contrast to calculations in
ref.[\cite{i14c}].

The results of calculations for the widths of exclusive decays (here we
consider the two-particle states) were performed in the models of
WSB, ISGW and [\cite{kitay}] and are presented in Tables \ref{td2.8} and
\ref{td2.9}. The Cabibbo nonsuppressed widths of $b$-spectator decays are
shown in Table \ref{td2.8}.

The $a_1$ and $a_2$ coefficients, accounting for the renormalization of the
four-fermion operators, are defined in the following way
\begin{eqnarray}
a_1 & = & C_+\frac{N_c+1}{2N_c}+C_-\frac{N_c-1}{2N_c},\\
a_2 & = & C_+\frac{N_c+1}{2N_c}-C_-\frac{N_c-1}{2N_c}.
\end{eqnarray}
In the limit $N_c\rightarrow \infty$ one has
\begin{eqnarray}
a_1 & \approx & 0.5\cdot(C_+ + C_-), \nonumber\\
a_2 & \approx & 0.5\cdot(C_+-C_-).
\end{eqnarray}

The $a_1$ and $a_2$ values, used in refs.[\cite{i14e,kitay}],
differ from each other because of  a different choices for the quark
masses and the $\Lambda_{QCD}$ parameter.

Note, that for the decays with the $B$ mesons in the final state, the
contribution of the annihilation and "penguin" diagrams is suppressed as
$O(\sin^{10}\theta_c)$. As one can see from Table \ref{td2.8}, the WSB results
agree with the ISGW model, and the sum of the two-particle decay widths
is equal to the total inclusive width of the $b$-spectator decay (see
Table \ref{td2.1}).
\begin{table}[t]
\caption{The widths (in units $10^{-6}$ eV) of two-particle hadronic
$\bar b$-spectator decays ($M_{B_c} = 6.27$ {\rm GeV}, $M_{B_s}=5.39$
{\rm GeV}, $M_{B_s^*}=5.45$ {\rm GeV})}
\label{td2.8}
\begin{center}
\begin{tabular}{||l|c|c|c|c|c|c||}
\hline
Decay    & WSB& $a_1=1.23$    & ISGW& $a_1=1.23$  & [\cite{kitay}]& $a_1 =
1.12$\\
mode     &    & $a_2=0.33$    &     & $a_2=0.33$  &            & $a_2 =
-0.26$\\
\hline
 $B_c^+ \to B_s+\pi^+$ & $a_1^2$ 31.1 & 47.8 & $a_1^2$ 44.0& 67.7 & $a_1^2$
58.4 & 73.3 \\

 $B_c^+ \to B_s+\rho^+$ & $a_1^2$ 12.5& 19.2&  $a_1^2$ 20.2& 3.1  & $a_1^2$
44.8 & 56.1\\

 $B_c^+ \to B_s^*+\pi^+$ & $a_1^2$ 25.6 & 39.4&  $a_1^2$ 34.7&53.4& $a_1^2$
51.6  &64.7  \\

 $B_c^+ \to B_s^*+\rho^+$ & $a_1^2$ 115.6 & 177.8 & $a_1^2$ 152.1&234& $a_1^2$
150 & 188  \\

 $B_c^+ \to B^++\bar K^0$ & $a_2^2$ 28.2 & 3.1 & $a_2^2$ 61.4 & 6.7 & $a_2^2$
96.5 & 4.25\\

 $B_c^+ \to B^++\bar K^{*0}$ & $a_2^2$ 10.0 & 1.1 & $a_2^2$ 24.1 & 2.6 &
$a_2^2$ 68.2 & 3.01 \\

 $B_c^+ \to B^{*+}+\bar K^0$ & $a_2^2$ 31.0 & 3.4  & $a_2^2$ 28.3 & 3.1 &
$a_2^2$ 73.3 & 3.23\\

 $B_c^+ \to B^{*+}+\bar K^{*0}$ & $a_2^2$ 147.1 & 16 & $a_2^2$ 163.8 & 18 &
$a_2^2$ 141 & 6.23\\

 $B_c^+ \to B^0+\pi^+$ & $a_1^2$ 0.97 & 1.49 & $a_1^2$ 1.89 & 2.9 & $a_1^2$
3.30 & 4.14  \\

 $B_c^+ \to B^0+\rho^+$ & $a_1^2$ 0.94 & 1.45 & $a_1^2$ 2.14 & 3.3 & $a_1^2$
5.97 & 7.48\\

 $B_c^+ \to B^{*0}+\pi^+$ & $a_1^2$ 1.58 & 2.42 &$a_1^2$ 1.28 & 2.0 & $a_1^2$
2.90 & 3.64  \\

 $B_c^+ \to B^{*0}+\rho^+$ & $a_1^2$ 8.82 & 13.6 & $a_1^2$ 8.86 &12. & $a_1^2$
11.9 & 15.0 \\

 $B_c^+ \to B^++\pi^0$ & $a_2^2$ 0.48 & 0.05 & $a_2^2$ 0.95 & 0.1 & $a_2^2$
1.65 & 0.074 \\

 $B_c^+ \to B^++\rho^0$ & $a_2^2$ 0.47 & 0.05 &$a_2^2$ 1.07 & 0.12 & $a_2^2$
2.98 & 0.132\\

 $B_c^+ \to B^++\omega$ & $a_2^2$ 0.38 & 0.04 & $a_2^2$ 0.87 & 0.09 & -- & --
\\

 $B_c^+ \to B^{*+}+\pi^0$ & $a_2^2$ 0.79 & 0.09 & $a_2^2$ 0.64 & 0.07 & $a_2^2$
1.45 & 0.064\\

 $B_c^+ \to B^{*+}+\rho^0$ & $a_2^2$ 4.41 & 0.48 & $a_2^2$ 4.43 & 0.48 &
$a_2^2$ 5.96 & 0.263\\

 $B_c^+ \to B^{*+}+\omega$ & $a_2^2$ 3.60 & 0.39 & $a_2^2$ 3.53 & 0.38 & -- &
--\\

 $B_c^+ \to B_s+K^+$ & $a_1^2$ 2.18 & 3.35 & $a_1^2$ 3.28 & 5. & $a_1^2$ 4.2 &
5.27\\

 $B_c^+ \to B_s^*+K^+$ & $a_1^2$ 1.71 & 2.6 & $a_1^2$ 2.52 & 3.9 & $a_1^2$ 2.96
& 3.72 \\

$B_c \to B^0+K^+$ & -- & -- & -- & -- & $a_1^2$ 0.255 & 0.32\\

$B_c \to B^0+K^{*+}$ & -- & -- & -- & -- & $a_1^2$ 0.180 & 0.226\\

$B_c \to B^{*0}+K^+$ & -- & -- & -- & -- & $a_1^2$ 0.195 & 0.244\\

$B_c \to B^{*0}+K^{*+}$ & -- & -- & -- & -- & $a_1^2$ 0.374 & 0.47\\
\hline
\end{tabular}
\end{center}
\end{table}

The widths in the model of ref.[\cite{kitay}] are slightly greater.
The reason for the deviation from the results of two other models can be
the fact, that for the $b$-spectator decays, there are
$B$  and $B_s$ mesons in the final state, so that these mesons are
relativistic systems because of the light quark presence and, hence, the
nonrelativistic approximation can poorly work.

Among the $c$-spectator decays, the widths, presented in Table \ref{td2.9},
are given for the decays,  where the WSB and ISGW models result in close values
and one can neglect contributions of the annihilation and "penguin"
diagrams. As one can see from Table \ref{td2.9}, the data of ISGW model well
agree with the results of model [\cite{kitay}].

The total inclusive nonleptonic width of the $B_c$ meson decay
with the $J/\psi$ particle in the final state can be obtained from the
corresponding width of the semileptonic decay
\begin{eqnarray}
\Gamma(B_c\rightarrow J/\psi X_{u\bar d(\bar s)}) =
3\cdot a_1^2 \Gamma(B_c\rightarrow J/\psi e\nu) |V_{ud(s)}|^2\;.
\end{eqnarray}
In the large $N_c$ limit, one has $3\cdot a_1^2=4.6$ ($a_1=1.18$) and
$$ \Gamma(B_c\rightarrow J/\psi X_{u\bar d(\bar s)})
   \approx 190\cdot10^{-6}\; {\rm eV}.
$$
The branching ratio of the $B_c$ decay with the $J/\psi$ particle in the
final state equals
$$ BR(B_c\rightarrow J/\psi+X) \approx 0.2 \;.
$$
\begin{table}[t]
\caption{The widths (in units $10^{-6}$ eV) of two-particle
hadronic $c$-spectator decays}
\label{td2.9}
\begin{center}
\begin{tabular}{||l|c|c|c|c||}
\hline
Decay mode     & ISGW & $a_1=1.18$ & [\cite{kitay}] & $a_1$ = 1.26\\
\hline
 $B_c^+ \to \eta_c+\pi^+$ & $a_1^2$ 1.71 & 2.63 & $a_1^2$ 2.07 & 3.29 \\

 $B_c^+ \to \eta_c+\rho^+$ & $a_1^2$ 4.04  & 6.2& $a_1^2$ 5.48 & 8.70\\

 $B_c^+ \to J/\psi+\pi^+$ & $a_1^2$ 1.79  & 2.75 & $a_1^2$ 1.97 & 3.14\\

 $B_c^+ \to J/\psi+\rho^+$ & $a_1^2$ 5.07  & 7.8 &  $a_1^2$ 5.95 & 9.45\\

 $B_c^+ \to \eta_c+K^+$ & $a_1^2$ 0.127  & 0.195 &  $a_1^2$ 0.161 & 0.256\\

 $B_c^+ \to \eta_c+K^{*+}$ & $a_1^2$ 0.203  &0.31 & $a_1^2$ 0.286 & 0.453\\

 $B_c^+ \to J/\psi+K^+$ & $a_1^2$ 0.130  &0.2 & $a_1^2$ 0.152 & 0.242 \\

 $B_c^+ \to J/\psi+K^{*+}$ & $a_1^2$ 0.263 &0.4 & $a_1^2$ 0.324 & 0.514  \\

$B_c \to \psi(2S) +\pi^+$ & -- & -- &  $a_1^2$ 0.251 & 0.398 \\

$B_c \to \psi(2S) +\rho^+$ & -- & -- &  $a_1^2$ 0.71 & 1.13 \\

$B_c \to \psi(2S) +K^+$ & -- & -- &  $a_1^2$ 0.018 & 0.029 \\

$B_c \to \psi(2S) +K^{*+}$ & -- & -- &  $a_1^2$ 0.038 & 0.060 \\
\hline
\end{tabular}
\end{center}
\end{table}

The WSB and ISGW models give close results for the two-particle
$B_c$ meson decays with the $B$ mesons in the final state. Unfortunately,
it is complex to detect the $B_c$ mesons in such decay modes, since
one has to reconstruct the $B$ mesons from the products of their weak decays.
For the $B_c$ meson observation and its mass determination, the
$B_c^+\rightarrow J/\psi \pi^+$ decay is more preferable [\cite{i14e}].
Its branching ratio is equal to
$$ BR(B_c^+\rightarrow J/\psi\pi^+) \approx 2\cdot10^{-3}.
$$

The $B_c$ meson decays, where the $CP$ violation can be observed,
$B_c^{\pm}\to (c\bar c)D^{\pm}$, $B_c\to D\rho(\pi)$ and
$B_c\to D^0 D_s$ are of a special interest.

The approximate estimates for the decay branching fractions and  the
asymmetry parameter of $CP$ violation, are obtained in ref.[\cite{15a}].
The corresponding results are presented in Table \ref{td2.10}.
The asymmetry parameter $A$ is defined in the following way
\begin{equation}
A = \frac{\Gamma(B_c^-\to \bar X)-\Gamma(B_c^+\to X)}
 {\Gamma(B_c^-\to \bar X)+\Gamma(B_c^+\to X)}.
\label{ass}
\end{equation}

A large value of the asymmetry is expected in the $B_c\to D^{*}_s D^{0}$
decays with the $D^0$ meson, decaying into the eigen $CP$ invariance state.
However, the branching ratio of such event is too low
$$ BR(B_c^+\to D_s^{*+} D^0) \approx 10^{-6}.
$$
The $D_s^{*+}$ meson identification is also complicated. As one can see from
Table \ref{td2.10}, the best mode for the $CP$ violation observation in the
$B_c$ meson decay can be $B_c^{\pm}\to(c\bar c) D^{\pm}$.
However, even at the expected statistics of the $B_c$ meson yield at the future
colliders (about $10^{9}$--$10^{11}$ events), it is difficult to observe
such events, if one takes into account the branching fractions of
the $(c\bar c)$ states and $D$ meson decays.
\begin{table}[b]
\caption{The branching ratios ($BR$) and asymmetries ($A$) for
the $CP$ violating $B_c$ decays}
\label{td2.10}
\begin{center}
\begin{tabular}{||l|c|c||}
\hline
$X$    & $BR(B_c^+\to X)$ & $A$ \\
\hline
 $ \eta_c D^{*+}$ & $1.0 \cdot 10^{-4}$ & $1.5 \cdot 10^{-2}$ \\

$ \eta_c D^+$ & $1.2 \cdot 10^{-4}$ & $-0.3 \cdot 10^{-2}$\\

$ J/\psi D^+$ & $0.5 \cdot 10^{-4}$  & $0.6 \cdot 10^{-2}$\\

 $D^0 \rho^+$ & $2.8 \cdot 10^{-5}$  & $1.9 \cdot 10^{-3}$ \\

 $D^+ \rho^0$ & $1.6 \cdot 10^{-5}$  & $3.0 \cdot 10^{-3}$ \\

 $D^{*0} \pi^+$ & $3.3 \cdot 10^{-5}$  & $1.3 \cdot 10^{-3}$ \\

 $D^{*+} \pi^0$ & $1.8 \cdot 10^{-5}$  & $2.0 \cdot 10^{-3}$ \\

 $D^0 \pi^+$ & $1.6 \cdot 10^{-6}$  & $-8.9 \cdot 10^{-3}$ \\

 $D^+ \pi^0$ & $0.4 \cdot 10^{-6}$  & $-13.8 \cdot 10^{-3}$ \\
\hline
\end{tabular}
\end{center}
\end{table}

It is difficult to estimate the decay widths, but it is worth to mention
about the $B_c$ meson decay modes such as $B_c\to 3 D X$ or
$B_c\to D_s \phi$ and $B_c\to \bar D K$. The $B_c\to \psi(3S) D$ decay
can be of a great interest, when  $\psi(3S)$ has the decay into the
$D$ meson pair. However, it is probable, that this decay width, as the width
of the decay into three $D$ mesons, is small because of a smallness of the
phase space, yet. The $B_c\to D_s \phi$ decay width can be roughly
estimated of the order of 2 \% [\cite{7ka}], but it will be very difficult
to observe the $B_c$ meson in such mode because of the complex reconstruction
of the $D_s$ meson.


\section{Production of $B_c$ mesons}

The electromagnetic and hadronic production of $B_c$ as the particle with the
mixed flavour supposes the joint production of the heavy quarks
$\bar b$ and $c$. This explains the low value of the $B_c$ production cross
section in comparison with the production cross sections of the particles
from the $\psi$ and $\Upsilon$ families. On the other hand, the absence of the
$B_c$ decays channels into light hadrons due to the strong interactions
leads to that all bound $(\bar b c)$ states basically transformed into
the lowest state with the probability close to unit, due to the radiative
transitions (see Section 2).

{}From the theoretical point of view, the production of $B_c$ meson,
having the small sizes, takes place with the virtualities of the order of
the heavy quark mass sum. This fact makes sure the perturbation theory
to be applicable to the processes of the $B_c$ production.
The nonperturbative part, related with the account for the
$B_c$ wave function, is quite reliably calculated in this case.

\subsection{Production of $B_c$ mesons in $e^+e^-$-annihilation}

The simplest example of the $B_c$ production in $e^+e^-$-annihilation
(in the region of $Z$ peak) is described by the diagrams on Figure \ref{fp1}.
\begin{figure}[t]
\vspace*{7.5cm}
\caption{The diagrams of the single $B_c$ meson production in
 $e^+e^ -$-annihilation}
\label{fp1}
\end{figure}

The matrix element of the $(\bar b c)$ quarkonium production is transformed
from the corresponding matrix element
$T(p_b,p_c)$ for four heavy quark production by the integration over
the relative momentum of the $\bar b$- and $c$-quarks
with the weight of the quarkonium wave function
\begin{equation}
T_j = \int\frac{{\rm d}^3{\bf {q}}}{(2\pi)^3} \Psi ({\bf{q}})
T(p_{\bar b}, p_{c})^{ab}_{\alpha\beta}
(-\hat p_{\bar b} + m_b)^{\alpha '\alpha}
(-\hat p_{c} + m_c)^{\beta '\beta} \Gamma_j^{\alpha '\beta '}
\frac{\sqrt{2M}}{\sqrt{2m_b 2m_c}} \frac{\delta^{ab}}{\sqrt{3}}\;,
\label{p1}
\end{equation}
where $M$ is the meson mass and
\begin{equation}
\Gamma^{\alpha\beta} =\frac{1}{\sqrt{2}} \gamma_5^{\alpha\beta}
\;,\; \;\;
\Gamma_\lambda^{\alpha\beta} =\frac{1}{\sqrt{2}}
\gamma_\mu^{\alpha\beta}\epsilon^\mu_\lambda
\end{equation}
for the pseudoscalar state and the vector one $(B_c, B^*_c)$, respectively.
The quark momenta are determined by the relations
\begin{equation}
p_{\bar b} =\frac{m_b}{M} p+q, \;,\;\;\;
p_{c} =\frac{m_c}{M} p-q\;,
\end{equation}
\begin{equation}
p\cdot q =0\;.
\end{equation}
For the heavy quarkonium one has $|{\bf q}| \ll m_b,\,m_c$ and eq.(\ref{p1})
can be simplified by the substitution for
$T^{ab}_{\alpha\beta}(p_{\bar b}, p_{c})$ by its value at $\bf q =0$. Then
\begin{equation}
\int \frac{{\rm d}^3{\bf{q}}}{(2\pi)^3}\Psi ({\bf{q}}) =\Psi ({\bf x})\left
|_{\bf x=0}\right.
\end{equation}
In papers of refs.[\cite{i17c}--\cite{3b}] the total cross sections of the
$B_c$  and $B^*_c$ mesons  and its distributions over the variable
$z={2E_{B_c}}/{\sqrt{s}}$ have been obtained. The result of the precise
numerical calculations in the technique of spiral amplitudes with the Monte
Carlo integration over the phase space is presented on Figure 8.
One can see that this distribution is rather hard with the maximum at
$z_{\rm max} ={M}/({M+m_c})\approx 0.8$. At this value of $z_{\rm max}$
the $B_c$ meson and $\bar c$-quark have zero relative velocity. If one
remembers, in the considered approximation the relative motion of the
$c$- and $\bar b$-quarks inside the $B_c$ meson is absent, then one
clearly finds, that the maximum in the distribution corresponds to the
configuration, when all quarks move as a whole with one and the same velocity.
In this case the minimal virtualities of the initial
$\bar b$-quark $p^2=(m_b +2m_c)^2$ and the gluon $k^2\approx 4m^2_c$ are
realized. At any other $z$ values, these virtualities increase. Note,
these speculations are correct only for the last two diagrams on
Figure \ref{fp1}, when one can neglect the contribution of the first and second
diagrams, suppressed up to two orders of magnitude in respect to the former.
In the asymptotic limit $s\rightarrow \infty$, when one can neglect terms
of the order of $M^2/s$ and the higher powers of this ratio, choosing the
special gauge condition (the axial gauge with the four-vector
$n=(1,0,0,-1)$ along the direction of the $b$-quark motion),
one can show that the contribution of the last diagram on Figure \ref{fp1}
survives only. In this case the expression for
$\sigma^{-1}{{\rm d}\sigma}/{{\rm d}z}$ acquires the sense of the fragmentation
function of the $\bar b$-quark into the $B_c$ meson,
if one chooses the $b$-quark production cross section $\sigma$ as the
normalization factor at the same energy.

The function of the $\bar b\rightarrow B_c$ fragmentation,
where $B_c$ is the pseudoscalar state, has the following form
\begin{eqnarray}
\dot D(z)_{\bar b\rightarrow B_c} & = &
\frac{8\alpha^2_S |\Psi (0)|^2}{81 m^3_c}
\frac{rz(1-z)^2}{(1-(1-r)z)^6}
(6-18(1-2r)z+ \nonumber \\
&&
(21-74r+68r^2)z^2 -
2(1-r)(6-19r+18r^2)z^3\nonumber \\
&&
+3(1-r)^2 (1-2r+2r^2)z^4)\;,
\label{p6}
\end{eqnarray}
and for the fragmentation into the vector state one has
\begin{eqnarray}
 D(z)_{\bar b\rightarrow B^*_c} & = &
\frac{8\alpha^2_s |\Psi (0)|^2 }{27 m^3_c}
\frac{rz(1-z)^2}{(1-(1-r)z)^6}
(2-2(3-2r)z+ \nonumber \\
&&
3(3-2r +4r^2)z^2 -
2(1-r)(4-r+2r^2)z^3 +\nonumber\\
&&
(1-r)^2(3-2r+2r^2)z^4)\;,
\label{p7}
\end{eqnarray}
where $r=m_c/(m_b+m_c)$.
As one can see from Figure  \ref{fp2}, $D_{\bar b\rightarrow B_c}(z)$ and
$D_{\bar b\rightarrow B^*_c}(z)$ are in a good agreement with the results of
precise calculations. The $\bar b\rightarrow B^*_c$ process
has slightly more hard
distribution in comparison with the $\bar b\rightarrow B_c$ one.
At $\alpha_S=0.22$, $|\Psi (0)|^2=f^2_{B_c}M_{B_c}/12$,
$f_{B_c} =560$ {\rm MeV} and $m_c =1.5$ {\rm GeV}, the corresponding integral
probabilities are equal to $3.8\cdot 10^{-4}$ for
$\bar b\rightarrow B_c$ and $5.4\cdot 10^{-4}$ for
$\bar b\rightarrow B^*_c$. The probabilities of the $c$-quark fragmentation
into $B_c$ are suppressed by two orders of magnitude in respect to the
values, given above. As the fraction of the $b\bar b$ production, the total
number of the produced $B_c(\bar B_c)$ mesons with the account of the
$B^*_c(\bar B^*_c)$ states and the first radial excitations is equal to
(at $\alpha_S=0.22$)
\begin{equation}
R_{B_c}=\frac{\sigma (e^+e^-\rightarrow B_c^++x)
+\sigma (e^+e^-\rightarrow B_c^-+x)}
{\sigma (e^+e^-\rightarrow b\bar b)} =2\cdot 10^{-3}
\end{equation}
\begin{figure}[b]
\vspace*{7.5cm}
\caption{The functions of the $\bar b$-quark fragmentation
into the $B_c$  and $B_c^*$  mesons}
\label{fp2}
\end{figure}

Due to the quark-hadron duality, there is the independent way of estimation
for this ratio. To reach this goal, one has to compare the obtained cross
section of the bound $\bar b c$ state production with the cross section for
the production of the colour-singlet $(\bar b c)$ pair in the process
$e^+e^-\rightarrow b\bar b c\bar c$ with the low values of invariant mass
$M_{\bar b c}$
\begin{equation}
\int\limits^{M^2_{{\rm th}}}_{m^2_0}
\frac{{\rm d}\sigma (e^+e^-\rightarrow b\bar b c\bar c)_{\bar b c-\mbox{singl}}
}{{\rm d}M^2_{\bar b c}}
 {\rm d}M^2_{\bar b c}\;,
\label{p9}
\end{equation}
where $m_0 = m_b+m_c \leq M_{\bar b c}\leq M_B+M_D+\Delta M=M_{{\rm th}}$
and $\Delta M \approx 0.5\div 1$ {\rm GeV}. Supposing $m_0 =6.1$ {\rm GeV} and
$M_{{\rm th}}$=8 {\rm GeV} as the threshold value, one gets the
$\bar b c$ system production cross section of the order of $7$ pb.
On the other hand, the sum of the cross sections for the production of
$B_c$ and its first excitations equals $9.3$ pb, as is seen from Table
\ref{tp1}.
\begin{table}[t]
\caption{The cross sections (in pb) for the production of the $S$-wave
states of $B_c$ mesons in the $Z$ boson peak}
\label{tp1}
\begin{center}
\begin{tabular}{|c|c|c|c|c|}\hline
State&$ 1^1S_0$ &$1^1S_1$ &$ 2^1S_0$ &$2^1S_1$\\ \hline
$\sigma(\alpha_S=0.22)$ & 3.14 & 4.37 & 0.805 & 1.078 \\
\hline
\end{tabular}
\end{center}
\end{table}

The comparison of these two independent estimates states, on the one hand,
about a good agreement. On the other hand, it means that the contribution of
the higher excitations is not large, and the total cross section is saturated
by the $S$-wave levels.

Recent direct calculations of the cross sections for the $P$ level production
[\cite{p1}] confirm this conclusion. According to the estimates of this paper,
the sum over the cross sections for the $P$-wave levels production is less
than 10 \% from the sum of the $S$-wave level contributions.

In ref.[\cite{i18}] the functions of the heavy quark fragmentation into the
heavy polarized vector quarkonium have been studied, so, for the longitudinally
polarized quarkonium, one has found the expression
\begin{eqnarray}
 D(z)^L_{\bar b\rightarrow B^*_c} & = &
\frac{8\alpha^2_s |\Psi (0)|^2 }{81 m^3_c}
\frac{rz(1-z)^2}{(1-(1-r)z)^6}
(2-2(3-2r)z+ \nonumber \\
&&
(9-10r +16r^2)z^2 -
2(1-r)(4-5r+6r^2)z^3 +\nonumber\\
&&
(1-r)^2(3-6r+6r^2)z^4)\;,
\label{p77}
\end{eqnarray}
that does not depend on the polarization of the fragmentating quark.
At $r=1/2$, expression (\ref{p77}) coincides with the result, obtained for
the heavy quarkonium with the hidden flavour $(\Upsilon,\;\psi)$ [\cite{i18}].

Fragmentation function (\ref{p77}) agrees with the consideration of the
heavy quark fragmentation into the heavy meson $(Q\bar q)$, where in the limit
of infinitely heavy quark, EHQT leads to the equal probability production
of the vector quarkonium with the arbitrary orientation of its spin, i.e.
to the absence of the spin alinement and to the ratio of the vector
and pseudoscalar state yields, equal to $V/P=3$ [\cite{p4}].

For the heavy quarkonium, the relative yield of the vector and pseudoscalar
mesons is close to unit, and the spin alinement of the vector state has a
notable value. For the $B_c^*$ meson, this can be observed in the angle
distribution of the $B_c^*\to B_c\gamma$ decay, which composes the total
$B_c^*$ width. This distribution has the form
\begin{equation}
\frac{{\rm d}\Gamma}{{\rm d}\cos \theta} \sim 1 -
\biggl(\frac{3\xi-2}{2-\xi}\biggr)
\cos^2 \theta\;,
\label{pr.2}
\end{equation}
where $\theta$ is the angle between the photon and the $B_c^*$ polarization
axis in the system of $B_c^*$ rest, and the asymmetry parameter $\xi$
determines the relative yield of the transversally polarized $B_c^*$ state
\begin{equation}
\xi = \frac{T}{L+T}\;.
\end{equation}
For the integral asymmetry at the small mass of the generated quark, entering
the meson, $r\ll 1$, one has
\begin{equation}
\xi = \frac{2}{3} + \frac{5}{16}r+O(r^2)\;.
\end{equation}
The anisotropy in the $B_c^*\to B_c\gamma$ decay is numerically equal to
6 \%.

In ref.[\cite{i18}] the vector quarkonium spin alinement has been studied
versus the transverse momentum in respect to the fragmentation axis.
One has derived quite bulky analytical expressions for the fragmentation
functions $D^{L,T}_{\bar b\to B_c^*}(p_{\rm t})$, that linearly tend to zero at
$p_{\rm t} \to 0$ and decrease as $1/p_{\rm t}^3$ at $p_{\rm t}\to \infty$.
It is interesting, that the average transverse momentum at the fragmentation
into the longitudinally polarized vector $B_c$ quarkonium is twice greater
than the average transverse momentum at the fragmentation
into the transversally polarized $B_c^*$ meson,
$\langle p_{\rm t}\rangle \approx 7$ {\rm GeV}.

The event with the $B_c$ meson has the characteristic signature. The hadron jet
from the $b$-quark must be produced in the direction, opposite to the
$B_c$ motion. The $B_c$ meson must be accompanied by the $\bar D$ meson
with the average ratio of the momenta
$\langle z_D\rangle/\langle z_{B_c}\rangle\approx 0.3$ and the average
angle between the momenta about $20^o $ [\cite{3b}].

The single $B_c$ meson production in $e^+e^-$-annihilation has been also
considered in refs.[\cite{i17c,7ka}].

In ref.[\cite{i17a}] the exclusive production of the
$B_c^{(*)+}B_c^{(*)-}$ pairs in $e^+e^-$-annihilation has been calculated at
low energies, where one can neglect the $Z$ boson contribution.
The total cross sections of the vector and pseudoscalar states have the form
\begin{eqnarray}
{}~ & ~ &\sigma(e^+e^- \to (Q_1 \bar Q_2)_P (\bar Q_1 Q_2)_P)  =
\frac{\pi^3 \alpha_S^2(4 m_2^2) \alpha^2_{{\rm em}}}{3^7\;4 m_2^6}\;
\frac{m_1^2}{M^2}\; f_P^4 (1-v^2)^3 v^3  \times \nonumber \\
{}~ & ~ &
\biggl(3 e_1 \biggl(\frac{2 m_2}{m_1} - 1+v^2\biggr)
 - 3e_2 \biggl(2 - (1-v^2)\frac{m_2}{m_1}\biggr)
\frac{m_2^3 \alpha_S(4 m_1^2)}{m_1^3 \alpha_S(4 m_2^2)}\biggr)^2\;, \\
{}~ & ~ & \sigma(e^+e^- \to (Q_1 \bar Q_2)_P (\bar Q_1 Q_2)_V)  =
\frac{\pi^3 \alpha_S^2(4 m_2^2) \alpha^2_{{\rm em}}}{3^7\;2 m_2^6}\;
f_P^2 f_V^2 (1-v^2)^4 v^3  \times \nonumber \\
{}~ & ~ & \biggl(3 e_1 - 3e_2
\frac{m_2^3 \alpha_S(4 m_1^2)}{m_1^3 \alpha_S(4 m_2^2)}\biggr)^2\;,
\\
{}~ & ~ & \sigma(e^+e^- \to (Q_1 \bar Q_2)_V (\bar Q_1 Q_2)_V)  =
\frac{\pi^3 \alpha_S^2(4 m_2^2) \alpha^2_{{\rm em}}}{3^7\;2 m_2^6}\;
f_V^4 (1-v^2)^3 v^3 \times \nonumber \\
{}~ & ~ & \biggl(3 e_1 - 3e_2
\frac{m_2^3 \alpha_S(4 m_1^2)}{m_1^3 \alpha_S(4 m_2^2)}\biggr)^2
[3 (1-v^2) + (1+v^2)(1-a)^2 + \nonumber \\
{}~ & ~ & \frac{a^2}{2} (1-v^2)
(1-3v^2)]\;,
\end{eqnarray}
where $v = \sqrt{1- 4M^2/s}$, $M=m_1+m_2$,
\begin{equation}
a = \frac{m_1}{M}\; \biggl(1 - \frac{e_2}{e_1} \frac{m_2^4}{m_1^4}
\frac{\alpha_S(4 m_1^2)}{\alpha_S(4 m_2^2)}\biggr)\bigg/
\biggl(1 - \frac{e_2}{e_1} \frac{m_2^3}{m_1^3}
\frac{\alpha_S(4 m_1^2)}{\alpha_S(4 m_2^2)}\biggr)\;.
\end{equation}
The relative yield of the $B_c$ meson pairs
$R=\sigma(B_c^+B_c^-)/\sigma(b\bar b)$
reaches its maximum at the energy $\sqrt{s}= 14$ {\rm GeV}, where it is
equal to $R\approx 10^{-4}$. This ratio rapidly decreases with the energy
growth, where the single $B_c$ production becomes to dominate.

As one can see, the study of the $B_c$ meson production in
$e^+e^-$-annihilation allows one to make the analytical researches of the
heavy quark dynamics.

Thus, in the $Z^0$ boson pole, where the $b$-quark production cross section
is large, one has to expect of the order of 2 events with the $B_c$ production
per each thousand $b\bar b$ pairs. As it is planned, in the experiments at the
LEP accelerator, the number about $2\cdot 10^7\;  Z^0$ bosons will be detected.
This means, that the total number of $B_c(\bar B_c)$ events has to be of the
order of $10^4$. Certainly, the real number of reconstructed events will be
less, if one takes into account the particular modes of the decay.

\subsection{Hadronic production of $B_c$ mesons}

As has been mentioned, the process of the $B_c$ meson production in
$e^+e^-$-annihilation at large energies can be reformulated as the process of
the $\bar b\rightarrow B_c (B^*_c)$ fragmentation, appearing with the
probability about $10^{-3}$.

The hadronic $B_c$ production turns out to be more complex. First, at
hadronic production the region of low partonic energies dominates, so that the
asymptotic regime with the cross section factorization
\begin{equation}
\frac{{\rm d}\sigma}{{\rm d}z}\sim\sigma_{b\bar b}\; D_{\bar b\rightarrow
B_c}(z)
\end{equation}
is not yet realized. Second, in the hadron interactions a new type of
diagrams appears, so we will further label the latter as the recombinational
diagrams, for which the factorization does not take place.

The contribution of such diagrams, dominating at low masses of the
$B_c^-\bar b c$ system, decreases with the growth of this mass, however,
it remains essential even at large masses and large transverse momenta.
The contribution of such type diagrams into the $B_c^{(*)}$ production has been
first calculated for the exclusive $B_c^{(*)}$ pair production in the
quark-antiquark annihilation at low energies [\cite{i17}].

The typical set of the QCD diagrams in the fourth order over
$\alpha_S$ is shown on Figure \ref{fp3}. Here, as in the case of the
$B_c$ production in $e^+e^-$-annihilation, the matrix element of the
$(\bar b c)$ quarkonium production is obtained from the corresponding matrix
element of four quarks production by the integration over the relative momentum
of the $c$- and $\bar b$-quarks with the weight, determined by the
quarkonium wave function.
\begin{figure}[t]
\vspace*{11.5cm}
\caption{The diagrams of the single $B_c$ meson production in
the gluon and quark subprocesses }
\label{fp3}
\end{figure}

At high energies, where the $B_c$ production cross sections are accessible
for the meson observation, the gluon-gluon contribution into the production
dominates.

The energetic spectra of the $B_c$  and $B_c^*$ mesons in the system of
mass centre for two colliding gluons are shown on Figure \ref{fp4}
at the different values of the total energy $\sqrt{s}=$20, 40 and
100 {\rm GeV}.
\begin{figure}[t]
\vspace*{8cm}
\caption{The differential ${\rm d}\sigma/{\rm d}z$
cross sections for
the single production of the $B_c$ mesons (a) and $B_c^*$ mesons (b) in
the gluon annihilation at different values of the total
energy}
\label{fp4}
\end{figure}

The $\sigma(gg\rightarrow B_c(B^*_c)\bar c b)$ values
are presented on Figure \ref{fp5} at the several energies of
the interacting gluons for $m_b=5.1$ {\rm GeV},
$m_c=1.5$ {\rm GeV} and $\alpha_S=0.2$. The ratio of the cross sections
$\sigma_{B_c^*}/\sigma_{B_c}$ is about 3 at the energies
20, 40 and 100 {\rm GeV} and it is about 2 at the energy 1 TeV, when in
$e^+e^-$-annihilation, where the
$\bar b\rightarrow B_c$ fragmentation dominates, this ratio is
$\sigma_{B_c^*}/\sigma_{B_c}\approx 1.3$.
\begin{figure}[b]
\vspace*{8cm}
\caption{The total cross sections for the single production
of $B_c$ mesons (empty triangles) and $B_c^*$ mesons (solid triangles) in the
gluon annihilation in comparison with the production cross section
(multiplied by the factor $2\cdot 10^{-3}$) of the $\bar b b$ quark pairs
(solid line)}
\label{fp5}
\end{figure}

The variation of the
$\sigma_{B_c^*}/\sigma_{B_c}$ ratio is the consequence of the change in the
production mechanism. The fragmentatonal component gives the low contribution
in comparison with the contribution of the recombination diagrams. This can
be noted from Figure \ref{fp4}, where the differential cross sections for the
$B_c$  and $B^*_c$ meson production, calculated by the Monte Carlo integration
of the exact expression for the matrix element squared, are presented in
comparison with the cross section, calculated under the fragmentation
formulae (\ref{p6}), (\ref{p7}).

The total cross section of the $B_c(B^*_c)$ mesons is obtained from the
partonic one $\sigma_{ij}(\hat s)$  by the convolution with the functions
of the parton distributions in the initial hadrons
\begin{equation}
\sigma_{\rm tot} (s) =
\int\limits^{s}_{4({m_b}+{m_c})^2}
\frac{{\rm d}\hat s}{s}
\int\limits^{1-{\hat s}/{s}}_{-1+{\hat s}/{s}}
\frac{{\rm d}x}{x^*}
\; \sum_{ij} f^i_a (x_1)f^j_b (x_2)
\hat \sigma_{ij}(\hat s)\;,\;\;
x^*=\biggl(x^2+\frac{4\hat s}{s}\biggr)^{1/2}\;.
\end{equation}
The cross sections, calculated with the account of the known parameterizations
for $f^{i,j}_{a,b}(x)$ [\cite{p2}], are presented in Table  \ref{tp3}.
\begin{table}[b]
\caption{The cross sections (in nb) of hadronic production of the
$B_c(B^*_c)$ mesons (the standard deviation in the last digit
is shown in the brackets)}
\label{tp3}
\begin{center}
\begin{tabular}{|c|c|c|c|c|}\hline
$n^{2S+1}L_j$ & $1^1S_0$ &$1^3S_1$ &$2^1S_0$ & $2^3S_1$ \\ \hline
$\sigma_{tot} (40\; \mbox{{\rm GeV}})\cdot 10^{5}$ &
 1.63(2) & 9.5(2) & 0.13(1) & 0.75(2) \\   \hline
$\sigma_{tot} (100\; \mbox{{\rm GeV}})\cdot 10^{3}$ &
7.8(2) & 36(1) & 1.1(2) & 5.2(2) \\ \hline
$\sigma_{tot} (1.8\; \mbox{TeV})$ &
13.3(8) & 53(3) & 2.7(2) & 10.4(5) \\ \hline
$\sigma_{tot} (16\; \mbox{TeV})\cdot 10^{-2}$ &
1.96(8) & 7.6(2) & 0.43(2) & 1.66(8) \\  \hline
\end{tabular}
\end{center}
\end{table}

The energy 40 {\rm GeV} is close to the c.m.s. energy for the carrying
out the fixed target experiments at the HERA accelerator.
At $\sqrt{s}=1.8$ TeV we present the cross section of the $B_c$ production
in $p\bar p$-collisions at Tevatron, and, finally, the energy
$\sqrt{s}=16$ TeV corresponds to the conditions of the $pp$-experiment at
the future LHC collider. The energetic dependence of the cross section,
summed over the particle and antiparticle $\bar B_c$ production, is shown on
Figure \ref{fp6}.
\begin{figure}[t]
\vspace*{8cm}
\caption{The total cross sections (in nb) for the single
production of $B_c$ mesons (empty circles) in $p\bar p$-interactions
at different energies and the cross sections (in mkb) for the beauty
particle production (solid circles)}
\label{fp6}
\end{figure}

{}From the values, presented in Table \ref{tp3}, it follows that at
$\sqrt{s}=40$ {\rm GeV}, the summed cross section
$\sigma_{\rm sum}$ for the meson production is about $10^{-4}$
of the total cross section of the $b\bar b$ production, so this makes the
$B_c$ study practically to be impossible in this experiment.
One has to note, that in this case we can not restrict ourselves
by the $gg\rightarrow B_c\bar c b$ contribution and we have taken into account
the contribution of the $q\bar q\rightarrow B_c\bar c b$ process.

The experiments at Tevatron and LHC, where ${\sigma_{\rm sum}}/
{\sigma_{b\bar b}}$ is about $10^{-2}$, will give the real possibility
for the observation of hadronic $B_c$ production. Therefore, at the energies
of these two facilities, we present the most interesting distributions of the
cross sections for the $1^1S_0$  and $1^3S_1$ states production
(note, that as our calculations show, the cross section at the energies
under consideration is completely determined by the gluon-gluon interaction,
since the quark-quark contribution is suppressed by two orders of magnitude,
$10^{-2}$).

The distributions for the $1^1S_0$ pseudoscalar and $1^3S_1$ vector mesons
are shown on Figures \ref{fp7} and \ref{fp8} at the energy of the interacting
hadrons 1.8 TeV.
\begin{figure}[t]
\vspace*{8cm}
\caption{The differential ${\rm d}\sigma/{\rm d}p_{\rm t}$ cross sections
for the single production of $B_c$  and $B_c^*$ mesons in $p\bar p$
interactions at the energy 1.8 TeV}
\label{fp7}
\end{figure}
\begin{figure}[b]
\vspace*{8cm}
\caption{The differential cross sections for the single
production of $B_c$  and $B_c^*$ mesons in $p\bar p$-interactions at the
energy 1.8 TeV; (a) ${\rm d}\sigma/{\rm d}y$, where $y$ is the particle
rapidity, (b)
${\rm d}\sigma/{\rm d}x$, where $x= 2E/s^{1/2}$}
\label{fp8}
\end{figure}

The distributions ${{\rm d}\sigma}/{{\rm d}x}$ (see Figure \ref{fp8}b)
show, that we deal with the central $B_c$ production, where the
complete cross section is collected in the interval from $-0.3$ to $0.3$.
The average transverse momentum of $B_c$ is about 6 {\rm GeV}, and
from the distribution over the angle between the directions of the $B_c$ and
$\bar c$-quark motions, one can conclude that $\bar c$-quark generally moves
in the direction close to the $B_c$ one [\cite{i17e}].

One has to note, that the considered diagrams of the QCD perturbation theory
are the diagrams of the fourth order over $\alpha_S$. This results in the
strong dependence of the cross section on the particular $\alpha_S$ choice.
The latter must be determined by the typical virtuality in the
production process. The analysis shows, that this virtuality is large in
the contributions, decreasing faster that $1/\hat s$, only.
In the remaining contributions, including the fragmentational one, it is not
large and about $4m_cm_b$. Under this reason the $\alpha_S=0.2$
value, chosen as the strong coupling constant, is the most reasonable
at this scale. The use of the running coupling constant
$\alpha_S (\hat s)$, for example, leads to the decrease of the
$B_c(B^*_c)$ production cross section by about 7 times. The pessimistic
estimates of the $B_c(B^*_c)$ production are presented in ref.[\cite{p3}],
where, as one can see, the $\alpha_S (\hat s)$ value has been used.

At low energies of hadron collisions, the quark-antiquark annihilation
with the $B_c$ production dominates in respect to the gluonic one, since,
in this case, the latter has a much lower luminosity, that decreases also with
the growth of the total energy of the partonic subprocess.
At the low energies of the quark-antiquark annihilation, the exclusive
$B_c^+B_c^-$ pair production can be essential. The total cross sections of the
vector and pseudoscalar $B_c$ meson production due to the quark-antiquark
annihilation have the form
\begin{eqnarray}
\sigma(1^-,1^-) &=& \frac{\alpha_S^4\pi^3}{8\cdot 3^8}\;
\frac{f_V^4}{\mu^6}\; \lambda^3 \sqrt{1-\lambda}\;
(1.3+1.4\lambda+0.3\lambda^2)\;,
\label{p88}\\
\sigma(1^-,0^-) &=& \frac{\alpha_S^4\pi^3}{16\cdot 3^8}\;
\frac{f_V^2f_P^2}{\mu^6}\; \frac{(m_b-m_c)^2}{M^2}
\lambda^3 \sqrt{1-\lambda}\;
(1+2\lambda)\;,
\label{p89}\\
\sigma(0^-,0^-) &=& \frac{\alpha_S^4\pi^3}{16\cdot 3^8}\;
\frac{f_P^4}{\mu^6}\; \lambda^3 \sqrt{1-\lambda}\;
(1-\lambda)^2\;,
\label{p90}
\end{eqnarray}
wherefrom one can see, that the vector state production dominates.
In eqs.(\ref{p88})--(\ref{p90})  we have introduced the notations
\begin{eqnarray}
\lambda &=& 4M^2/s\;,\nonumber \\
\mu &=& \frac{m_bm_c}{m_b+m_c}\;.\nonumber
\end{eqnarray}

The numerical estimates of the total cross sections for the $B_c$ production
in $p\bar p$ interactions are presented in Table \ref{tp4}.
\begin{table}[t]
\caption{The total cross sections (in units $10^{-2}$ pb) for the pair
production of $B_c$ mesons due to the quark-antiquark annihilation in
$p\bar p(pp)$ interactions at low energies}
\label{tp4}
\begin{center}
\begin{tabular}{||c|l|l|l||}
\hline
$\sqrt{s}$, {\rm GeV} & $\sigma(1^-,1^-)$ & $\sigma(1^-,0^-)$ &
$\sigma(0^-,0^-)$\\
\hline
30 &0.9 (0.08) &0.24 (0.022) & 0.006 (0.0004) \\
40 &5.8 (0.94) &1.6 (0.25) & 0.054 (0.007)\\
50 &15.8 (3.5) & 4.3 (0.95)&0.18 (0.034) \\
\hline
\end{tabular}
\end{center}
\end{table}

Summing up the consideration of the hadronic production, one can draw
the conclusions.

1. The mechanism of the hadronic $B_c(B^*_c)$ production strongly differs
from the production in $e^+e^-$-annihilation.

2. The relative fraction of the fragmentation contribution is low even in
the region of large transverse momenta.

3. The vector state production is enforced in respect to $e^+e^-$-annihilation.

Thus, the hadronic $B_c$ production requires the analysis for the large
number of diagrams and its detailed study opens the possibility for the
investigation of effects of the heavy quark dynamics in the higher orders of
the QCD perturbation theory. As for the $B_c$ yield at the real
physical facilities, it is quite high, but the registration of the $B_c$ events
is essentially determined by the detector acceptance (cuts of the transverse
momenta of particles, characteristics of the vertex detector an so on).

\subsection{$B_c$ meson production in $\nu N$-, $ep$- and $\gamma\gamma$-
collisions}

In the previous sections we have considered the $B_c$ meson production
in the processes, where one has the maximal current statistics for the
production of hadrons with heavy quarks, i.e. at Fermilab and LEP colliders.
In the present section we consider the estimates for the $B_c$ meson
production in the processes of the deep inelastic scattering of neutrino
and electrons by nucleons and in $\gamma\gamma$-interactions at future
facilities.

\subsubsection{$B_c$ production in $\nu N$-interactions}

The diagrams of the neutrino-production of the $B_c$ mesons on quarks
and gluons are show on Figure \ref{n1}.
\begin{figure}[t]
\vspace*{3.5cm}
\caption{The diagrams of the $B_c$ meson production
in the processes of the neutrino scattering on gluons (a) and quarks (b)}
\label{n1}
\end{figure}

Note, that for the $B_c$ production in the neutrino collisions with
gluons, the suppression of the partonic subprocess cross section by the factor
$|V_{bc}|^2$ in comparison with the partonic subprocess of the $B_c$
neutrino-production on light quarks, is compensated by the more high luminosity
of the gluonic subprocess in comparison with the quark one, so that
the both mechanisms of the $B_c$ production in the neutrino-nucleon
scattering give the comparable contributions, and
$\sigma(\nu N\to B_c X)\approx 10^{-43}$ cm$^2$ at the neutrino energy
$E_\nu \approx$ 500--1000 {\rm GeV} in the laboratory system.
\begin{figure}[b]
\vspace*{3.5cm}
\caption{The diagrams of the $B_c^*$ meson production in
the model of vector meson dominance}
\label{n2}
\end{figure}

After the integration over the valent parton  $d$ distribution, the
$c$-quark production in the $W^{*+}d \to c$ process, suppressed as
$\sin^2{\theta_c}$, has the value, comparable with the $c$-quark
production in the $W^{*+}s\to c$ process, since the strange quark "sea" is
suppressed in respect to both the valent quark distribution and the "sea"
of the more light $d$-quark.

The estimates of the $B_c$ meson production cross sections,
calculated on the basis of the diagrams on Figure \ref{n1},
agree with the estimates, obtained in the model of the vector meson
dominance (Figure \ref{n2}) and in the model of the soft gluon emission
of the $(\bar b c)$ pair, that in the colour-singlet state and with the
low invariant mass $M(\bar b c) < M_B+M_D$, transforms, in accordance with
the quark-hadron duality, into the $(\bar b c)$ bound state, which radiatively
decays into the basic $1S_0$ state in a cascade way with the probability,
equal to 1.

As a result, one can reliably state that the total cross section for the
$B_c$ meson production in $\nu N$-collisions is of the order of $10^{-6}$
from the total cross section of the $\nu N$-scattering, so that,
at a characteristic statistics about $10^6$ events in neutrino experiments,
one can expect only several events with the $B_c$ meson production.

\subsubsection{Production of $B_c$ mesons in $ep$-scattering}

In contrast to $\nu N$-scattering, in $ep$-collisions in addition to the
processes of the weak charged current exchange, the main contribution into
the $B_c$ meson production will give the processes with the virtual
$\gamma$-quanta exchange (Figure \ref{n3}).

The exact calculation of the diagrams on Figure \ref{n3} is not yet performed
at present. However, one can think that the estimate, made in the Monte Carlo
simulation system for hadron production HERWIG [\cite{i19}], is quite reliable,
since the HERWIG parameters have been chosen to get correct values for
total hadronic cross sections of the charmed and beauty particles production,
being in an agreement with the experimental values.
Moreover, the HERWIG estimates of the $B_c$ production cross sections in
$e^+e^-$ and hadronic interactions agree with the values, obtained in the exact
calculation of the diagrams in the QCD perturbation theory.
\begin{figure}[b]
\vspace*{3.5cm}
\caption{The diagrams of the $B_c$ meson production
in parton process of $\gamma^*g$-scattering}
\label{n3}
\end{figure}

Thus, in accordance with the estimates in the HERWIG system, one can expect
about $10^3$ events per year with the $B_c$ production at the HERA facilities.
This $B_c$ yield is comparably close to that of at LEP. However, the
extraction of $B_c$ events at HERA is complicated by the presence of a
hadronic background, that is essentially lower at LEP.

\subsubsection{Photonic production of $B_c$ mesons}

At present, the future $\gamma \gamma$-colliders with the high
luminosity ($\sim 10^{34}$ $cm^{-2}s^{-1}$) are intensively discussed.
In this section we calculate the cross section of single $B_c$ production
at energies $\sqrt{s}$ about 30 GeV in accordance with the diagrams, shown on
Figure \ref{f1ph}. The calculation technique coincides with that of described
in the section on the hadronic production of $B_c$.

The total cross sections of the $B_c$ and $B_c^*$ production are presented in
Table \ref{sig}, where one takes $\alpha_S\approx 0.2 $.
One can see that near threshold the pseudoscalar state production is suppressed
in comparison with the production of the vector one, so at $\sqrt{s}=15$ GeV
one has $\sigma_{B_c^*}/ \sigma_{B_c} \sim 55$.
Such behavior of the
$\sigma_{B_c^*}/ \sigma_{B_c}$ ratio has been noted in [6], where the
strong suppression of the pseudoscalar meson pair production in respect
to the vector one takes place in the quark-antiquark annihilation.
At large energies of the initial photons this ratio decreases and becomes
$\sigma_{B_c^*}/ \sigma_{B_c} \sim 4$. The
inclusive cross sections $\sigma_{B_c}$ and $\sigma_{B_c^*}$ have the
maximum at $\sqrt{s}=20-30$~GeV and with the $s$ growth they
fall like the total cross section for the heavy quarks $\sigma_{b \bar b}$
production.
\begin{figure}[t]
\vspace*{3.5cm}
\caption{The types of diagrams in the photonic production
of $B_c$ mesons}
\label{f1ph}
\end{figure}

\begin{table}[b]
\caption{The cross section (in pb) of the photonic production
of $B_c\; (B_c^*)$}
\label{sig}
\begin{center}
\begin{tabular}{|c|c|c|c|c|}    \hline
 $ \sqrt{s}, {\rm GeV}     $ & 15  & 20  & 40  & 100  \\ \hline
 $\sigma_{B_c}$&$5.1\cdot 10^{-3}$&  $3.8\cdot 10^{-2}$&
 $6.7\cdot 10^{-2}$ &  $2.5\cdot10^{-2} $   \\   \hline
 $ \sigma_{B_c^*}$&$ 2.8\cdot 10^{-1}    $
&$6.0\cdot 10^{-1}$& $4.0 \cdot 10^{-1}$& $1.1\cdot 10^{-1}$\\
  \hline
\end{tabular}
\end{center}
\end{table}

The distributions
${\sigma^{-1}}{\rm d}\sigma/{\rm d}z$ over
the variable $z={2|{\bf p}|}/{\sqrt{s}}$, with $\bf p$ being
the meson momentum, are shown on Figures \ref{f2ph} and \ref{f3ph} for the
$B_c$  and $B_c^*$ mesons.
As follows from these figures the scaling in these distributions is
broken: with the energy growth the shift onto the low $z$ values takes
place. Note, the analogous picture has been observed in the gluonic
production of $B_c$ mesons.

Note, that the detailed consideration shows, that in the matrix element of the
$\gamma \gamma \rightarrow b \bar b c \bar c$ process and hence in the
$\gamma \gamma \rightarrow B_c \bar b  c$ matrix element one can distinguish
three groups of contributions, which are separately gauge invariant under
both the gluon field transformation and the photon one. The first group
of contributions is composed of the diagrams when the quark production
is independent (we will label these diagrams as the recombination diagrams),
the second group consist of the diagrams, where the ($c \bar c$) pair is
produced from the $b$-quark line (we will mark these diagrams as the
$b$-quark fragmentation diagrams, their contribution will be denoted as
$\sigma^{b-frag}$), the third group contains the diagrams with the
($b\bar b$) pair
production from the $c$-quark line, so that they are $c$-fragmentation
diagrams with the corresponding contributions denoted as $\sigma^{c-frag}$.
\begin{figure}[t]
\vspace*{8cm}
\caption{The cross section distributions, normalized to the
unit, over $z$ for the $B_c$ meson production at different energies}
\label{f2ph}
\end{figure}

\begin{figure}[b]
\vspace*{8cm}
\caption{The cross section distributions, normalized to the
unit, over $z$ for the $B_c^*$ meson production at different energies}
\label{f3ph}
\end{figure}

In refs.[\cite{2b,p1,p3}]  the assumption was offered that the
$b$-fragmentation
contribution has to dominate at large values of the $B_c$ transverse
momentum, independently of the type of the process.
So the approximate equation has to be valid:
\begin{equation}
\label{pt}
\frac{d\sigma^{q-frag}_{B_c}}{dP_t}=
\int \limits_{{2P_t}/{\sqrt{s}}}^{1}
\frac{d\sigma_{q \bar q}}{d k_t}(\frac{P_t}{z}) \cdot
\frac{D_{q\rightarrow B_c}(z)}{z}dz,
\end{equation}
where ${{\rm d}\sigma_{q \bar q}}/{\rm d}k_t$ is the differential cross section
for the production of the fragmenting $q$-quark in the Born approximation,
$k_t$ is its transversal momentum, and
$D_{q\rightarrow B_c}(z)$ is the function of the $q\rightarrow B_c+X$
fragmentation.

Remind that in the $e^+e^-$-annihilation the $b$-quark fragmentation dominates
and the $c$-quark fragmentation contribution is suppressed by two orders of
magnitude. In the $\gamma \gamma$-interactions, the $c$-quark fragmentation
contribution is enlarged due to the quark charge ratio $(Q_c/Q_b)^4=16$
and, therefore we can not neglect it (as one does in $e^+e^-$-annihilation).
 Note further, that the $c$-quark fragmentation
contribution and the $b$-quark fragmentation one are related to each other
by the simple permutation of the quark masses and charges
($m_c\leftrightarrow m_b$ and $Q_c\leftrightarrow Q_b$) (\ref{pt}).

The distributions ${{\rm d}\sigma}/{{\rm d}p_{\rm t}}$,
${{\rm d}\sigma^{\rm c-frag}}/{{\rm d}p_{\rm t}}$ and
${{\rm d}\sigma^{\rm b-frag}}/{{\rm d}p_{\rm t}}$ at 100 {\rm GeV}
for the $B_c$  and $B_c^*$ meson production are shown on Figures \ref{f4ph} and
\ref{f5ph}. The distributions, predicted in accordance with eq.(\ref{pt})
for the $b$-fragmentation (curve $1$) and $c$-fragmentation (curve $2$)
are also shown. One can see that, as in the hadronic production,
the contribution of the recombinational type diagram is essential at any
reasonable values of the transverse momentum of $B_c$ meson and
it can not be neglected, when one calculates the cross sections even at the
large transverse momenta. One can see from the figure, that
for the $b$-fragmentation contribution over the
$p_{\rm t}$, greater than about 30 {\rm GeV}, the fragmentational mechanism
gives correct predictions.
\begin{figure}[t]
\vspace*{8cm}
\caption{The ${\rm d}\sigma/{\rm d}p_{\rm t}$,
${\rm d}\sigma^{\rm c-frag}/{\rm d}p_{\rm t}$ and
${\rm d}\sigma^{\rm b-frag}/{\rm d}p_{\rm t}$
distributions over the transverse momentum for the invariant contributions
into the cross section of the $B_c$ meson production at 100 GeV. The curvers
$1$ and $2$ correspond to the prediction of the fragmentational mechanism
(217) for the $b$-quark ($1$) and $c$-quark ($2$)}
\label{f4ph}
\end{figure}
\begin{figure}[b]
\vspace*{8cm}
\caption{The ${\rm d}\sigma/{\rm d}p_{\rm t}$,
${\rm d}\sigma^{\rm c-frag}/{\rm d}p_{\rm t}$ and
${\rm d}\sigma^{\rm b-frag}/{\rm d}p_{\rm t}$
distributions over the transverse momentum for the invariant contributions
into the cross section of the $B_c^*$ meson production at 100 GeV. The curvers
$1$ and $2$ correspond to the prediction of the fragmentational mechanism
(217) for the $b$-quark ($1$) and $c$-quark ($2$)}
\label{f5ph}
\end{figure}
Thus, in its maximum at the energy 20-30 GeV the total cross section, including
the $B_c^*$ and corresponding antiparticle production, is about 1 pb.
This corresponds to $10^5$ $B_c$, produced at the $\gamma \gamma$-collider
with luminosity of $10^{34} cm^{-2}s^{-1}$.
At large energies, the cross section
falls like the $b\bar b$ pair production one. The $B_c$ production
mechanism is close to that of in the gluon-gluon interactions,
and it does not come to the simple $b$-quark fragmentation.

\section{Conclusion}

The discovery and study of the family of $(\bar b c)$ heavy quarkonium
with the open charm and beauty will allow one significantly to specify
the notion of the dynamics of the heavy quark interactions and the
parameters of the Standard Model of elementary particles (such values
as the $b$- and $c$-quark masses, the coupling of the $b$- and
$c$-quarks -- $|V_{bc}|$ etc.). The present review tends to the aim of
a creation of a theoretical basis for the object-directed experimental
search and study of the $(\bar b c)$ heavy quarkonium family.

Summing the considered problems, one can note the following.

We have shown that below the threshold of the ($\bar b c$) system decay
into the $BD$ meson pair, there are 16 narrow states of the $B_c$ meson
family, whose masses can be reliably calculated in the framework of the
nonrelativistic potential models of the heavy quarkonia. The flavour
independence of the QCD-motivated potentials in the region of average
distances between the quarks in the ($\bar b b$), ($\bar c c$) and
($\bar b c$) systems and their scaling properties allow one to find
the regularity of the spectra for the levels, nonsplitted by the spin-dependent
forces: in the leading approximation the state density of the system
does not depend on the heavy quark flavours, i.e. the distances between
the nL-levels of the heavy quarkonium do not depend on the heavy quark
flavours.

We have described the spin-dependent splittings of the ($\bar b c$) system
levels, i.e. the splittings, appearing in the second order over the inverse
heavy quark masses, $V_{SD}\approx O(1/m_bm_c)$, with account of the
variation of the effective Coulomb coupling constant of the quarks
(the interaction is due to relativistic corrections, coming from the one gluon
exchange).

The approaches, developed to describe emission by the heavy quarks, have been
applied to the description of the radiative transitions in the ($\bar b c$)
family, whose states have no electromagnetic or gluonic channels of
annihilation. The last fact means that, due to the cascade processes with the
emission of photons and pion pairs, the higher excitations decay
into the lightest pseudoscalar $B_c$ meson, decaying in the
weak way. Therefore, the excited states of the ($\bar b c$) system have the
widths, essentially less (by two orders of magnitude) than those in the
charmonium and bottomonium systems.

As for the value of the leptonic decay constant $f_{B_c}$, it can be the
most reliably estimated from the scaling relation for the leptonic constants
of the heavy quarkonia, due to the relation, obtained in the framework
of the QCD sum rules in the specific scheme.
In the other schemes of the QCD sum rules, it is necessary to do an
interpolation of the scheme parameters (the hadronic continuum threshold
and the number of the spectral density moment or the Borel parameter) into
the region of the ($\bar b c$) system, so this procedure leads to the essential
uncertainties. The $f_{B_c}$ estimate from the scaling relation agrees with the
results of the potential models, whose accuracy for the leptonic constants
is notably lower. The value of $f_{B_c}$ essentially determines the decay
widths and the production cross sections of the $B_c$ mesons.

The theoretical consideration of semileptonic $B_c$ decays shows, that
the results of the potential quark models agree with the predictions
of the QCD sum rules, if one accounts for the Coulomb-like
$\alpha_S/v$-corrections. In this case, the approximate spin symmetry in
the sector of heavy quarks allows one to derive the relations for the
form factors of semileptonic $B_c$ decays in the rest point of the recoil
meson.

The $B_c$ meson production allows in some cases the description on the level
of analytical expressions, such as the universal functions of the heavy quark
fragmentation into the heavy quarkonium. The fragmentational mechanism
dominates
in the $B_c$ production in the $e^+e^-$-annihilation at high energies
(in the peak of $Z$ boson) and it can be studied at the LEP facilities.

The hadronic production of $B_c$ is basically determined by the processes of
the $\bar b$- and $c$-quark recombination, since the partonic subprocesses
have the most large luminosity in the region of low invariant masses of the
produced system $(b\bar b c\bar c)$.
The $B_c$ meson yield in respect to the production of the beauty hadrons is
of the order\footnote{In the present review we do not consider in details
the $B_c$ production in the neutrino-nucleon interactions, where one can
expect only several events with the $B_c$ production per year, since the
coupling constant of the $b$- and $c$-quarks is low [\cite{7ka}],
so that these processes have no practical significance for the experimental
search of $B_c$.}
{}~of $10^{-3}$.

Modes of $B_c \to \psi X$ decays with the characteristic signature
of the $J/\psi$ particle have the quite large probability
$$ BR(B_c^+ \to \psi X) \approx 0.2\;.
$$
Therefore, the $B_c$ particle search can start from the separation of
the events, containing the $J/\psi$ particle, whose production vertex is beyond
the primary intersection point. The selected set of the events will, of course,
contain the background from decays of ordinary heavy-light
$B$ mesons $(\bar b u,\; \bar b d,\; \bar b s)$, since the probability of
the $B\to J/\psi K X$ decay is about 1 \%, and the heavy-light $B$ meson
yield is three orders of magnitude greater than the $B_c$ one.
The background separation requires the down-cut over the effective mass
of the $J/\psi X$ system, where $X$ denotes the charged particles, having
tracks from the $J/\psi$ vertex. The most preferable channel for the
$B_c$ extraction is that of the
$B_c^+ \to \psi l^+ \nu_l$ decay, since $B_c$ is the only heavy particle
with the three lepton vertex of the decay $\psi l^+ \to l'^+l'^-l^+$.
The probability of this channel is equal to
$$ BR(B_c^+\to \psi l^+ \nu_l) \approx 8\; \%\;,\;\; l=e,\; \mu,\; \tau\;.
$$
At a quite large statistics\footnote{The CDF facility with the vertex detector
at the Tevatron FNAL has, in this sense, a preferable position.},
the events with the decay $B_c^+ \to \psi l^+ \nu$  can give the possibility
for the determination of the $B_c$ mass value under the $\psi l$ mass spectrum
or the missed transverse momentum of neutrino in respect to the direction
of the $B_c$ motion (see Figure \ref{z1}). The necessary condition for the such
measurement is a quite high separation of charged hadrons and leptons.
\begin{figure}[t]
\vspace*{8cm}
\caption{The distribution over the invariant masses
of the $\psi l$ (a) and $\psi l \nu_{mis}$ (b) systems in the
$B_c^+\to \psi l^+ \nu_l$ decay, where $\nu_{mis}$ is the neutrino with
the momentum, equal to the missed transverse momentum in respect to the
direction of the $B_c$ meson motion}
\label{z1}
\end{figure}

The straightforward measurement of the $B_c$ mass can be made in the mode of
the $B_c^+ \to J/\psi \pi^+$ decay, having the branching ratio, equal to
$$ BR(B_c^+ \to \psi \pi^+) \approx 0.2\; \%\;.
$$
The mode of the $B_c^\pm \to J/\psi \pi^\pm \pi^\pm \pi^\mp$ decay,
where three $\pi$ mesons can compose the $a_1$ meson, also is of the interest.
This mode must have a significantly greater probability, than the
$B_c \to J/\psi \pi$ decay.

Since the $B_c$ production at the colliding $e^+e^-$ beams takes, as
mentioned, the fragmentational character, in general (see Figure  \ref{fp1}),
it must be accompanied by the $D$ meson presence \underline{in the same jet},
where the $B_c$ candidate is being observed. Such signature of the event
would turn out to give a large advantage for the $B_c$ meson search at
$e^+e^-$ colliders in respect to the search at hadron colliders, where
the recombinational mechanism dominates in the $B_c$ meson production at the
accessible energies in the nearest future (see Figure  \ref{fp3}).
However, one must take into account the possibility of that the probability
of the $b$-quark fragmentation production of the free
$c\bar c$-quark pair is one order of magnitude greater than the
probability of the fragmentation into the $B_c$ meson and the single free
$c$-quark. This means, that with account for the branching ratios for the
$B$  and $B_c$ decays into $J/\psi  X$, the events with the $B_c$ decay and the
single $D$ meson will appear only two times more often, than the decay of the
heavy-light $B$ meson into  $J/\psi X$ with the instantaneous production of
two $D$ mesons in the same jet. It is not clear, whether one can quite
effectively separate these two processes at the present of the vertex
detectors, i.e. whether one can not lose the vertex of the second $D$ meson.

It is evident, that the progress in the experimental study of the $B_c$ meson
and general physics of the heavy quarks will be mainly related with the
development of the vertex detectors, so that the latter would give the
possibility of a reliable observation of several heavy quarks instantaneously
(to search the cascade decays, for example). However, since at the present
statistics of LEP and Fermilab, several dozens of the $B_c$ meson production
events must be observed, one can think, that the practical registration of
$B_c$ will be realized in the nearest future.

In the conclusion the authors express their gratitude to O.P.Youshchenko
for the discussions and a help in the preparing of the paper.
One of the authors (AKL) thanks his collaborators A.V.Berezhnoy and
M.V.Shevlyagin.

This work was partially supported by the ISF grant NJQ000.

\section{Appendices}

\makeatletter
  \@addtoreset{equation}{subsection}
  \def\thesubsection{\Roman{subsection}}
  \def\theequation{\thesubsection.\arabic{equation}}
\makeatother
\setcounter{subsection}{0}
\setcounter{equation}{0}
\subsection{Covariant quark model}

Consider the general statements of the covariant description of the
composed quarkonium model.

By definition, the energy fraction, carrying out by
the quark $i$ in the ($Q \bar Q'$) meson, is its constituent mass $m_i$,
so that
\begin{equation}
M = m + m'\;,
\end{equation}
where $M$ is the meson mass, $m$ and $m'$ are the fixed values.
For the four-momenta, one has
\begin{eqnarray}
k = \frac{m}{M}\;P + q\;, \nonumber \\
k' = \frac{m'}{M}\;P - q\;,
\label{2ape}
\end{eqnarray}
where $P$ is the meson momentum, $q$ is the relative momentum of quarks
inside the meson.

For the quark propagator, one has
\begin{equation}
S(k) = (k_{\mu} \gamma^{\mu} + m)\;D(k)\;.
\label{3ape}
\end{equation}
The constituent quark has, in fact, the fixed energy, so that in the
$D(k)$ function, only the imaginary part gives the contribution. In the meson
rest frame, one has
\begin{equation}
\Im m\;D(k) = \frac{\pi}{m}\;\delta(|k_0| - m)\;.
\label{4ape}
\end{equation}
Eq.(\ref{4ape}) with account for eq.(\ref{2ape}) can be rewritten in the
covariant form
\begin{equation}
\Im m\;D(k) = \frac{\pi M}{m}\;\delta (Pq) \;.
\label{5ape}
\end{equation}

The quark-meson vertex can be represented as
\begin{equation}
L_{q\bar q M} = \bar v(k) \Gamma v'(k')\;D^{-1}(k)\;D^{-1}(k')\;\chi (P;q)\;,
\label{6ape}
\end{equation}
where $v$ and $v'$ are the quark spinors, the $D(k)$ function is defined in
eq.(\ref{3ape}),  $\Gamma$ is the spinor matrix, determining the quantum
numbers of meson.

The nonrelativistic description of the meson means, that the form factor
is determined by the expression
\begin{equation}
\chi(P;q) = 2\pi\;\delta(Pq)\;\phi (q^2)\;.
\label{8}
\end{equation}
In the following, we suppose
\begin{equation}
\phi (q^2) = N\;\exp\biggl(\frac{q^2}{\omega^2}\biggr)\;.
\label{9ape}
\end{equation}

The choice (\ref{9ape}) reflects the typical form of the $S$-wave
functions of the charmonium and bottomonium, and it allows one to
perform the analytical calculation of the semileptonic decay widths for
$B_c$ meson.

Let us define the decay constants $f$ for the pseudoscalar and vector mesons
\begin{eqnarray}
\langle 0|{ J}_{5\mu}(x)|{ P}(q)\rangle & = & i f_{ P}\;q_\mu \exp\{iqx\}\;,\\
\langle 0|{ J}_{\mu}(x)|{ V}(q,\lambda)\rangle & = & i f_{ V}\;M_{ V}\;
\epsilon_{\mu}^{(\lambda)}\;\exp\{iqx\}\;,
\end{eqnarray}
where $\lambda$ is the vector meson polarization, and the quark currents are
\begin{eqnarray}
J_{5\mu}(x) & = & \bar Q (x) \gamma_5\gamma_\mu Q'(x)\;,
\label{10ape}
\\
J_\mu (x) & =  & \bar Q (x) \gamma_\mu Q'(x)\;,
\label{11ape}
\end{eqnarray}
In the nonrelativistic potential model, one has
\begin{equation}
f_P \approx f_V = f\;,
\end{equation}
so that
\begin{equation}
f = 2 \sqrt{\frac{3}{M}}\;\Psi ({\bf 0})\;,
\label{13ape}
\end{equation}
where $\Psi ({\bf 0})$ is the quarkonium wave function at origin.
The oscillator function, resulting in eq.(\ref{9ape}), has the form
\begin{equation}
\Psi ({\bf{r}}) = \biggl(\frac{\omega^2}{2 \pi}\biggr)^{3/4}\;
\exp\biggl(-\frac{r^2\omega^2}{4}\biggr)\;.
\label{14}
\end{equation}
Condition (\ref{13ape}) means that the normalization constant $N$ in
eq.(\ref{9ape}) equals
\begin{equation}
N = \frac{M}{m\;m'}\;\frac{\sqrt{6}}{f}\;.
\end{equation}
Thus, for the quark-meson form factor, one finds
\begin{equation}
\chi (P;q) = 2 \pi\;\delta(Pq)\;\frac{M}{m\;m'}\;\frac{\sqrt{6}}{f}\;
\exp\biggl(\frac{q^2}{\omega^2}\biggr)\;,
\label{16}
\end{equation}
where $\omega$ is determined by eqs.(\ref{13ape}) and (\ref{14}),
so that the only free parameter of the model is the constant $f$.
For the $\psi$ particle, $f_{\psi}$ can be, for example, related with the
width of the $\psi \to e^+ e^-$ decay
\begin{equation}
\Gamma (\psi \to e^+ e^-) = \frac{4 \pi}{3}\;\alpha_{{\rm em}}^2\;
e_c^2\;\frac{f_{\psi}^2}{M_{\psi}}\;,
\label{17}
\end{equation}
where $e_c=2/3$ is the $с$-quark electric charge. From eq.(\ref{17}), the
experimental value of the leptonic width [\cite{1k}] gives
\begin{equation}
f_{\psi} = 410 \pm 15\;{\rm MeV}\;.
\label{18}
\end{equation}

As for the $f_{B_c}$ and  $f_{B_s}$ values, these constants are determined
theoretically in the framework of the QCD sum rules and in the
potential models.

Note, that the stated model of composed quarkonium gives, for instance,
the exact formula of the nonrelativistic M1-transition for the
electromagnetic decay of the vector state into the pseudoscalar one
$V \to P \gamma$
\begin{equation}
\Gamma(V \to P \gamma) = \frac{16}{3}\;\mu^2\;\omega_{\gamma}^3\;,
\end{equation}
where $\omega_\gamma$ is the $\gamma$-quantum energy, and the magnetic
moment $\mu$ equals
\begin{equation}
\mu = \frac{1}{2}\;\sqrt{\alpha_{{\rm em}}}\;
\biggl(\frac{e}{2 m}+\frac{e'}{2 m'}\biggr)\;,
\end{equation}
where $e$ and $e'$ are the quark electric charges in the units of the
electron charge.

\subsection{Spectral densities for three-particle functions}

The spectral densities for the three-particle functions are
determined in the following way [\cite{21k}]

\begin{eqnarray}
\rho_+(s_1,s_2,Q^2) & = & \frac{3}{2
k^{3/2}}\biggl\{\frac{k}{2}(\Delta_1+\Delta_2)
-k\bigl[m_3(m_3-m_1)+ \nonumber \\
& ~ & m_3(m_3-m_2)\bigr] -
\bigl[2(s_1\Delta_2+s_2\Delta_1)-u(\Delta_1+\Delta_2)\bigr] \nonumber \\
& ~ & \biggl[m_3^2-\frac{u}{2}+m_1 m_2 - m_2 m_3 - m_1 m_3\biggr]\biggr\}\;
\end{eqnarray}
\begin{eqnarray}
\rho_V(s_1,s_2,Q^2) & = & \frac{3}{k^{3/2}}\bigl\{(2 s_1\Delta_2 -
u\Delta_1)(m_3 - m_2) + (2 s_2\Delta_1- u \Delta_2) \nonumber \\
& ~ & (m_3-m_1) + m_3 k\biggr\}\;
\end{eqnarray}
\begin{eqnarray}
\rho_0^A(s_1,s_2,Q^2) & = & \frac{3}{k^{1/2}}\biggl\{(m_1-m_2)\biggl[m_3^2
+\frac{1}{k}(s_1\Delta^2_2 +
s_2\Delta_1^2-u\Delta_1\Delta_2)\biggr]- \nonumber   \\
& ~ & m_2\biggl(m_3^2-\frac{\Delta_1}{2}\biggr) - m_1\biggl(m_3^2
-\frac{\Delta_2}
{2}\biggr)+ \nonumber \\
& ~ & m_3\biggl[m_3^2-\frac{1}{2}(\Delta_1+\Delta_2-u) + m_1
m_2\biggr]\biggr\}\;
\end{eqnarray}
\begin{eqnarray}
&&\rho_+^A(s_1,s_2,Q^2) = \frac{3}{k^{3/2}}\biggl\{m_1\biggl[2s_2\Delta_1-
u \Delta_2+4\Delta_1\Delta_2+2\Delta_2^2\biggr] +~~~~~~~~~~~~~
{}~~~~~~~~~~~~\nonumber \\
&&~~~~~m_1m_3^2\biggl[4s_2-2u\biggr] + m_2\biggl[2s_1\Delta_2-u\Delta_1\biggr]
-
 m_3\biggl[2(3s_2\Delta_1 + s_1\Delta_2)- \nonumber \\
&&~~~~~u(3\Delta_2 + \Delta_1)
+k+4\Delta_2 \Delta_1 + 2 \Delta_2^2+m_3^2(4s_2-2u)\biggr] +\nonumber \\
&&~~~~~\frac{6}{k}(m_1 - m_3)
\biggl[4s_1s_2\Delta_1 \Delta_2 - u(2s_2\Delta_1\Delta_2+s_1\Delta_2^2 +
s_2\Delta_1^2) + \nonumber \\
&&~~~~~2s_2(s_1 \Delta_2^2 + s_2 \Delta_1^2)\biggr]\biggr\}.
\end{eqnarray}
where
\begin{eqnarray}
k & = & (s_1+s_2+Q^2)^2-4s_1s_2\;, \nonumber \\
u & = & s_1+s_2+Q^2\;, \nonumber \\
\Delta_1 & = & s_1-m_1^2+m_3^2\;, \nonumber \\
\Delta_2 & = & s_2-m_2^2+m_3^2\;. \nonumber
\end{eqnarray}
In the $B_c\rightarrow\eta_c(J/\psi)e\nu$ decays, one has
$m_1=m_b$ and $m_2=m_3=m_c$ for the masses.

\subsection{QCD sum rule scheme for three-point correlators}

Let us consider the sum rules for the $f_+ (Q^2)$ form factor
\begin{eqnarray}
 \sum_{i,j=1}^{\infty}f_{B_c}^i\frac{{M_{B_c}^i}^2}{m_b+m_c}
f_{\eta_c}^j\frac{{M_{\eta_c}^j}^2}{2m_c}f_+^{ij}(Q^2)
\frac{1}{(M^{i~2}_{B_c}-p_1^2)(M_{\eta_c}^{j~2}-p_2^2)} =  \nonumber \\
\frac{1}{(2\pi)^2}\int\frac{\rho_+(s_1,s_2,Q^2)}
{(s_1-p_1^2)(s_2-p_2^2)}{\rm d}s_1 {\rm d}s_2.
\label{B1}
\end{eqnarray}

Applying the Borel operators $\hat L_{\tau_1}(-p_1^2)$ and
$\hat L_{\tau_2}(-p_2^2)$, defined in Section 2, to eq.(\ref{B1}), one
derives the following sum rules
\begin{eqnarray}
&&\sum_{i,j=1}^{\infty}f_{B_c}^i{M_{B_c}^i}^2f_{\eta_c}^j{M_{\eta_c}^j}^2
f_+^{ij}(Q^2)
\exp({-{M_{B_c}^i}^2\tau_1-{M_{\eta_c}^j}^2\tau_2})=~~~~~~~~~~~~~~~~~~~~~\nonumber \\
&&~~~~~~~~~~~~~~=\frac{2(m_b+m_c)m_c}{(2\pi)^2}\int {\rm d}s_1{\rm
d}s_2\rho_+(s_1,s_2,Q^2)
\exp({-s_1\tau_1-s_2\tau_2})
\label{B2}
\end{eqnarray}
Introduce the notation
\begin{eqnarray}
S_i = \sum_{j=1}^{\infty}f_{\eta_c}^j{M_{\eta_c}^j}^2f_+^{ij}(Q^2)
\exp({-{M_{\eta_c}^j}^2 \tau_2})
\label{B3}
\end{eqnarray}
and transform the left hand side of eq.(\ref{B2}) with the use of
the formula by Euler--MacLaurin [\cite{AS}]
\begin{eqnarray}
&&\sum_{i=1}^{\infty}f_{B_c}^i{M_{B_c}^i}^2 S_i\exp({-{M_{B_c}^i}^2 \tau_1}) =
\int \limits_{M_{B_c}^k}^{\infty}{\rm d}M_{B_c}^n\frac{{\rm d}n}{{\rm
d}M_{B_c}^n}
f_{B_c}^n{M_{B_c}^n}^2 S_n \exp({-{M_{B_c}^n}^2 \tau_1}) \nonumber \\
&&+\sum_{n=0}^{n=k-1}f_{B_c}^n{M_{B_c}^n}^2 S_n \exp({-{M_{B_c}^n}^2 \tau_1})
+\cdots
\label{B4}
\end{eqnarray}
Acting by $\hat L_{\tau'}((M_{B_c}^k)^2)$ to eq.(\ref{B2}) and accounting for
eq.(\ref{B4}), one gets
\begin{eqnarray}
&& \sum_{j=1}^{\infty}f_{\eta_c}^j{M_{\eta_c}^j}^2f_+^{kj}(Q^2)
\exp({-{M_{\eta_c}^j}^2 \tau_2}) = ~~~~~~~~~~~~~~~~~\nonumber \\
&&~~~~~~= \frac{2m_c(m_b+m_c)}{(2\pi)^2}
\frac{{\rm d}M_{B_c}^k}{{\rm d}k}\frac{2}{M_{B_c}^kf_{B_c}^k}\int
\rho({M_{B_c}^k}^2,s_2,Q^2)
\exp(-{s_2\tau_2})
\end{eqnarray}

Making the analogous procedure for the sum of the $\eta_c^i$
resonances, one obtains
\begin{eqnarray}
f_+^{kl}(Q^2) = \frac{8m_c(m_b+m_c)}{M_{B_c}^k M_{\eta_c}^lf_{B_c}^k
f_{\eta_c}^l}\frac{{\rm d}M_{B_c}^k}{{\rm d}k}\frac{{\rm d}M_{\eta_c}^l}{{\rm
d}l}\frac{1}{(2\pi)^2}
\rho_+({M_{B_c}^k}^2,{M_{\eta_c}^l}^2,Q^2)
\end{eqnarray}
Here we have used the property of the Borel operator
$$ \hat L_\tau(x)(x^n \exp({- b x})) \rightarrow \delta_+^{(n)}(\tau - b)\;.
$$
It is not complex to generalize this procedure for the remaining
form factors.


\vspace*{0.3cm}
\hfill {\it Received September, 1994}
\newpage
\tableofcontents
\end{document}